\documentclass[a4paper,10pt,openany]{article}
\usepackage{setspace}
\usepackage{hyperref}
\usepackage{pst-3d}
\usepackage{pst-3dplot}
\usepackage{etex}
\usepackage[utf8]{inputenc}
\usepackage[english]{babel}
\usepackage{amsmath,amssymb,amsthm,mathrsfs,amsfonts,dsfont}
\usepackage{graphicx}
\usepackage[small, skip=10pt]{caption}
\usepackage{epigraph}
\usepackage{indentfirst}
\usepackage{booktabs}
\usepackage{fancyhdr}
\usepackage{vmargin}
\usepackage{feynmf}
\usepackage{yfonts}
\usepackage{lscape}
\usepackage{subfig}
\usepackage{textcomp}
\usepackage{pstricks}
\usepackage[metapost]{mfpic}
\usepackage{fancybox}
\usepackage{slashed}
\usepackage[verbose]{wrapfig}
\usepackage{pifont}
\usepackage{keystroke}
\usepackage{appendix}
\usepackage{enumitem}
\usepackage{array}
\usepackage{colortbl}
\usepackage{alltt}
\usepackage{boxedminipage}
\usepackage{tikz}
\usepackage{calc}
\usepackage{subfloat}
\usepackage{multirow}
\usepackage{cmll}
\usepackage{multicol}
\definecolor{color1}{RGB}{204,0,51}
\definecolor{color2}{RGB}{159,182,205}
\usepackage{bookmark}

\usepackage{verbatim}
\usepackage[vcentermath]{youngtab}
\usepackage{young}
\usepackage{pst-grad} 
\usepackage{pst-plot}
\usepackage{transparent}
\usepackage{color,soul}
\usepackage{cancel}
\usepackage{bm}
\usepackage{pifont}
\usepackage{placeins}
\usepackage{mathtools}
\usepackage{rotating}
\usepackage[refpage]{nomencl}
\usepackage{hhline}
\usepackage{marginnote}
\usepackage{upgreek}
\usepackage{stackengine}
\usepackage{ytableau}
\usepackage{listings}
\usepackage{calligra}
\usepackage{afterpage}
\usepackage{cite}
\usepackage{cleveref}
\usepackage{makecell}
\usepackage{accents}
\usepackage{breqn}
\usepackage{environ}

\usepackage{tikz-cd}
\usetikzlibrary{cd,decorations.markings, arrows.meta}
\tikzset{
    marrow/.style={decoration={markings,mark=at position 0.5 with {\arrow{#1}}}, postaction=decorate}
}

\usepackage[final]{showlabels}

\def\rme{\mathrm{e}}
\def\rmd{\mathrm{d}}
\def\rmR{\mathrm{R}}

\newcommand{\lr}{\left(}
\newcommand{\rr}{\right)}

\definecolor{darkergreen}{rgb}{0.0, 0.5, 0.0}

\allowdisplaybreaks

\sethlcolor{yellow}
\definecolor{boh}{RGB}{79,47,79}

\hypersetup{
        linkbordercolor={red}
}

\hypersetup{
unicode=false,          
pdftoolbar=true,        
pdfmenubar=true,        
pdffitwindow=false,     
pdfstartview={FitH},    
pdftitle={Non-Relativistic Ten-Dimensional Minimal Supergravity},    
pdfauthor={Eric Bergshoeff, Johannes Lahnsteiner, Luca Romano, Jan Rosseel, Ceyda Simsek},     
pdfsubject={},   
pdfnewwindow=true,      
colorlinks=false,       
linkcolor=red,          
citecolor=red,        
filecolor=red,      
urlcolor=red           
linkbordercolor={red},
citebordercolor={red},
urlbordercolor={red}
}

\makeatletter

\newcommand{\Rmnum}[1]{\expandafter\@slowromancap\romannumeral #1@}
\makeatother

\usepackage{alphalph}

\makeatletter
\newalphalph{\aalphalph}[mult]{\alphalph@alph}{26}
\newcommand{\alphalphval}[1]{%
\@ifundefined{c@#1}{
\aalphalph{#1}
}{%
\aalphalph{\value{#1}}
}
}
\makeatother

\usepackage{color}

\def\chapterautorefname~#1\null{Chap.~(#1)\null}
\def\sectionautorefname~#1\null{Sec.~(#1)\null}
\def\subsectionautorefname~#1\null{sub--Sec.~(#1)\null}
\def\figureautorefname~#1\null{Fig.~(#1)\null}
\def\tableautorefname~#1\null{Tab.~(#1)\null}
\def\equationautorefname~#1\null{eq.~(#1)\null}

\def\equationautorefname~#1\null{eq.~(#1)\null}

\NewEnviron{multieq}[1][2]{

\begin{multicols}{#1}
\begin{subequations}
\setlength{\abovedisplayskip}{-15pt}
\allowdisplaybreaks
\begin{align}
\BODY
\end{align}
\end{subequations}
\end{multicols}
\setlength{\parindent}{0pt}
}

\DeclareMathAlphabet\mathbfcal{OMS}{cmsy}{b}{n}

\usepackage[yyyymmdd]{datetime}

\title{\bf Non-Relativistic Ten-Dimensional Minimal Supergravity}
\date{}

\begin{document}

\begin{flushright}
\small
August 24\textsuperscript{th}, 2021\\
\normalsize
\end{flushright}
{\let\newpage\relax\maketitle}
\maketitle
\def\equationautorefname~#1\null{eq.~(#1)\null}
\def\tableautorefname~#1\null{tab.~(#1)\null}

\vspace{0.8cm}

\begin{center}
\renewcommand{\thefootnote}{\alph{footnote}}
{\sl\large E.~A.~Bergshoeff$^{~1}$}\footnote{Email: {\tt e.a.bergshoeff[at]rug.nl}},
{\sl\large J.~Lahnsteiner$^{~1}$}\footnote{Email: {\tt j.m.lahnsteiner[at]outlook.com}},
{\sl\large L.~Romano$^{~1}$}\footnote{Email: {\tt lucaromano2607[at]gmail.com}},\\[.1truecm]
{\sl\large J.~Rosseel$^{~2}$}\footnote{Email: {\tt jan.rosseel[at]univie.ac.at}} and
{\sl\large C.~\c Sim\c sek$^{~1}$}\footnote{Email: {\tt c.simsek[at]rug.nl}}
\setcounter{footnote}{0}
\renewcommand{\thefootnote}{\arabic{footnote}}

\vspace{0.5cm}
${}^1${\it Van Swinderen Institute, University of Groningen\\
Nijenborgh 4, 9747 AG Groningen, The Netherlands}\\
\vskip .2truecm
${}^2${\it Faculty of Physics, University of Vienna,\\
Boltzmanngasse 5, A-1090, Vienna, Austria }\\

\vspace{1.8cm}


{\bf Abstract}
\end{center}
\begin{quotation}
{\small
We construct a non-relativistic limit of ten-dimensional $\mathcal{N}=1$ supergravity from the point of view of the symmetries, the action, and
the equations of motion. This limit can only be realized in a supersymmetric way provided we impose by hand a set of geometric constraints, invariant under all the symmetries of the non-relativistic theory, that define a so-called `self-dual' Dilatation-invariant String Newton-Cartan geometry. The non-relativistic action exhibits three emerging symmetries: one local scale symmetry and two local conformal supersymmetries. Due to these emerging symmetries the Poisson equation for the Newton potential and two partner fermionic equations do not follow from a variation of the non-relativistic action but, instead, are obtained by a supersymmetry variation of the other equations of motion that do follow from a variation of the non-relativistic action.
We shortly discuss the inclusion of the Yang-Mills sector that would lead to a non-relativistic heterotic supergravity action.
}
\end{quotation}

\newpage

\tableofcontents

\section{Introduction}\label{sec:intro}

\noindent Recently, major progress has been made in understanding the formulation of non-relativistic (NR) string theory in a general curved background thereby generalizing the original proposal for NR string theory in a flat background \cite{Gomis:2000bd,Danielsson:2000gi} and its early extensions to special curved backgrounds \cite{Gomis:2005pg}. These new developments have taken place both in terms of a description via a two-dimensional non-linear sigma model as well as from the point of view of a target space action and equations of motion for the background fields. For the closed bosonic string these results have been obtained either by  taking a NR limit \cite{Bergshoeff:2018yvt,Bergshoeff:2019pij,Bergshoeff:2021bmc,Bidussi:2021ujm} or by applying a null reduction \cite{Harmark:2017rpg,Harmark:2018cdl,Harmark:2019upf,Bidussi:2021ujm}.
Moreover, the relation between the sigma model beta-functions and the target space equations of motion has been clarified, both  for closed and open strings,  proving the one-loop quantum consistency of the NR string theory \cite{Gomis:2019zyu,Yan:2019xsf,Gomis:2020fui,Gomis:2020izd,Gallegos:2019icg}.
There is also an intriguing relationship with Double Field Theory \cite{Gallegos:2020egk,Cho:2019ofr,Ko:2015rha,Morand:2017fnv,Blair:2020gng, Park:2020ixf}. For other  recent work on NR string theory in a curved background, see \cite{Kluson:2018egd,Kluson:2018vfd,Kluson:2019ifd,Roychowdhury:2019qmp,Hartong:2021ekg}.

At first sight, the natural target space geometry of the NR string theory of \cite{Gomis:2000bd,Danielsson:2000gi} generalized to arbitrary backgrounds is given by a Newton-Cartan-like geometry with co-dimension two foliation that is characterized by the following `zero torsion constraint' on the longitudinal Vielbein $\tau_\mu{}^A$:\footnote{In this paper we only consider `stringy limits'  where the  longitudinal directions are scaled differently from the  transverse directions. We will not consider `particle' limits like in \cite{Batlle:2016iel, Bergshoeff:2017btm}.}
\begin{equation}\label{stringconstraint}
D_{[\mu}(\omega)\tau_{\nu]}{}^A=0\,.
\end{equation}
Here, the index $A=0,1$ refers to the directions longitudinal to the string and the derivative $D_\mu(\omega)$ is covariant with respect to longitudinal Lorentz transformations. Since we are working in the second order formalism part of the constraints \eqref{stringconstraint} are identically satisfied. To obtain the genuinely geometric constraints one should project \eqref{stringconstraint} onto those components where the spin connection cancels out:
\begin{align} \label{stringconstraint2}
  e_{A^\prime}{}^\mu \tau_{(A|}{}^\nu \partial_{[\mu} \tau_{\nu]| B)} =0\,,\qquad\mathrm{and}\qquad e_{A^\prime}{}^\mu e_{B^\prime}{}^\nu \partial_{[\mu} \tau_{\nu]}{}^A =0\,.
\end{align}
Here $A'$ refers to the directions transverse to the string, $\tau_A{}^\mu$, $e_{A^\prime}{}^\mu$ are (projective) inverses of the longitudinal and transverse Vielbeine $\tau_\mu{}^A$,\,$e_\mu{}^{A^\prime}$ and $(AB)$ indicates the symmetric part of $AB$.
The geometry defined  by the zero torsion constraint \eqref{stringconstraint} is referred to as
String Newton-Cartan (SNC) geometry \cite{Andringa:2012uz}.\,\footnote{For earlier work on SNC geometry, see \cite{Gomis:2005pg,Brugues:2004an,Brugues:2006yd}.}
The NR string then couples to the SNC Vielbeine, as well as to a Kalb-Ramond (KR) and dilaton background field.

In our recent paper \cite{Bergshoeff:2021bmc}, we studied the target space action and equations of motion of the NS-NS sector of NR string theory, from the viewpoint of taking a NR limit of the relativistic action and equations of motion. The resulting NR NS-NS action has also been derived from a Double Field Theory point of view in \cite{Gallegos:2020egk}. We showed in particular that in the NR case a natural geometric constraint, consistent with (part of) the target space equations of motion, is not given by the  constraints \eqref{stringconstraint2} of SNC geometry, but by the weaker dilatation-invariant geometric constraints\,\footnote{These constraints can formally be obtained by replacing the covariant derviative in \eqref{stringconstraint} by a dilatation covariant derivative $D_\mu(\omega,b)$ where $b_\mu$ is the dilatation gauge field. The explicit form \eqref{TTSintro} then follows after projecting to those components where both gauge fields cancel out. The second constraint in \eqref{TTSintro} can also be written as $\epsilon_{AB}\tau^A\wedge\tau^B\wedge\rmd\tau^C=0$, which shows that it is sufficient to define an integrable co-dimension two foliation, see e.g. \cite{frankel_2011}.}
\begin{align} \label{TTSintro}
  e_{A^\prime}{}^\mu \tau_{\{A|}{}^\nu \partial_{[\mu} \tau_{\nu]| B\}} =0\,,\qquad\mathrm{and}\qquad e_{A^\prime}{}^\mu e_{B^\prime}{}^\nu \partial_{[\mu} \tau_{\nu]}{}^A =0\,,
\end{align}
where   $\{AB\}$ indicates the symmetric traceless part of $AB$. The constraints \eqref{TTSintro} characterize what we called a `Dilatation invariant String Newton-Cartan' (DSNC) geometry in \cite{Bergshoeff:2021bmc}. We stress that these constraints are not imposed by hand neither do they follow from taking the NR limit  but they follow as natural solutions of (part of) the target space equations of motion. The geometry before imposing any constraints is referred to as Torsional String Newton-Cartan (TSNC) geometry \cite{Bergshoeff:2021bmc,Bidussi:2021ujm}.

The NR limit of the relativistic equations of motion for the metric, Kalb-Ramond and dilaton background fields can be taken such that it preserves the number of independent equations of motion. However, not all of the resulting NR equations follow from a variation of the NR action that is obtained by taking the NR limit of the relativistic one. A distinguishing feature of this NR NS-NS target space action  is that it is invariant under an emerging local scale symmetry that is absent in the relativistic case. Due to this emerging symmetry there is one equation, the Poisson equation for the Newton potential, that does not follow from the NR target space action. Instead, this equation constitutes, together with all the other equations that do follow from the NR target space action, a so-called reducible but indecomposable representation of the NR symmetries. For the Poisson equation this means that, by varying it under Galilean boosts, one can generate the full set of equations but none of these other equations transforms back under Galilean boosts to the Poisson equation. A similar story applies to the beta-functions. The NR string sigma model is also invariant under an emerging dilatation symmetry and consequently the number of beta-functions that one can calculate is one less than the number of relativistic equations of motion. Setting these beta-functions to zero one finds that a certain, purely nonlinear equation that has a dilatation weight opposite to that of the Poisson equation is missing. For an interpretation of this missing nonlinear equation, see the recent paper \cite{Yan:2021lbe}. Schematically, the situation can be summarized as follows:
\begin{eqnarray}
\textrm{NR e.o.m.} &\rightarrow &\ \ \textrm{ common equations \ + Poisson\ +\ Non-linear}\,,\\[.1truecm]
\textrm{NR NS-NS action} &\rightarrow &\ \ \textrm{ common equations \ +\ Non-linear}\,,\\[.1truecm]
\textrm{NR}\ \beta\textrm{-functions} &\rightarrow &\ \ \textrm{ common equations \ + Poisson}\,.
\end{eqnarray}
One might worry that the theory becomes overdetermined by changing the number of independent degrees of freedom---through the emergence of the dilatation gauge symmetry---while leaving the number of independent equations of motion unchanged. This, however, is avoided since the linearization of one of the equations becomes trivial. It would be interesting to get a systematic understanding of the interplay of the emergence of symmetries and the differential structure of the equations of motion after taking the non-relativistic limit. The fact that the NR NS-NS action does not lead to the Poisson equation for the Newton potential is consistent with the fact that no action principle is known for NC gravity based upon the Bargmann algebra.\,\footnote{For suggestions of such an action based on a larger algebra, see \cite{Hansen:2019pkl}.}

So far, most calculations have been performed for the bosonic NR string only.\,\footnote{For earlier work on NR strings and supersymmetry, see \cite{Gomis:2004pw,Gomis:2005pg}.} This work is a companion to our previous paper where we enlarge our investigations of the NR NS-NS gravity background to the case of a NR minimal supergravity background.\,\footnote{The NR supergravity theory is minimal in the sense that, although it contains {\it two} independent supersymmetries, both are needed to obtain translations along the longitudinal directions as the result of an anti-commutator of two supercharges.} To be specific, we will present the  NR limit of the ten-dimensional  $\mathcal{N}=1$  supergravity action and equations of motion defining  the dynamics of the background fields. This sector is common to all superstring theories. We have a heterotic superstring in mind but we will postpone a discussion of the Yang-Mills sector to the conclusions. To obtain the results of this paper, we will follow the same strategy as used in \cite{Bergshoeff:2021bmc} but there are notable new features in the supersymmetric case. One complication is that, unlike in the bosonic case, there is no direct connection between a two-dimensional sigma model description and the NR target space effective action. A Green-Schwarz sigma model formulation for the  NR superstring has been given for a flat background \cite{Gomis:2004pw} but not for a NR minimal supergravity background. Like in the relativistic case, this will probably require a superspace formulation. Alternatively, starting from a NR  sigma model with (1,0) world-sheet supersymmetry\,\footnote{For the recent construction of a NR sigma model with (1,1) worldsheet supersymmetry, see \cite{Blair:2019qwi}.}, the target space supersymmetry of the background fields is not manifest. In both cases we cannot use the sigma model description to read off the emergent target space fermionic St\"uckelberg symmetries that we expect to team up with the emergent local scale symmetry that we found in the bosonic case.

Another complication, not encountered in the bosonic case,  is that taking the naive NR limit of the supersymmetry rules leads to divergent terms in these transformation rules. Concerning the action, we find that, like in the bosonic case, a NR limit of the $\mathcal{N}=1$ supergravity action can be defined due to a miraculous cancellation of divergent terms when  taking this  limit.
By performing a careful analysis of the NR limit, we will show in this paper that the dangerous divergent terms in the supersymmetry rules are controlled by two facts about the theory. Firstly, we are making use of the fact that the NR action is invariant under two emergent local fermionic St\"uckelberg symmetries arising as partners of the emergent local scale symmetry that we already found in the bosonic case. We will call the two emerging fermionic symmetries $S$- and $T$-supersymmetry where the $S$-supersymmetry is of a type that is also encountered in conformal supergravity. Secondly, we are imposing by hand the following constraints on the geometry:
\begin{align} \label{selfdualintro}
  e_{A^\prime}{}^\mu \tau_{+}{}^\nu \partial_{[\mu} \tau_{\nu]}{}^- =0\,,\qquad\mathrm{and}\qquad e_{A^\prime}{}^\mu e_{B^\prime}{}^\nu \partial_{[\mu} \tau_{\nu]}{}^- =0\,.
\end{align}
Here, $\pm$ refer to (anti-)selfdual projections $\tau_\mu{}^\pm = 2^{-1/2}(\tau_\mu{}^0\pm\tau_\mu{}^1)$ in the two longitudinal directions. The constraints \eqref{selfdualintro} constitute half of the constraints \eqref{TTSintro} defining a DSNC geometry and define  what we will call a `self-dual' DSNC geometry. Importantly these constraints are invariant under all the symmetries of the NR theory---including supersymmetry---and therefore do not lead to additional constraints. They can be substituted into the equations of motion but not into the NR action. In that sense the NR minimal supergravity action is a so-called pseudo-action.

As we will show in this paper, the self-dual constraints \eqref{selfdualintro} are in fact needed to show the consistency of the NR theory. More precisely, they are a necessary requirement for the closure of the non-relativistic superalgebra and for making sure that the set of NR equations of motion is closed under supersymmetry. To better understand the geometric meaning of \eqref{selfdualintro}, it is useful to rewrite them as
\begin{align}\label{eq:selfdualform}
    \tau_{[\mu}{}^-\partial_{\nu}^{}\tau_{\rho]}{}^- = 0\,,
\end{align}
which shows that \eqref{selfdualintro} defines an integrable co-dimension one foliation along the lightcone direction $\tau_-{}^\mu\partial_\mu$. This, in turn, implies that one can choose coordinates such that $\tau_\mu{}^- = \rme^{\kappa}\partial_\mu t$ for some $\kappa=\kappa(x^\mu)$ and $t=t(x^\mu)$. This corresponds to the twistless torsional constraints of ordinary Newton-Cartan geometry, encountered in the literature, see e.g. \cite{Christensen:2013lma}. We note, however, that the equations of motion for the background fields can lead to further torsion constraints on the curls of $\tau_\mu{}^-$ and $\tau_\mu{}^+$. The final background geometry can only be determined after the constraints that follow from these equations of motion have been taken into account and can take the form of a co-dimension two foliation.

Due to the emergent bosonic and fermionic local symmetries we find that the NR  action does not give rise to the full set of equations of motion. There is a  Poisson equation for the Newton potential and there are  two additional fermionic equations that do not follow from the variation of the NR action. However, unlike in the bosonic case, the NR action knows indirectly about these three missing equations in the sense that they can be obtained by varying the other equations of motion that follow from the NR action under supersymmetry.

The organization of this paper is as follows. In section 2 we give a brief review of the relativistic $\mathcal{N}=1$  supergravity theory together with the transformation rules of all fields in a new basis of the fields that contains powers of $c$ for finite $c$, i.e., before taking the actual NR limit. In the next section we discuss in detail the NR limit of the relativistic  supergravity action  ending up with a NR action that has emergent dilatations plus an emerging $S$- and $T$-supersymmetry.  In section 4, we take the  NR limit of the  equations of motion and show that we obtain the same equations of motion that follow from varying the NR action derived in the previous section  plus three more equations: the Poisson equation for the Newton potential together with two fermionic equations. Furthermore, we show how these three missing equations of motion are connected to the  ones that do follow from the variation of the NR action  via supersymmetry. In the final section we discuss our results. In particular, we mention  a few subtleties when including the Yang-Mills sector of a heterotic supergravity theory. There are 5 appendices.
Our notations and conventions particular to the supersymmetric case are given in appendix \ref{sec:conventions}. In appendix \ref{sec:TSNC} we collect a few useful formulae describing Torsional String Newton-Cartan Geometry. This is the generic background geometry of non-relativistic string theory, and the self-dual DSNC geometry \eqref{TTSintro} is a special case. To make this paper more user-friendly for those  who wish to investigate compactifications or solutions of the NR superstring we have summarized in appendix \ref{sec:KSeqs} the bosonic  equations of motion with the fermions set equal to zero together with the Killing spinor equations. Appendix \ref{sec:algebra} contains details on the supersymmetry algebra that underlies the NR supergravity theory of this paper. Finally, in the last appendix \ref{sec:NRSYM} we show how the NR limit can be defined for the special case of a supersymmetric Yang-Mills theory in a flat background.

\section{\boldmath $D=10$, $\mathcal{N}=1$ Supergravity} \label{sec:Nisonesugra}

\noindent In this section, we will briefly review ten-dimensional $\mathcal{N}=1$ supergravity \cite{Bergshoeff:1981um,Chamseddine:1980cp}, i.e., the common part of the effective theories for the massless modes of all superstrings. We will summarize its fields and their transformation rules, as well as its action. In order to define the NR limit, one performs an invertible field redefinition that involves a (dimensionless) parameter $c$, such that the NR limit corresponds to sending $c \rightarrow \infty$. To facilitate taking this limit in the next sections, we will here also give the field redefinition that is involved and apply it to the transformation rules of $\mathcal{N}=1$ supergravity. The index, spinor, and Clifford algebra conventions that we use throughout this paper are collected in appendix \ref{sec:conventions}.

The bosonic field content of ten-dimensional $\mathcal{N}=1$ supergravity consists of the Vielbein $E_\mu{}^{\hat A}$, the Kalb-Ramond (KR) two-form field $B_{\mu\nu}$, and the dilaton $\Phi$. The fermionic fields are given by the gravitino $\Psi_\mu$, and the dilatino $\uplambda$. Here, $\Psi_\mu$ is a left-handed Majorana-Weyl spinor, while $\uplambda$ is a right-handed one. The action of $\mathcal{N}=1$ supergravity is then given by
\begin{align} \label{eq:Nisoneaction}
S = \frac{1}{2\kappa^2}\int \rmd^{10}x\,E\,\rme^{-2\Phi}\bigg\{&\mathcal R+ 4\,\partial_\mu\Phi\,\partial^\mu\Phi - \frac{1}{12}\,\mathcal H_{\mu\nu\rho}\mathcal H^{\mu\nu\rho}-2\,\bar\Psi_\mu\Gamma^{\mu\nu\rho}D_\nu\Psi_\rho - 4\,\bar\uplambda\,\Gamma^{\mu\nu}D_\mu\Psi_\nu \notag\\
& + 2\,\bar\uplambda\,\slashed D\uplambda +\frac{1}{24}\mathcal H^{\rho\sigma\tau}\Big(2\,\bar\Psi_\mu\Gamma^{[\mu}\Gamma_{\rho\sigma\tau}\Gamma^{\nu]}\Psi_\nu - 4\,\bar\Psi_\mu\Gamma^\mu{}_{\rho\sigma\tau}\uplambda - 2\,\bar\uplambda\,\Gamma_{\rho\sigma\tau}\uplambda\Big)\notag\\
&-4\,\bar\Psi_\mu\slashed\partial\Phi\Gamma^\mu\uplambda - 4\,\bar{\Psi}_\mu\Gamma^\mu\Psi_\nu\,\partial^\nu\Phi \ \ ( + \text{ quartic fermion terms})\bigg\}\,,
\end{align}
where $\kappa$ denotes the gravitational coupling constant and $E = \mathrm{det}(E_\mu{}^{\hat{A}})$. The Ricci scalar $\mathcal{R}$ is constructed from the Levi-Civita spin connection $\Omega_\mu{}^{\hat{A}\hat{B}}$ and
\begin{align}
\label{eq:defH}
\mathcal{H}_{\mu\nu\rho} = 3 \partial_{[\mu} B_{\nu\rho]} \,,
\end{align}
is the field strength of the KR field. We moreover defined the (anti-symmetrized when necessary) covariant derivatives of $\Psi_\mu$ and $\uplambda$ by
\begin{align}
D_{[\mu} \Psi_{\nu]} &= \partial_{[\mu} \Psi_{\nu]} - \frac14 \Omega_{[\mu|}{}^{\hat{A}\hat{B}} \Gamma_{\hat{A} \hat{B}} \Psi_{|\nu]} \,, \qquad \qquad D_\mu \uplambda = \partial_\mu \uplambda - \frac14 \Omega_{\mu}{}^{\hat{A}\hat{B}} \Gamma_{\hat{A} \hat{B}} \uplambda \,.
\end{align}
Note that $\Omega_\mu{}^{\hat{A}\hat{B}}$, $\mathcal{R}$ and $\mathcal{H}_{\mu\nu\rho}$ in the action \eqref{eq:Nisoneaction} do not contain any fermionic contributions (such as supercovariantizations). The first three terms of \eqref{eq:Nisoneaction} are thus purely bosonic, and only the remaining terms contain fermions. We have not explicitly written the quartic fermion terms that are present in \eqref{eq:Nisoneaction}. In this paper, we will consistently truncate quartic fermion terms in actions, and we will only give the terms in the supersymmetry transformation rules that are consistent with this truncation.

The fields of ten-dimensional $\mathcal{N}=1$ supergravity transform as follows under local Lorentz transformations with parameter $\Lambda^{\hat{A}\hat{B}}$, a one-form symmetry of the KR field with parameter $\Theta_\mu$ and supersymmetry with a left-handed Majorana-Weyl spinor parameter $\varepsilon$:
\begin{subequations}\label{eq:trafosrel}
\begin{align}
\delta E_\mu{}^{\hat{A}} &= \Lambda^{\hat{A}}{}_{\hat{B}} E_\mu{}^{\hat{B}} + \bar{\varepsilon}\,\Gamma^{\hat{A}}\Psi_\mu\,, \qquad \qquad
\delta B_{\mu\nu} = 2 \partial_{[\mu} \Theta_{\nu]} + 2\,\bar{\varepsilon}\,\Gamma_{[\mu}\Psi_{\nu]}\,,  \qquad \qquad
\delta \Phi = \frac{1}{2}\,\bar{\varepsilon}\,\uplambda\,, \label{eq:trafosVielBDil} \\
\delta \Psi_\mu &= \frac{1}{4} \Lambda^{\hat{A}\hat{B}} \Gamma_{\hat{A}\hat{B}} \Psi_\mu + D_\mu(\Omega^{(+)}) \varepsilon \ \, (+ \text{ terms quadratic in } \Psi_\mu \text{ and } \uplambda) \,, \label{eq:trafosGravo} \\
\delta \uplambda &= \frac{1}{4} \Lambda^{\hat{A}\hat{B}} \Gamma_{\hat{A}\hat{B}} \uplambda + \Gamma^\mu\varepsilon\,\mathcal \partial_\mu \Phi - \frac{1}{12}\,\Gamma^{\hat{A}\hat{B}\hat{C}}\varepsilon\, \mathcal{H}_{\hat{A}\hat{B}\hat{C}}\ \, (+ \text{ terms quadratic in } \Psi_\mu \text{ and } \uplambda) \,, \label{eq:trafosDilo}
\end{align}
\end{subequations}
where we have defined the following torsionful covariant derivative of $\varepsilon$
\begin{align} \label{eq:DOmegapm}
D_\mu(\Omega^{(+)})\varepsilon &= \partial_\mu\varepsilon - \frac14\,\Omega^{(+)}_\mu{}^{\hat A\hat B}\Gamma_{\hat A\hat B}\varepsilon\,, \qquad \text{with} \qquad \Omega^{(+)}_\mu{}^{\hat A\hat B} = \Omega_\mu{}^{\hat A\hat B} + \frac12 \mathcal{H}_\mu{}^{\hat A\hat B} \,.
\end{align}

In order to take the NR limit in the following sections, we introduce a (dimensionless) parameter $c$ and perform the following field redefinition
\begin{alignat}{4}\label{eq:rescale}
\tau_\mu{}^A &= c^{-1}\,E_\mu{}^A\,, \qquad \ \ & e_\mu{}^{A'} &= E_\mu{}^{A'}\,, \qquad \ \ & b_{\mu\nu} &= B_{\mu\nu}+\epsilon_{AB}\,E_\mu{}^A E_\nu{}^B\,, \qquad \ \ & \phi &= \Phi - \log c\,,\notag\\
\psi_{\mu\pm} &= c^{\mp 1/2}\Pi_{\pm}\Psi_\mu\,,& \lambda_\pm &= c^{\mp 1/2}\Pi_\pm\uplambda\,,
\end{alignat}
where we have split the Lorentz index $\hat{A}$ into a longitudinal index $A=0,1$ and a transversal index $A'=2,\cdots,9$. Note that the redefinition of the spinor fields involves the `worldsheet chirality' projection operators $\Pi_\pm$, that are defined in \eqref{eq:Piproj} \cite{Gomis:2004pw}. We refer to appendix \ref{ssec:fermconventions} for various properties that are obeyed by worldsheet chirality projected spinors and that are used throughout this paper. For the bosonic fields, the above redefinition coincides with the one used in \cite{Bergshoeff:2021bmc} to derive the NR limit of NS-NS gravity.

As will be seen in the next sections, the fields $\tau_\mu{}^A$, $e_\mu{}^{A'}$, $b_{\mu\nu}$, $\phi$, $\psi_{\mu\pm}$ and $\lambda_{\pm}$ correspond, after taking the limit $c \rightarrow \infty$, to the fields of NR ten-dimensional minimal supergravity. As explained in \cite{Bergshoeff:2021bmc}, in order to calculate the transformation rules of these fields in the NR theory, it is important that the field redefinition \eqref{eq:rescale} is invertible. This is the case and the inverse of \eqref{eq:rescale} is given by
\begin{alignat}{3}\label{eq:rescaleinv}
E_\mu{}^A &= c\,\tau_\mu{}^A\,,\qquad  E_\mu{}^{A'} = e_\mu{}^{A'}\,,\qquad
& B_{\mu\nu} &= -c^2 \, \epsilon_{AB}\,\tau_\mu{}^A \tau_\nu{}^B + b_{\mu\nu}\,,\qquad & \Phi &= \phi + \log c\,,\notag\\
\Psi_{\mu} &= c^{1/2} \psi_{\mu +} + c^{-1/2} \psi_{\mu -} \,,\qquad & \uplambda &= c^{1/2} \lambda_+ + c^{-1/2} \lambda_- \,,
\end{alignat}
where it is understood that $\psi_{\mu\pm}$ and $\lambda_\pm$ are worldsheet chirality projected spinors (i.e., obey $\psi_{\mu\pm} = \Pi_\pm \psi_{\mu\pm}$ and $\lambda_\pm = \Pi_\pm \lambda_\pm$). It is also useful to introduce objects $\tau_A{}^\mu$ and $e_{A'}{}^\mu$ as the following (invertible) redefinitions of components of the inverse Vielbein $E_{\hat{A}}{}^\mu$:
\begin{align}
\label{eq:invErescale}
\tau_A{}^\mu = c\, E_A{}^\mu \,, \qquad e_{A'}{}^\mu = E_{A'}{}^\mu\,.
\end{align}
The $\tau_\mu{}^A$, $e_\mu{}^{A'}$, $\tau_A{}^\mu$, $e_{A'}{}^\mu$ then satisfy the following `projective invertibility' relations:
\begin{align}
\label{eq:projinv}
\tau_B{}^\mu \tau_\mu{}^A &= \delta_B{}^A \,, \qquad \qquad  e_{B'}{}^\mu e_{\mu}{}^{A'} = \delta_{B'}{}^{A'} \,, \qquad \qquad  \tau_A{}^\mu e_\mu{}^{A'} = 0 \,, \nonumber \\
e_{A'}{}^\mu \tau_\mu{}^A &= 0 \,, \qquad \qquad \tau_\mu{}^A \tau_A{}^\nu + e_\mu{}^{A'} e_{A'}{}^\nu = \delta^\nu_\mu \,.
\end{align}

To set the stage for our derivation of the NR limit of the action and equations of motion of ten-dimensional $\mathcal{N}=1$ supergravity, we will end this section by applying the above field redefinition \eqref{eq:rescale}, \eqref{eq:rescaleinv} to the transformation rules \eqref{eq:trafosrel}. To do this, we introduce an analogous invertible redefinition of the parameters $\Lambda^{\hat{A}\hat{B}} = \left(\Lambda^{AB}, \Lambda^{A A'}, \Lambda^{A' B'}\right)$, $\Theta_\mu$ and $\varepsilon$ of local Lorentz transformations, the KR one-form symmetry and supersymmetry:
\begin{align}\label{eq:parresc}
\lambda^{A'B'} &= \Lambda^{A'B'}\,, \qquad \quad  \lambda^{AA'} = c\,\Lambda^{AA'}\,,\qquad \quad \lambda_M\epsilon^{AB} = \Lambda^{AB}\,, \qquad \quad \theta_\mu = \Theta_\mu \notag\\
\epsilon_\pm &= c^{\mp 1/2}\Pi_{\pm}\varepsilon \quad \Leftrightarrow \quad \varepsilon = c^{1/2} \epsilon_+ + c^{-1/2} \epsilon_- \quad (\text{with } \Pi_\pm \epsilon_\pm = \epsilon_\pm) \,.
\end{align}
After taking the $c\rightarrow \infty$ limit, $\lambda_M$ will correspond to the parameter of longitudinal Lorentz transformations, $\lambda^{AA'}$ to the parameter of Galilean boosts and $\lambda^{A'B'}$ to the parameter of transversal rotations, while the parameters $\epsilon_\pm$ will be those of non-relativistic supersymmetry.

Using the redefinitions \eqref{eq:rescale}, \eqref{eq:rescaleinv} and \eqref{eq:parresc}, one can easily find how $\tau_\mu{}^A$, $e_\mu{}^{A'}$, $b_{\mu\nu}$, $\phi$, $\psi_{\mu\pm}$ and $\lambda_{\pm}$ transform under the symmetries with parameters $\lambda_M$, $\lambda^{AA'}$, $\lambda^{A'B'}$, $\theta$ and $\epsilon_\pm$. Considering first the transformation rules under the bosonic symmetries, one finds
\begin{subequations}\label{eq:LorsymmBOS}
\begin{align}
\delta\tau_\mu{}^A & = \lambda_M\epsilon^A{}_B\tau_\mu{}^B + \frac{1}{c^2}\,\lambda^A{}_{A'}e_\mu{}^{A'}\,, & \delta e_\mu{}^{A'} &= \lambda^{A'}{}_{B'}e_\mu{}^{B'} - \lambda_{A}{}^{A'} \tau_\mu{}^A\,,\\
\delta b_{\mu\nu} &= 2\,\partial_{[\mu}\theta_{\nu]} - 2\,\epsilon_{AB}\lambda^A{}_{A'}\tau_{[\mu}{}^B\,e_{\nu]}{}^{A'}\,, & \delta\phi &= 0\,.
\end{align}
\end{subequations}
for the bosonic fields and
\begin{subequations}\label{eq:LorsymmFER}
\begin{align}
\delta\psi_{\mu+} &= \frac14\,\big(\lambda^{A'B'}\Gamma_{A'B'}-2\,\lambda_M\big)\psi_{\mu +} + \frac{1}{2\,c^2}\,\lambda^{AA'}\Gamma_{AA'}\psi_{\mu -}\,, \\
\delta\psi_{\mu-} &= \frac14\,\big(\lambda^{A'B'}\Gamma_{A'B'}+2\,\lambda_M\big)\psi_{\mu -} + \frac{1}{2}\,\lambda^{AA'}\Gamma_{AA'}\psi_{\mu +}\,,\\
\delta\lambda_+ &= \frac14\,\big(\lambda^{A'B'}\Gamma_{A'B'}-2\,\lambda_M\big)\lambda_+ + \frac{1}{2\,c^2}\,\lambda^{AA'}\Gamma_{AA'}\lambda_-\,,\\
\delta\lambda_- &= \frac14\,\big(\lambda^{A'B'}\Gamma_{A'B'} + 2\,\lambda_M\big)\lambda_- + \frac{1}{2}\,\lambda^{AA'}\Gamma_{AA'}\lambda_+\,,
\end{align}
\end{subequations}
for the fermionic fields. Note that for both sets of transformation rules \eqref{eq:LorsymmBOS}, \eqref{eq:LorsymmFER}, the limit $c \rightarrow \infty$ is well-defined.

One can similarly find the transformation rules under supersymmetry (with parameters $\epsilon_\pm$). For the bosonic fields $\tau_\mu{}^A$, $e_\mu{}^{A'}$, $b_{\mu\nu}$, $\phi$, one finds, upon using that certain bilinears with spinors of definite worldsheet chirality are identically zero (see e.g. \eqref{eq:projbil}), that
\begin{subequations} \label{eq:bossusy}
\begin{align}
\delta \tau_\mu{}^A &=\bar{\epsilon}_{+}\Gamma^{A}\psi_{\mu +} + \frac{1}{c^2}\,\bar\epsilon_-\Gamma^A\psi_{\mu -}\,,\\
\delta e_{\mu}{}^{A'} &= \bar{\epsilon}_{+}\Gamma^{A'}\psi_{\mu -}+\bar{\epsilon}_{-}\Gamma^{A'}\psi_{\mu +}\,,\\
\delta \phi &=\frac{1}{2}(\bar{\epsilon}_{+}\lambda_{-}+\bar{\epsilon}_{-}\lambda_{+})\,,\\
\delta b_{\mu\nu}&= 4\,\tau_{[\mu}{}^{A}\bar{\epsilon}_{-}\Gamma_{A}\psi_{\nu]-} + 2\Big(e_{[\mu}{}^{A'}\bar\epsilon_+\Gamma_{A'}\psi_{\nu]-}+e_{[\mu}{}^{A'}\bar{\epsilon}_{-}\Gamma_{A'}\psi_{\nu]+}\Big)\,.
\end{align}
\end{subequations}
As for \eqref{eq:LorsymmBOS} and \eqref{eq:LorsymmFER}, the $c \rightarrow \infty$ limit of these transformations is regular. The derivation of the $\epsilon_\pm$ supersymmetry transformation rules of $\psi_{\mu\pm}$ and $\lambda_{\pm}$ is straightforward, but leads to more lengthy expressions that involve powers of $c^2$, $c^0$ and $c^{-2}$. We collect terms with like powers of $c$ as follows:
\begin{subequations}\label{eq:susyexpansion}
\begin{align}
\delta\psi_{\mu\pm} &= c^2\, \delta^{(2)}\psi_{\mu\pm} + c^0\,\delta^{(0)}\psi_{\mu\pm} + c^{-2}\,\delta^{(-2)}\psi_{\mu\pm}\,,\\
\delta\lambda_{\pm} &= c^2\,\delta^{(2)}\lambda_{\pm} + c^0\,\delta^{(0)}\lambda_{\pm} + c^{-2}\,\delta^{(-2)}\lambda_{\pm}\,.
\end{align}
\end{subequations}
Explicitly, the terms that appear at order $c^2$ are given by
\begin{subequations}\label{eq:divsupersymmetry}
\begin{align}
\delta^{(2)}\psi_{\mu + } &= \frac{1}{2}\,\tau_\mu{}^+\tau^{A'B'-}\Gamma_{A'B'}\epsilon_+\,,\\
\delta^{(2)}\psi_{\mu - }  &= \frac{1}{2}\,\tau_\mu{}^+\big(\tau^{A'B'-}\Gamma_{A'B'}\epsilon_- - \tau^{A'--}\Gamma_{-A'}\epsilon_+ \big)\,,\\
\delta^{(2)}\lambda_+ &= 0\,,\\
\delta^{(2)}\lambda_- &= -\frac{1}{2}\tau^{A'B'-}\Gamma_{A'B'-}\epsilon_+ \,,
\end{align}
\end{subequations}
where $\tau_{\mu\nu}{}^A = \partial_{[\mu} \tau_{\nu]}{}^A$ and we refer to appendix \ref{ssec:bosconventions} for details on how curved $\mu$, $\nu$ indices are converted into flat longitudinal and transversal ones and on how flat light-cone indices ($A$, $B = +,-$) are used to denote longitudinal directions (as an alternative to $A$, $B = 0,1$).

The terms in \eqref{eq:susyexpansion} at order $c^0$ can be written in terms of composite fields $b_\mu$, $\omega_\mu$, $\omega_\mu{}^{AA'}$, $\omega_\mu{}^{A'B'}$ that depend on the bosonic fields $\tau_\mu{}^A$, $e_\mu{}^{A'}$, $b_{\mu\nu}$ and $\phi$. Their explicit expressions can be found in appendix \ref{sec:TSNC}. These composite fields correspond to the dependent dilatation and spin connections of the Torsional SNC (TSNC) geometry, that was introduced in \cite{Bergshoeff:2021bmc}. In particular, after taking the NR $c \rightarrow \infty$ limit, $\omega_\mu$, $\omega_\mu{}^{AA'}$ and $\omega_\mu{}^{A'B'}$ will correspond to spin connections for longitudinal SO$(1,1)$ Lorentz transformations, Galilean boosts and SO$(8)$ transversal rotations, while $b_\mu$ will act as a gauge field for an emerging local dilatation symmetry.
In terms of these dependent gauge fields, we then find that
\begin{subequations} \label{eq:deltasusy0}
\begin{align}
\delta^{(0)}\psi_{\mu + } &= \delta_+\psi_{\mu+} + \delta_-\psi_{\mu-} + \frac12\,\tau_\mu{}^+\Gamma_+\eta_-\,,\\
\delta^{(0)}\psi_{\mu - } &= \delta_+\psi_{\mu-} + \delta_-\psi_{\mu-} + \tau_\mu{}^+\,\rho_-\,,\\
\delta^{(0)}\lambda_+ &= \delta_+\lambda_+ + \delta_-\lambda_+ \,,\\
\delta^{(0)}\lambda_-  &= \delta_+\lambda_- + \delta_-\lambda_- + \eta_-\,,
\end{align}
\end{subequations}
where
\begin{subequations} \label{eq:STrest}
\begin{align}
&\eta_- = \left(\partial_+\phi\,\Gamma^+ - \frac14\,h^{-A'B'}\Gamma_{-A'B'}\right)\epsilon_+ + 2\,b^{A'}\Gamma_{A'}\epsilon_- \ \, (\text{+ terms quadratic in $\psi_{\mu\pm}$, $\lambda_{\pm}$})\,,\\
&\rho_- = \left(-2\,\partial_+\phi +\frac14\,h^{-A'B'}\Gamma_{A'B'}\right)\epsilon_- + \frac12\,W_+{}^{-A'}\Gamma_{-A'}\epsilon_+  \ \, (\text{+ terms quadratic in $\psi_{\mu\pm}$, $\lambda_{\pm}$})\,.
\end{align}
\end{subequations}
and
\begin{subequations}
\label{eq:NRsupersymmetry}
\begin{align}
&\delta_+\psi_{\mu+} =\mathcal D_\mu\epsilon_+ - \frac18\,e_{\mu C'}h^{C'A'B'}\Gamma_{A'B'}\epsilon_+ && \quad(\text{+ terms quadratic in $\psi_{\mu\pm}$, $\lambda_{\pm}$})\,,\\
&\delta_-\psi_{\mu+} = \big(e_{\mu B'}\tau^{B'A'+} + \tau_\mu{}^-\tau^{A'++}\big)\Gamma_{A'+}\epsilon_- && \quad(\text{+ terms quadratic in $\psi_{\mu\pm}$, $\lambda_{\pm}$})\,,\\
&\delta_+\psi_{\mu-} = -\frac12\,\omega_\mu{}^{-A'}\Gamma_{-A'}\epsilon_+ && \quad(\text{+ terms quadratic in $\psi_{\mu\pm}$, $\lambda_{\pm}$})\,,\\
&\delta_-\psi_{\mu-} = \mathcal D_\mu\epsilon_- - \frac18\,e_{\mu C'}h^{C'A'B'}\Gamma_{A'B'}\epsilon_- && \quad(\text{+ terms quadratic in $\psi_{\mu\pm}$, $\lambda_{\pm}$})\,,\\
&\delta_+\lambda_+ = \big(\mathcal \partial_{A'}\phi\,\Gamma^{A'} - b_{A'}\,\Gamma^{A'} - \frac{1}{12}\,h^{A'B'C'}\Gamma_{A'B'C'}\big)\epsilon_+&& \quad(\text{+ terms quadratic in $\psi_{\mu\pm}$, $\lambda_{\pm}$})\,,\\
&\delta_-\lambda_+ = \frac12\,\tau^{A'B'+}\Gamma_{A'B'+}\epsilon_-&& \quad(\text{+ terms quadratic in $\psi_{\mu\pm}$, $\lambda_{\pm}$})\,,\\
&\delta_+\lambda_- = 0&& \quad(\text{+ terms quadratic in $\psi_{\mu\pm}$, $\lambda_{\pm}$})\,,\\
&\delta_-\lambda_- = \big(\mathcal \partial_{A'}\phi\,\Gamma^{A'} - b_{A'}\,\Gamma^{A'}- \frac{1}{12}\,h^{A'B'C'}\Gamma_{A'B'C'}\big)\epsilon_-&& \quad(\text{+ terms quadratic in $\psi_{\mu\pm}$, $\lambda_{\pm}$})\,,
\end{align}
\end{subequations}
where $h_{\mu\nu\rho} = 3 \partial_{[\mu} b_{\nu\rho]}$, $\mathcal D_\mu \epsilon_\pm$ is given by
\begin{align}
\mathcal D_\mu\epsilon_\pm = \bigg(\partial_\mu - \frac14\,\omega_\mu{}^{A'B'}\Gamma_{A'B'}\pm\frac12\,\omega_\mu\mp\frac12\,b_\mu\bigg)\epsilon_\pm\,,
\end{align}
and $W_+{}^{-A'}$ in \eqref{eq:STrest} refers to components of the spin connections that are not fully determined in TSNC geometry, but that do not play a role in the rest of this paper (see also appendix \ref{sec:TSNC} and \cite{Bergshoeff:2021bmc}). Note that the redefined supersymmetry transformation rules \eqref{eq:susyexpansion} also contain non-trivial terms at order $c^{-2}$. We will not give the explicit expressions for $\delta^{(-2)} \psi_{\mu\pm}$, $\delta^{(-2)} \lambda_\pm$ here, as we will not need them in what follows.

Let us finish this section, by commenting on the appearance of terms of order $c^2$ in the supersymmetry transformation rules of the fermionic fields $\psi_{\mu\pm}$ and $\lambda_\pm$. Since we wish to identify $\tau_\mu{}^A$, $e_\mu{}^{A'}$, $b_{\mu\nu}$, $\phi$, $\psi_{\mu\pm}$ and $\lambda_{\pm}$ as fields in the NR theory that is obtained after taking $c \rightarrow \infty$, one would hope that the redefinitions \eqref{eq:rescale}, \eqref{eq:rescaleinv} and \eqref{eq:parresc} lead to transformation rules for these fields that take the form of an expansion in powers of $c^{-2}$ that starts at order $c^0$. That way, the $c\rightarrow \infty$ limit of these transformation rules is well-defined and can be identified with the transformation rules of the NR theory. Clearly, the terms of order $c^2$ in the supersymmetry transformation rules of $\psi_{\mu\pm}$ and $\lambda_\pm$ are potentially problematic in this regard. In order to explain how to deal with these `divergent' terms in the next section, let us make the following useful observations here. One can isolate the divergent terms of order $c^2$ in the supersymmetry transformation rules of $\psi_{\mu\pm}$, $\lambda_\pm$ by performing the following field redefinition:
\begin{align} \label{eq:NRfermionstilde}
\tilde\psi_{\mu+} &\equiv \psi_{\mu +} - \frac12\,\tau_\mu{}^+\Gamma_+\lambda_-\,,\qquad
\tilde\psi_{\mu -} \equiv \psi_{\mu -} - \tau_\mu{}^+\, \tilde\psi_{-}  \quad \text{with} \quad \tilde\psi_{-} \equiv \tau_+{}^\mu\psi_{\mu -}\,.
\end{align}
Using \eqref{eq:divsupersymmetry}, one can then easily see that the parts of the supersymmetry transformation rules of $\tilde{\psi}_{\mu \pm}$, $\lambda_+$, $\tilde\psi_{-}$ and $\lambda_-$ at order $c^2$ are given by:
\begin{align}\label{eq:susydivergencered}
\delta^{(2)}\tilde\psi_{\mu+}&=0\,, \qquad \qquad \quad
\delta^{(2)}\tilde\psi_{\mu-}=0\,, \qquad \qquad \quad \delta^{(2)}\lambda_{+}=0 \,, \nonumber \\
\delta^{(2)}\tilde\psi_{-}&=\frac12\,\big(\tau^{A'B'-}\Gamma_{A'B'}\epsilon_- - \tau^{A'--}\Gamma_{-A'}\epsilon_+ \big)\,, \nonumber \\
\delta^{(2)}\lambda_- &= -\frac12\,\tau^{A'B'-}\Gamma_{A'B'-}\epsilon_+\,.
\end{align}
The supersymmetry rules of $\tilde{\psi}_{\mu\pm}$ and $\lambda_+$ thus do not contain any divergent terms at order $c^2$ and their $c \rightarrow \infty$ limit is well-defined. Note furthermore that the parts of the supersymmetry transformations of $\tilde{\psi}_{\mu\pm}$ at order $c^0$ then also no longer depend on the quantities $\eta_-$, $\rho_-$, defined in \eqref{eq:STrest}. We will see the significance of these observations in what follows.

\section{The NR Limit of the Action} \label{sec:limitaction}

\noindent In the previous section, we reviewed the action and transformation rules of relativistic $D=10$, $\mathcal{N}=1$ supergravity, introduced a field redefinition that involves a parameter $c$ and expressed all transformation rules for the redefined fields as expansions in powers of $c^{-2}$. Starting from this section, we wish to discuss the NR limit $c \rightarrow \infty$, that should lead to NR minimal ten-dimensional supergravity, similar to how NR NS-NS gravity was obtained in \cite{Bergshoeff:2021bmc}, by performing the bosonic part of the field redefinition \eqref{eq:rescale}, \eqref{eq:rescaleinv} and taking the $c\rightarrow \infty$ limit. In this section, we will apply this limit to the action of ten-dimensional $\mathcal{N}=1$ supergravity, while the limit of its equations of motion will be discussed in the next section.

Ordinarily, the NR limit of quantities (such as an action or equations of motion) is performed by applying a $c$-dependent field redefinition to the quantities under consideration, expanding the result in powers of $c^{-2}$ and retaining only the terms at leading order. In case the transformation rules of the redefined fields assume the form of expansions in powers of $c^{-2}$ that start at order $c^0$, this procedure guarantees that one ends up with quantities that are invariant or covariant under the NR transformation rules that are given by the $c\rightarrow \infty$ limit of those of the redefined fields.

We wish to apply a similar limit procedure to the action \eqref{eq:Nisoneaction} of $D=10$, $\mathcal{N}=1$ supergravity. In particular, we still wish to define the NR limit of \eqref{eq:Nisoneaction} as the leading order term in the $c^{-2}$--expansion of \eqref{eq:Nisoneaction}, after performing the field redefinition \eqref{eq:rescale}, \eqref{eq:rescaleinv}. Similarly, we still want to identify the transformation rules of the NR theory as the part at order $c^0$ in the $c^{-2}$--expansions of the relativistic transformation rules \eqref{eq:LorsymmBOS}, \eqref{eq:LorsymmFER}, \eqref{eq:bossusy} and \eqref{eq:susyexpansion} for the redefined fields. Note, however, that presently the $c^{-2}$--expansion of some of the supersymmetry transformation rules of the redefined fermionic fields starts at order $c^2$, instead of at order $c^0$. As a consequence, the interpretation of the $c^0$ part of these transformation rules as NR ones is no longer straightforward. Moreover, invariance or covariance under these NR transformation rules of the leading order of an expansion of a quantity in powers of $c^{-2}$ is also no longer guaranteed. Remarkably, even though some transformation rules diverge in the $c\rightarrow \infty$ limit, it turns out that the $c\rightarrow \infty$ limit can be taken in a smooth way upon imposition of a constraint. Importantly, after taking the limit, the resulting NR action exhibits invariance under three emerging symmetries: one dilatation symmetry and two fermionic shift symmetries. The emerging dilatation symmetry was already encountered when taking the NR limit of NS-NS gravity \cite{Bergshoeff:2021bmc}. In this paper, we find that it extends to a symmetry of the NR limit of the action of $D=10$, $\mathcal{N}=1$ supergravity and that it is accompanied by two fermionic symmetries as supersymmetric counterparts.

The emergence of the two fermionic shift symmetries in the NR limit of \eqref{eq:Nisoneaction} can be understood on general grounds. In order to see this, let us first apply the redefinition \eqref{eq:rescaleinv} to the action \eqref{eq:Nisoneaction} and expand the result in powers of $c^{-2}$. This gives a sum of three terms, at orders $c^0$, $c^{-2}$ and $c^{-4}$ respectively:
\begin{align} \label{eq:actionexpansion}
S = S^{(0)} + c^{-2} S^{(-2)} + c^{-4} S^{(-4)} \,,
\end{align}
where each of the $S^{(i)}$ now depends on the fields $\tau_\mu{}^A$, $e_\mu{}^{A'}$, $b_{\mu\nu}$, $\phi$, $\psi_{\mu\pm}$ and $\lambda_\pm$. It is important to note that it is non-trivial that the expansion \eqref{eq:actionexpansion} of $S$ starts at order $c^0$. Indeed, examining all terms of \eqref{eq:Nisoneaction} separately, one finds that some of them can contribute terms at order $c^2$ in the expansion \eqref{eq:actionexpansion}, so that there can in principle be a $c^2 S^{(2)}$ term on the right-hand-side of \eqref{eq:actionexpansion}. It turns out however that all such contributions cancel identically. For the bosonic part of the action \eqref{eq:Nisoneaction}, this relies on an order $c^2$ contribution from the Ricci scalar cancelling against a similar contribution from the kinetic term of the KR field, as was explained in \cite{Bergshoeff:2021bmc}. One can check that this cancellation of order $c^2$ terms extends to the full $D=10$, $\mathcal{N}=1$ supergravity action \eqref{eq:Nisoneaction}, so that
\begin{align} \label{eq:S2zero}
S^{(2)} = 0
\end{align}
identically.

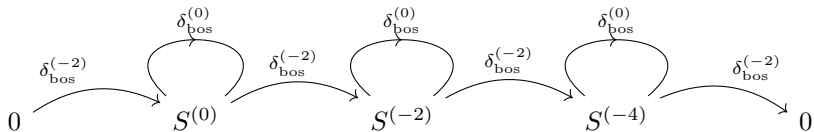
\begin{figure}[t]
  \centering
  \begin{tikzcd}
     0 \arrow[rr, bend left, "\delta_{\mathrm{bos}}^{(-2)}"] & &
     S^{(0)}\arrow[loop, distance=4em, dash, marrow=>, out=north west, in=north east, "\delta_{\mathrm{bos}}^{(0)}"]\arrow[rr, bend left, "\delta_{\mathrm{bos}}^{(-2)}"]& &
     S^{(-2)}\arrow[loop, distance=4em, out=north west, dash, marrow=>, in=north east, "\delta_{\mathrm{bos}}^{(0)}"]\arrow[rr, bend left, "\delta_{\mathrm{bos}}^{(-2)}"] & &
     S^{(-4)}\arrow[loop, distance=4em, out=north west, dash, marrow=>, in=north east, "\delta_{\mathrm{bos}}^{(0)}"] \arrow[rr, bend left, "\delta_{\mathrm{bos}}^{(-2)}"]& &
     0
  \end{tikzcd}
  \caption{Schematic representation of the symmetry transformation of the different terms in \eqref{eq:actionexpansion} under generic Lorentzian bosonic symmetries $\delta_{\mathrm{bos}}$. We see that only the leading order is invariant under Lorentzian boosts by itself. This is equivalent to the statement of manifest Galilei invariance of $S_{NR}=\lim_{c\to\infty}S$.}
  \label{fig:symmetrymixingboost}
\end{figure}

Note that the $c \rightarrow \infty$ limit of $S$ is then well-defined and gives $S^{(0)}$, which we identify as the action that results from taking the NR limit. Let us now examine how $S^{(0)}$ transforms under the NR symmetry transformation rules, that correspond to the parts at order $c^0$ of the relativistic transformation rules \eqref{eq:LorsymmBOS}, \eqref{eq:LorsymmFER}, \eqref{eq:bossusy} and \eqref{eq:susyexpansion}, after performing the redefinition \eqref{eq:rescale}, \eqref{eq:rescaleinv}. In order to do this, we will not yet take the $c \rightarrow \infty$ limit, but rather require that the full relativistic action $S$, written as the $c^{-2}$--expansion \eqref{eq:actionexpansion}, is invariant under the full relativistic transformation rules \eqref{eq:LorsymmBOS}, \eqref{eq:LorsymmFER}, \eqref{eq:bossusy} and \eqref{eq:susyexpansion}. Expanding the symmetry variation of $S$ in powers of $c^{-2}$ and requiring that terms at different order in this expansion vanish separately, then indicates how $S^{(0)}$, $S^{(-2)}$ and $S^{(-4)}$ transform into each other under the different $c^{-2}$ orders of the relativistic transformation rules. Let us do this first for the bosonic symmetries \eqref{eq:LorsymmBOS}, \eqref{eq:LorsymmFER}. The infinitesimal action $\delta_{\mathrm{bos}}$ of a generic bosonic symmetry leads to two variations $\delta_{\mathrm{bos}}^{(0)}$ and $\delta_{\mathrm{bos}}^{(-2)}$, according to
\begin{align}
\delta_{\mathrm{bos}} F = \delta_{\mathrm{bos}}^{(0)} F + c^{-2} \delta_{\mathrm{bos}}^{(-2)} F \,,
\end{align}
where $F$ is any of the fields $\tau_\mu{}^A$, $e_\mu{}^{A'}$, $b_{\mu\nu}$, $\phi$, $\psi_{\mu\pm}$, $\lambda_\pm$. As a consequence
\begin{align} \label{eq:bosvarSexp}
\delta_{\mathrm{bos}} S = \delta_{\mathrm{bos}}^{(0)} S^{(0)} + c^{-2} \left( \delta_{\mathrm{bos}}^{(0)} S^{(-2)} + \delta_{\mathrm{bos}}^{(-2)} S^{(0)} \right)  + \mathcal{O}(c^{-4})\,.
\end{align}
The requirement that $\delta_{\mathrm{bos}} S = 0$, then imposes that every $c^{-2}$ order in \eqref{eq:bosvarSexp} is separately zero. One thus in particular finds that
\begin{align}
\delta_{\mathrm{bos}}^{(0)} S^{(0)} = 0 \,,
\end{align}
or in other words, that the NR action $S^{(0)}$ is as expected invariant under the NR bosonic symmetries, whose transformation rules are given by $\delta_{\mathrm{bos}}^{(0)}$. See figure \ref{fig:symmetrymixingboost} for a schematic representation of the above statements.

We can apply a similar reasoning to the supersymmetries \eqref{eq:bossusy}, \eqref{eq:susyexpansion}. In this case, the supersymmetry transformations can contain terms at order $c^2$, so that the infinitesimal action $\delta_Q$ of a generic supersymmetry $Q$ leads to three variations $\delta_Q^{(2)}$, $\delta_Q^{(0)}$ and $\delta_Q^{(-2)}$
\begin{align}
\delta_Q F = c^2 \delta_Q^{(2)} F + \delta_Q^{(0)} F + c^{-2} \delta_Q^{(-2)} F \,,
\end{align}
where again $F$ is any of the fields $\tau_\mu{}^A$, $e_\mu{}^{A'}$, $b_{\mu\nu}$, $\phi$, $\psi_{\mu\pm}$, $\lambda_\pm$. The supersymmetry variation $\delta_Q S$ of the action can then be expanded as
\begin{align} \label{eq:susyvarSexp}
\delta_Q S = c^2 \delta_Q^{(2)} S^{(0)} + c^0 \left( \delta_Q^{(0)} S^{(0)} + \delta_Q^{(2)} S^{(-2)} \right) + \mathcal{O}(c^{-2}) \,.
\end{align}
Requiring invariance of $S$ again imposes that every order of $c^{-2}$ in \eqref{eq:susyvarSexp} is separately zero. This in particular leads to the following two requirements
\begin{align} \label{eq:deltaQconstraints}
\delta_Q^{(2)} S^{(0)} = 0  \quad \qquad \text{and} \quad \qquad \delta_Q^{(0)} S^{(0)} = - \delta_Q^{(2)} S^{(-2)} \,.
\end{align}
From \eqref{eq:susydivergencered}, we see that only $\delta_Q^{(2)} \tilde\psi_{-}$ and $\delta_Q^{(2)} \lambda_-$ are non-zero and that these two variations moreover have the effect of shifting the two fields $\tilde\psi_{-}$ and $\lambda_-$ independently. The only way, in which the variation $\delta_Q^{(2)} S^{(0)}$ can vanish, is then if $S^{(0)}$ does not depend on $\tilde\psi_{-}$ and $\lambda_-$. We can alternatively state this in terms of the fields $\psi_{\mu\pm}$ and $\lambda_{\pm}$. The requirement that $\delta_Q^{(2)} S^{(0)}$ vanishes then boils down to saying that $S^{(0)}$ is invariant under two fermionic shift symmetries, that we call the $S$- and $T$-symmetries and whose non-trivial action on $\psi_{\mu\pm}$ and $\lambda_\pm$ is as follows:
\begin{align}
\label{eq:STsymm}
\delta_S \psi_{\mu+} &= \frac12 \tau_\mu{}^+ \Gamma_+ \eta_- \,, \qquad \qquad \delta_S \lambda_- = \eta_- \,, \nonumber \\
\delta_T \psi_{\mu-} &= \tau_\mu{}^+ \rho_- \,,
\end{align}
where $\eta_-$ and $\rho_-$ are the parameters of the $S$- and $T$-symmetry respectively. As mentioned above and as can be verified in the explicit expression for $S^{(0)}$ given below, the NR action $S^{(0)}$ is also invariant under an emerging dilatation symmetry that has the following non-trivial action on the fields in $S^{(0)}$:
\begin{alignat}{2}
\label{eq:dilsymm}
\delta_D \phi &= \lambda_D \,, \qquad \qquad & \delta_D \tau_\mu{}^A &= \lambda_D \tau_\mu{}^A \,, \nonumber \\
\delta_D \psi_{\mu\pm} &= \pm \frac12 \lambda_D \psi_{\mu\pm} \,, \qquad \qquad & \delta_D \lambda_{\pm} &= \pm \frac12 \lambda_D \lambda_{\pm} \,.
\end{alignat}
Note that the dilatation weights of the NR fields are the same as the exponents of the powers of $c$ in the redefinition \eqref{eq:rescaleinv} of the relativistic fields in terms of the NR ones (for the dilaton, this rule holds when considering, e.g., $\exp(\Phi) = c\, \exp(\phi)$).

\begin{figure}[t]
  \centering
	\begin{tikzcd}
     0 \arrow[rr, bend left, "\delta_Q^{(-2)}"] & &
     S^{(0)}\arrow[loop, distance=4em, dash, marrow=>, out=north west, in=north east, "\delta_Q^{(0)}"]\arrow[rr, bend left, "\delta_Q^{(-2)}"]\arrow[ll,bend left, "\delta_Q^{(2)}"] & &
     S^{(-2)}\arrow[loop, distance=4em, out=north west, dash, marrow=>, in=north east, "\delta_Q^{(0)}"]\arrow[rr, bend left, "\delta_Q^{(-2)}"]\arrow[ll,bend left, "\delta_Q^{(2)}"]& &
     S^{(-4)}\arrow[loop, distance=4em, out=north west, dash, marrow=>, in=north east, "\delta_Q^{(0)}"] \arrow[rr, bend left, "\delta_Q^{(-2)}"]\arrow[ll,bend left, "\delta_Q^{(2)}"]& &
     0\arrow[ll,bend left, "\delta_Q^{(2)}"]
  \end{tikzcd}
  \caption{Schematic representation of \eqref{eq:susyvarSexp}. The diagram shows that different orders in the expansion \eqref{eq:actionexpansion} mix under supersymmetry $\delta_Q$, according to the rules: the sum of all arrows ending at a certain order $S^{(i)}$ vanishes. This gives rise to \eqref{eq:deltaQconstraints}.}
  \label{fig:symmetrymixingsusy}
\end{figure}
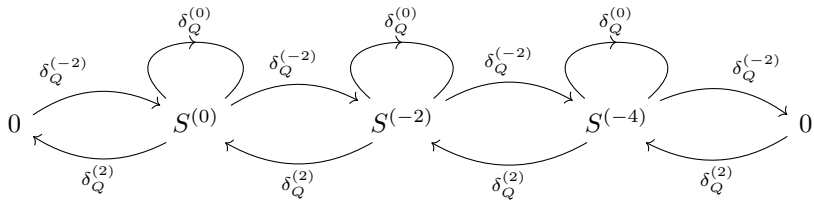

We have seen above that general considerations allow us to conclude that the non-relativistic action $S^{(0)}$ is invariant under the fermionic shift symmetries \eqref{eq:STsymm}. One might wonder whether a similar, general argument exists for the bosonic dilatation-shift symmetry of the NR action $S^{(0)}$. As we will show now the answer is yes. However, the argument is slightly more subtle than the one for the fermionic St\"uckelberg shift symmetries. Instead of expanding the supersymmetry invariance of the relativistic action as in \eqref{eq:susyvarSexp} we have to consider the commutator of two supersymmetries, acting on the relativistic action---which of course gives zero $[\delta_Q(\varepsilon_1),\delta_Q(\varepsilon_2)]S=0$---and extract information about the NR action $S^{(0)}$ from the different orders in the expansion. Moreover, we use some information about the algebra, in particular the commutator between $S$-symmetry and supersymmetry \eqref{eq:Ssusy}, from which we conclude that
\begin{align}
\big[\delta_Q^{(2)}(\varepsilon_1),\delta_Q^{(0)}(\varepsilon_2)\big] = \delta_D(\lambda_D')+\cdots\,,
\end{align}
where the ellipses denote terms involving symmetries of $S^{(0)}$ that do not play a role in the present discussion, see appendix \ref{sec:algebra} for more details. The parameters on the right-hand-side are dependent expressions  $\lambda_D' = -1/4\,\tau^{A'B'-}\,\bar\epsilon_{2+}\Gamma_{A'B'-}\epsilon_{1-}$. Furthermore we can use the fact that the fermionic St\"uckelberg symmetries commute, see section \ref{sec:algstst}, to show that $[\delta_Q^{(2)}(\varepsilon_1),\delta^{(2)}_Q(\varepsilon_2)]=0$. Taking all of the above into account one can then show that the terms in $[\delta_Q(\varepsilon_1),\delta_Q(\varepsilon_2)]S=0$ at order $\mathcal O(c^2)$ vanish if and only if
\begin{align}
\delta_D S^{(0)}=0\,.
\end{align}
This shows that the non-relativistic action is indeed dilatation invariant as a consequence of the divergence structure in the supersymmetry rules and the particular form of the commutator between supersymmetry and fermionic shift symmetries.

Since we wish to identify $\delta_Q^{(0)}$ as the NR supersymmetry transformation rules, the second requirement of \eqref{eq:deltaQconstraints} tells us that the NR action $S^{(0)}$ is not necessarily invariant under these NR supersymmetries, but is rather given by the variation $\delta_Q^{(2)} S^{(-2)}$ of the $c^{-2}$ order of the expansion of \eqref{eq:Nisoneaction} under the leading $c^2$ order of the relativistic supersymmetry transformation rules. From \eqref{eq:divsupersymmetry} we see however that all terms in $\delta_Q^{(2)} S^{(-2)}$ are proportional to $\tau_{A'B'}{}^-$ or $\tau_{A'+}{}^{-}$. We thus find that the variation $\delta_Q^{(0)} S^{(0)}$ of the NR action $S^{(0)}$ under the NR supersymmetry transformation rules gives zero when the following constraints on the torsion $\tau_{\mu\nu}{}^{A}$ are imposed
\begin{align}\label{eq:DSNC-}
\tau_{A'B'}{}^- = 0\,,\qquad \qquad \tau_{A'+}{}^- =0\,.
\end{align}
These constraints are invariant under the dilatation symmetry \eqref{eq:dilsymm}, and we will refer to SNC geometry, in which these constraints are imposed, as `self-dual Dilatation invariant SNC geometry' or self-dual DSNC geometry for short. The constraints \eqref{eq:DSNC-} are not only invariant under dilatations; they are invariant under all non-relativistic transformation rules, and in particular, their variation under NR supersymmetry vanishes identically. This relies on the fact that the self-dual longitudinal Vielbein $\tau_\mu{}^-$ is a singlet\footnote{For a general account on supersymmetry singlets and the conditions for such fields to exist, see \cite{Bergshoeff:1989qh}.} under NR supersymmetry, i.e., $\delta^{(0)}_Q\tau_\mu{}^-=0$, which follows from the chirality properties of the non-relativistic spinors (see \eqref{eq:selfduality}), in particular, $\Gamma^-\psi_{\mu+}=0$. One can thus impose the constraints \eqref{eq:DSNC-} by hand in the theory and still maintain supersymmetry without having to impose extra constraints.

In fact, the self-dual DSNC constraints are a necessary requirement for the consistency of the theory. Above, we have already seen glimpses of that when discussing the supersymmetry of the action. We will see more (and stronger) evidence for this crucial fact when discussing the consistency of the non-relativistic equations of motion in section \ref{sec:consistencyofEOM}. Here, we will consider parts of the supersymmetry algebra and show that it closes if and only if \eqref{eq:DSNC-} are imposed. In other words, we show that the self-dual DSNC constraints are a necessary requirement for the existence of a supergravity multiplet. More details on the algebra are given in appendix \ref{sec:algebra}. In the following, we will, unless mentioned otherwise, slightly abuse notation and denote by $\delta_Q^{(0)}$ the NR supersymmetry transformation rules, given in the $c \rightarrow \infty$ limit of \eqref{eq:bossusy} and in \eqref{eq:deltasusy0}, not including the parts that involve the parameters \eqref{eq:STrest}. The latter correspond to field-dependent $S$- and $T$-transformations and their omission will not change our arguments. For all practical purposes, this is equivalent to indentifying \eqref{eq:NRsupersymmetry} as the non-relativistic supersymmetry rules for the fermions. The commutator of two such supersymmetries on $\tau_\mu{}^+$ then gives:
\begin{align}
\big[\delta_Q^{(0)}(\eta_+),\delta_Q^{(0)}(\epsilon_+)\big]\tau_\mu{}^+ &= \xi^\nu_{(++)}D_\nu\tau_\mu{}^+ +\big(\partial_\mu\xi^\nu_{(++)}\big)\tau_\nu{}^+ +\xi^\nu_{(++)}\rmR_{\mu\nu}(H^+)\notag\\
&= \big(\mathcal L_{(++)} - \delta_M(\xi^\nu_{(++)}\omega_\nu) - \delta_D(\xi^\nu_{(++)}b_\nu)\big)\tau_\mu{}^+\,,
\end{align}
where $\xi^\mu_{(++)} = \bar\epsilon_+\Gamma^+\eta_+\,\tau_+{}^\mu$, $\mathcal L_{(++)}$ denotes the usual Lie derivative along $\xi^\mu_{(++)}$, and $\rmR_{\mu\nu}(H^A)$ denotes the fully covariant torsion $2-$form (see \eqref{eq:torsion2Form}). To show closure we have used the conventional constraints $\tau_+{}^\nu\,\rmR_{\mu\nu}(H^+)=0$---similar to how one uses $\rmR_{\mu\nu}(P^{\hat A})=0$ in the analogous relativistic calculation. Let us now turn to the closure on the other longitudinal Vielbein $\tau_\mu{}^-$, which is a singlet under supersymmetry $\delta_Q^{(0)}\tau_\mu{}^-=0$. Hence it is clear that $[\delta_Q^{(0)}(\eta_+),\delta_Q^{(0)}(\epsilon_+)]\tau_\mu{}^-=0$, and consequently we have to require that
\begin{align}
\big(\mathcal L_{(++)} - \delta_M(\xi^\nu_{(++)}\omega_\nu) - \delta_D(\xi^\nu_{(++)}b_\nu)\big)\tau_\mu{}^- = -\xi_{(++)}^\nu\,\rmR_{\mu\nu}(H^-) = 0\,.
\end{align}
Using the conventional constraints \eqref{eq:torsion2Form} it is not hard to see that this is equivalent to setting $\tau_{A'}{}^{--}=0$, and by requiring consistency with Galilean boosts $\tau_{A'B'}{}^-=0$. This proves that the self-dual DSNC constraints are a necessary requirement for closure of the algebra.

We can summarize the above discussion as follows. The NR limit $S^{(0)}$ of the ten-dimensional $\mathcal{N}=1$ supergravity action is obtained as the leading order term in the $c^{-2}$--expansion of  \eqref{eq:Nisoneaction}, after performing the field redefinition \eqref{eq:rescale}. This NR action $S^{(0)}$ is invariant under two emerging fermionic $S$- and $T$-shift symmetries \eqref{eq:STsymm}, an emerging dilatation symmetry \eqref{eq:dilsymm}, as well as under the $c\rightarrow\infty$ limit of the bosonic transformation rules \eqref{eq:LorsymmBOS}, \eqref{eq:LorsymmFER}. The NR supersymmetry transformation rules are identified as the order $c^0$ part in the relativistic transformation rules \eqref{eq:susyexpansion}. The action $S^{(0)}$ is then only invariant under NR supersymmetry, if one assumes that the self-dual DSNC geometry constraints \eqref{eq:DSNC-} hold.\footnote{Note that one needs to treat $S^{(0)}$ as a pseudo-action, when checking its invariance under NR supersymmetry, i.e., one should only impose the constraints \eqref{eq:DSNC-} after performing a general variation.}

Let us finish this section by giving the explicit expression for the NR action $S_{NR} = S^{(0)}$. It is useful to split $S_{NR}$ into a part $S_B$ that is purely bosonic, a part $S_{\psi\psi}$ that is quadratic in the gravitini $\psi_{\mu\pm}$, a part $S_{\lambda\lambda}$ that is quadratic in the dilatini $\lambda_{\pm}$ and a remaining quadratic fermion part $S_{\lambda\psi}$ that contains both a gravitino and a dilatino:
\begin{align}\label{eq:NRaction}
S_{NR} = S_B + S_{\lambda\lambda} + S_{\lambda\psi} + S_{\psi\psi} + \text{quartic fermion terms} \,.
\end{align}
As mentioned above, we will ignore all quartic fermion terms and only require supersymmetry up to cubic fermion terms. The bosonic part of the action has been given in \cite{Bergshoeff:2021bmc} and reads:
\begin{subequations} \label{eq:NRaction2}
\begin{align} \label{eq:BosAction}
S_B = \frac{1}{2\,\kappa^2}\int \rmd^{10}x \,e\,\rme^{-2\,\phi}\bigg(&\rmR(J)+4\,\partial_{A'}\phi\,\partial^{A'}\phi-\frac{1}{12}\,h_{A'B'C'}h^{A'B'C'} \notag\\
&-4\,e_{A'}{}^{\mu}(\partial_{\mu} b^{A'}-\omega_{\mu}{}^{A'B'}b_{B'}-\omega_{\mu}{}^{AB'}\tau^{A'}{}_{B'A}) \notag\\&- 4\,b_{A'}b^{A'}- 4\,\tau_{A'\{AB\}}\tau^{A'\{AB\}}  \bigg)\,,
\end{align}
where $e = \mathrm{det}(\tau_\mu{}^A, e_\mu{}^{A'})$ and $\rmR(J)$ and other geometric quantities are defined in appendix \ref{sec:TSNC}, see in particular \eqref{eq:RJ}. We refer to \cite{Bergshoeff:2021bmc} for a detailed explanation of the notation.
The part of the action that is quadratic in the dilatini reads
\begin{align} \label{eq:SNR2}
S_{\lambda\lambda} = \frac{1}{2\kappa^2}\int \rmd^{10}x \,e\,\rme^{-2\,\phi}\bigg( &2\,\bar{\lambda}_{\pm}\Gamma^{A'}D_{A'}\lambda_{\mp}+
2\bar{\lambda}_{+}\Gamma^{A}D_{A}\lambda_{+}\notag\\
&-\frac{1}{6}h_{A'B'C'}(\bar{\lambda}_{+}\Gamma^{A'B'C'}\lambda_{-})+\tau_{B'C'A}(\bar{\lambda}_{-}\Gamma^{B'C'A}\lambda_{-})\bigg)\,,
\end{align}
where the covariant derivatives are covariant with respect to Galilean symmetries and dilatations, see \eqref{eq:LorsymmFER} and \eqref{eq:galspinconn}. The notation $\bar\lambda_\pm\Gamma \lambda_\mp$ is a shorthand for $\bar\lambda_+\Gamma\lambda_- + \bar\lambda_-\Gamma\lambda_+$, and will be used also below. The off-diagonal terms read
\begin{align} \label{eq:SNR3}
S_{\lambda\psi} = \frac{1}{2\kappa^2}\int \rmd^{10}x \,e\,\rme^{-2\,\phi}\bigg(&-4\,\bar{\lambda}_{\pm}\Gamma^{A'B'}e_{A'}{}^\mu e_{B'}{}^\nu D_{[\mu}\psi_{\nu]\mp}-8\,\bar{\lambda}_{+}\Gamma^{AB'}\tau_A{}^{\mu}e_{B'}{}^{\nu}D_{[\mu}\psi_{\nu]+}\nonumber\\
&-4\,\bar{\lambda}_{\pm}\Gamma^{A'B'}\psi_{A'\mp}\,D_{B'}\phi -4\,\bar{\lambda}_{+}\Gamma^{AB'}\psi_{A+}\,D_{B'}\phi \nonumber\\
&+\frac{1}{6}\,h_{A'B'C'}(\bar{\lambda}_{\pm}\Gamma^{A'B'C'D'}\psi_{D'\mp})+\frac{1}{2}h_{A'B'C'}(\bar{\lambda}_{+}\Gamma^{A'B'C'D}\psi_{D+})\nonumber\\
&-(\eta^{DA}+\epsilon^{DA})\tau_{B'C'D}(\bar{\lambda}_{-}\Gamma^{B'C'}\psi_{A+}-\bar{\lambda}_{+}\Gamma^{B'C'}\psi_{A-})\nonumber\\
&+2\,\tau_{B'C'}{}^A\bar{\lambda}_{\pm}\Gamma^{B'C'}\psi_{A\mp} + 2\,\tau^{C'\{AB\}}\bar{\lambda}_{+}\Gamma_{C'A}\psi_{B+}\nonumber\\
&-2\tau_{B'C'A}\bar{\lambda}_{-}\Gamma^{AB'C'D'}\psi_{D'-}\bigg)\,.
\end{align}
The pure gravitino terms are given by
\begin{align} \label{eq:SNR4}
S_{\psi\psi} = \frac{1}{2\kappa^2}\int \rmd^{10}x \,e\,\rme^{-2\,\phi}\bigg(&
-2\,\bar{\psi}_{A+}\Gamma^{AB'C'}e_{B'}{}^\mu e_{C'}{}^\nu D_{[\mu}\psi_{\nu] +}-4\,\bar{\psi}_{A'+}\Gamma^{A'B'C}e_{B'}{}^\mu \tau_C{}^\nu D_{[\mu}\psi_{\nu] +}\nonumber\\
&-2\,\bar{\psi}_{A'\pm}\Gamma^{A'B'C'}e_{B'}{}^\mu e_{C'}{}^\nu D_{[\mu}\psi_{\nu]
\mp}+\frac{1}{2}\,h^{A'B'C'}(\bar{\psi}_{A'\pm}\Gamma_{B'}\psi_{C'\mp})\nonumber\\
&-\frac{1}{6}h_{A'B'C'}\big(\bar{\psi}_{D'+}\Gamma^{A'B'C'D'E}\psi_{E+}+ \frac12\,\bar{\psi}_{D'\pm}\Gamma^{A'B'C'D'E'}\psi_{E'\mp}\big)+\nonumber\\
&-4\,\big(\bar{\psi}_{A'\pm}\Gamma^{A'}\psi_{B'\pm}+\,\bar{\psi}_{A+}\Gamma^{A}\psi_{B'+}\big)D^{B'}\phi \nonumber\\
&-2\,(\eta^{AD}+\epsilon^{AD})\tau^{B'C'}{}_D\,\bar{\psi}_{C'\pm}\Gamma_{B'}\psi_{A\mp} +2\,\tau^{B'C'}{}^A(\bar{\psi}_{B'-}\Gamma_{A}\psi_{C'-})\notag\\
&- 2\big(\eta_{BC} -\epsilon_{BC}\big)\tau^{C'\{AB\}}\bar\psi^C{}_+\Gamma_A\psi_{C'+}\notag\\
&+(\eta_{AB}+\epsilon_{AB})\tau_{B'C'}{}^A\,\bar{\psi}_{D'\pm}\Gamma^{BB'C'D'E}\psi_{E\mp}\nonumber\\
&+\tau_{B'C'}{}^A\,\bar{\psi}_{D'-}\Gamma_A\Gamma^{B'C'D'E'}\psi_{E'-}\bigg)\,.
\end{align}
\end{subequations}
Finally, the NR supersymmetry transformation rules that leave $S_{NR}$ invariant (up to cubic fermion terms), upon imposition of the constraints \eqref{eq:DSNC-}, are found in \eqref{eq:deltasusy0} and \eqref{eq:NRsupersymmetry}, as well as in the $c\rightarrow \infty$ limit of \eqref{eq:bossusy}. Note that we can leave out the parts in \eqref{eq:deltasusy0} that involve $\eta_-$ and $\rho_-$ (whose explicit expressions are given in \eqref{eq:STrest}) from these NR supersymmetry transformation rules as these take the form of $S$- and $T$-symmetries.

\section{The NR Limit of the Equations of Motion} \label{sec:limiteoms}

\noindent In the previous section, we discussed the NR limit of the action of ten-dimensional $\mathcal{N}=1$ supergravity. We saw that the resulting NR action is only invariant under NR supersymmetry after the (supersymmetric) self-dual DSNC geometry constraints \eqref{eq:DSNC-} have been imposed by hand. Nevertheless, one can derive equations of motion from it by treating it as a pseudo-action, i.e., by applying the usual unconstrained variational principle and imposing the constraints \eqref{eq:DSNC-} only after variation. In this section, we will examine these equations of motion in more detail. We will see that they can be derived as NR limits of a subset of the equations of motion of relativistic $D=10$, $\mathcal{N}=1$ supergravity. As we will explain, this subset is a proper one due to the fact that the NR action is invariant under the emerging $S$-, $T$- and dilatation symmetries \eqref{eq:STsymm}, \eqref{eq:dilsymm}. The NR limit of the remaining relativistic equations of motion leads to extra `missing equations of motion'. These missing equations of motion consist of two fermionic equations, as well as a bosonic one that can be identified as a supersymmetric generalization of the Poisson equation for the Newton potential of NR gravity. Due to the fact that the NR action is only invariant under NR supersymmetry up to the self-dual DSNC geometry constraints, the equations of motion derived from it do not form a closed set under NR supersymmetry but can also transform to the missing equations of motion. The full set of missing equations of motion and equations of motion derived from the NR action does, however, form a supersymmetric set only if the self-dual DSNC geometry constraints \eqref{eq:DSNC-} are imposed by hand.

\subsection{Equations of Motion from the NR Action and Missing NR Equations of Motion}

\noindent Viewing the NR action $S_{NR}$, given in \eqref{eq:NRaction}, \eqref{eq:NRaction2} as a pseudo-action, we can derive equations of motion for $\tau_\mu{}^A$, $e_\mu{}^{A'}$, $\phi$, $b_{\mu\nu}$, $\lambda_\pm$ and $\psi_{\mu \pm}$, by computing Euler-Lagrange derivatives, denoted here by $\langle \tau \rangle_A{}^\mu$, $\langle e \rangle_{A'}{}^\mu$, $\langle \phi \rangle$, $\langle b \rangle^{\mu\nu}$, $\langle \lambda_\pm \rangle$ and $\langle \psi_\pm \rangle^\mu$ respectively, with respect to these fields. Explicitly, we define these functional derivatives as the result of performing an unconstrained variation of $S_{NR}$ as follows:
\begin{align} \label{eq:NReomvar}
\delta\,S_{NR} = \frac{1}{2\,\kappa^2}\int\,\rmd^{10}x\,e\,&\rme^{-2\,\phi}\Big\{\,\langle\tau\rangle_A{}^\mu\,\delta\tau_\mu{}^A + \langle e\rangle_{A'}{}^\mu\,\delta e_\mu{}^{A'} - 8\,\langle\phi\rangle\,\delta\phi +\frac12\,\langle b\rangle^{\mu\nu}\delta b_{\mu\nu}\notag\\
& \quad +4\,\delta\bar\lambda_+\,\langle\lambda_-\rangle + 4\,\delta\bar\psi_{\mu+}\langle\psi_-\rangle^\mu +4\,\delta\bar\lambda_-\,\langle\lambda_+\rangle + 4\,\delta\bar\psi_{\mu-}\langle\psi_+\rangle^\mu\Big\}\,,
\end{align}
where the coefficients of the different terms have been chosen for later convenience. The equations of motion derived from $S_{NR}$ are then given by setting $\langle \tau \rangle_A{}^\mu$, $\langle e \rangle_{A'}{}^\mu$, $\langle \phi \rangle$, $\langle b \rangle^{\mu\nu}$, $\langle \lambda_\pm \rangle$ and $\langle \psi_\pm \rangle^\mu$ to zero and supplementing the resulting set of equations by hand with the self-dual DSNC geometry constraints \eqref{eq:DSNC-}.

As can be expected, the equations thus found can also be obtained as a NR limit of the equations of motion of relativistic $10D$, $\mathcal{N}=1$ supergravity. To clarify this, we denote convenient combinations of the Euler-Lagrange derivatives with respect to $E_{\mu}{}^{\hat{A}}$, $B_{\mu\nu}$, $\Phi$, $\uplambda$ and $\Psi_\mu$ of the relativistic action \eqref{eq:Nisoneaction} by $[G]_{\hat{A}}{}^{\mu}$, $[B]^{\mu\nu}$, $[\Phi]$, $[\uplambda]$ and $[\Psi]^\mu$ respectively. We define these combinations via the following variation:
\begin{align}\label{eq:releom}
\delta S = \frac{1}{2\kappa^2}\int \rmd^{10} x \,E\,\rme^{-2\Phi}\Big\{ & -2\,[G]_{\hat{A}}{}^\mu \delta E_\mu{}^{\hat{A}} + \frac12\,[B]^{\mu\nu}\delta B_{\mu\nu}  - 8[\Phi]\delta\Phi \notag\\
& + 4\,\delta\bar\uplambda\,[\uplambda]+ 4\,\delta\bar\Psi_\mu\,\big([\Psi]^\mu + \Gamma^\mu\,[\uplambda]\big) \Big\}\,.
\end{align}
Up to the order in fermions we are working in, $[G]_{\hat{A}}{}^{\mu}$, $[B]^{\mu\nu}$, $[\Phi]$, $[\uplambda]$ and $[\Psi]^\mu$ are explicitly given by
\begin{subequations} \label{eq:releoms}
\begin{align}
[G]_{\hat{A}\mu} & \equiv \mathcal R_{\hat{A}\mu} + 2\,\nabla_{\hat{A}}\partial_\mu\Phi - \frac14 \mathcal{H}_{\hat{A}\rho\sigma} \mathcal{H}_\mu{}^{\rho\sigma} - 2\,E_{\mu\hat A}\,[\Phi] \ \ (+ \text{ quadratic fermion terms})\,, \label{eq:EOM1}\\
[B]_{\mu\nu} & \equiv \nabla^\rho \mathcal{H}_{\rho\mu\nu} - 2\lr\partial^\rho\Phi\rr \mathcal{H}_{\rho\mu\nu} \ \ (+ \text{ quadratic fermion terms}) \,, \label{eq:EOM2} \\
[\Phi] &\equiv \nabla^\mu\partial_\mu\Phi + \frac14 \mathcal{R}- \partial^\mu\Phi\partial_\mu\Phi - \frac{1}{48}\, \mathcal{H}_{\mu\nu\rho} \mathcal{H}^{\mu\nu\rho} \ \ (+ \text{ quadratic fermion terms}) \,, \label{eq:EOM3}\\
[\uplambda]&\equiv\slashed D\uplambda - \Gamma^{\mu\nu}D_\mu\Psi_\nu - \slashed\partial\Phi\uplambda - \Gamma^\mu\slashed\partial\Phi\Psi_\mu - \frac{1}{24}\mathcal H^{\mu\nu\rho}\big(\Gamma^\sigma{}_{\mu\nu\rho}\Psi_\sigma + \Gamma_{\mu\nu\rho}\uplambda\big) \,,\\
[\Psi]_\mu&\equiv 2\,\Gamma^\nu\big(D_{[\mu}\Psi_{\nu]}-\frac18\,\mathcal H_{\rho\sigma[\mu}\Gamma^{\rho\sigma}\Psi_{\nu]}\big) -  \big(D_\mu - \frac18\,\mathcal H_{\mu\nu\rho}\Gamma^{\nu\rho}\big)\uplambda \notag\\
&\quad+ \slashed\partial\Phi\Psi_\mu- \frac{1}{12} \Gamma^{\hat A\hat B\hat C}\Psi_\mu\mathcal H_{\hat A\hat B\hat C}  \,,
\end{align}
\end{subequations}
where (as in the relativistic action \eqref{eq:Nisoneaction}) the Ricci tensor $\mathcal{R}_{\mu\nu}$ and all covariant derivatives are constructed from the relativistic Levi-Civita (spin) connection. For brevity, we have not explicitly given the quadratic fermion terms in $[G]_{\hat{A}}{}^\mu$, $[B]_{\mu\nu}$ and $[\Phi]$.

The equations of motion, obtained by putting the Euler-Lagrange derivatives $\langle \tau \rangle_A{}^\mu$, $\langle e \rangle_{A'}{}^\mu$, $\langle \phi \rangle$, $\langle b \rangle^{\mu\nu}$, $\langle \lambda_\pm \rangle$ and $\langle \psi_\pm \rangle^\mu$ to zero, can then be obtained from a NR limit, in the sense that they correspond to the leading order terms in a $c^{-2}$--expansion of particular combinations of their relativistic counterparts $[G]_{\hat{A}}{}^{\mu}$, $[B]^{\mu\nu}$, $[\Phi]$, $[\uplambda]$ and $[\Psi]^\mu$. To quickly find out which combinations of the relativistic Euler-Lagrange derivatives in this way lead to the Euler-Lagrange derivatives for the NR fields, we note that we can use \eqref{eq:rescale}, \eqref{eq:rescaleinv} to write
\begin{align}
\delta E_\mu{}^A &= c\, \delta \tau_{\mu}{}^A \,,  \qquad \delta E_\mu{}^{A'} = \delta e_\mu{}^{A'} \,, \qquad \delta B_{\mu\nu} = -2\, c\, \epsilon_{AB} \delta \tau_{[\mu}{}^A E_{\nu]}{}^B + \delta b_{\mu\nu} \,, \qquad \delta \Phi = \delta \phi \,, \nonumber \\
\delta \bar{\uplambda} &= c^{1/2} \delta \bar{\lambda}_+ + c^{-1/2} \delta \bar{\lambda}_- \,,  \qquad \qquad \qquad \delta \bar{\Psi}_\mu = c^{1/2} \delta \bar{\psi}_{\mu +} + c^{-1/2} \delta \bar{\psi}_{\mu -} \,.
\end{align}
Using this in \eqref{eq:releom}, we can rewrite the variation of the relativistic action $S$ as follows
\begin{align}\label{eq:releom2}
\delta S = \frac{1}{2\kappa^2}\int & \rmd^{10}x \,E\,\rme^{-2\Phi}\Big\{ c\, [E]_A{}^\mu \delta \tau_\mu{}^A  + [E]_{A'}{}^\mu \delta e_\mu{}^{A'}  + \frac12\,[B]^{\mu\nu}\delta b_{\mu\nu}  - 8[\Phi]\delta\phi \notag\\
& + 4\,c^{1/2}\, \delta\bar\lambda_+ \,\Pi_- [\uplambda] +4\, c^{1/2}\, \delta\bar\psi_{\mu +}\, \Pi_- \big([\Psi]^\mu + \Gamma^\mu\,[\uplambda]\big) \notag \\ &+ 4\,c^{-1/2}\, \delta\bar\lambda_- \,\Pi_+[\uplambda] + 4\, c^{-1/2}\, \delta\bar\psi_{\mu -} \,\Pi_+ \big([\Psi]^\mu + \Gamma^\mu\,[\uplambda]\big) \Big\}\,,
\end{align}
where we have introduced the notation\footnote{The quantities $[E]_A{}^\mu$ and $[E]_{A'}{}^\mu$ then correspond to the Euler-Lagrange derivatives of the relativistic action \eqref{eq:Nisoneaction} with respect to $E_\mu{}^A$ and $E_\mu{}^{A'}$, after viewing \eqref{eq:Nisoneaction} as a functional of $b_{\mu\nu} = B_{\mu\nu} + \epsilon_{AB} E_\mu{}^A E_\nu{}^B$ (instead of as a functional of $B_{\mu\nu}$).}
\begin{align}
[E]_A{}^\mu &\equiv -2\,[G]_A{}^\mu - \epsilon_{AB} E_\nu{}^B [B]^{\mu\nu} \,, \qquad \qquad \qquad
[E]_{A'}{}^\mu \equiv -2\,[G]_{A'}{}^\mu \,.
\end{align}
Note that in \eqref{eq:releom2}, the quantities that multiply $\delta \tau_\mu{}^A$, $\delta e_\mu{}^{A'}$, $\delta b_{\mu\nu}$, $\delta \phi$, $\delta \bar{\lambda}_\pm$ and $\delta \bar{\psi}_{\mu \pm}$ have not yet been expanded in powers of $c^{-2}$ and are thus still given in terms of the relativistic fields $E_\mu{}^{\hat{A}}$, $B_{\mu\nu}$, $\Phi$, $\uplambda$ and $\Psi_\mu$. Performing a $c^{-2}$--expansion of these quantities in $\delta S$, noting that the result should take the form $\delta S = \delta S_{NR} + \mathcal{O}(c^{-2})$ (according to \eqref{eq:actionexpansion} with $S^{(0)} = S_{NR}$) and comparing with \eqref{eq:NReomvar}, we see that
\begin{align}\label{eq:nrEOM}
&\langle \tau\rangle_A{}^\mu=\big([E]_A{}^\mu\big)^{(-1)} \,, && \langle \phi\rangle=\big([\Phi]\big)^{(0)}\,, \notag\\
& \langle e\rangle_{A'}{}^\mu=\big([E]_{A'}{}^\mu\big)^{(0) \,, }&& \langle b\rangle^{\mu\nu}=\big([B]^{\mu\nu}\big)^{(0)}\,,\notag\\
&\langle \psi_-\rangle^\mu=\big(\Pi_-([\Psi]^\mu+\Gamma^\mu[\uplambda])\big)^{(-1/2)}\,,&& \langle \lambda_-\rangle=\big(\Pi_-[\uplambda])^{(-1/2)}\,,\notag\\
&\langle \psi_+\rangle^\mu=\big(\Pi_+([\Psi]^\mu+\Gamma^\mu[\uplambda])\big)^{(+1/2)}\,,&& \langle \lambda_+\rangle=\big(\Pi_+[\uplambda])^{(+1/2)}\,,
\end{align}
where here and in the following, the notation $(X)^{(n)}$ is used to denote all terms of order $c^n$ in the expression obtained by expanding a relativistic quantity $X$ in powers of $c^{-2}$, after performing the field redefinition \eqref{eq:rescaleinv}. In particular, all quantities on the right-hand-side of the equations in \eqref{eq:nrEOM} refer to the leading order terms in these $c^{-2}$--expansions.

The tensors $\langle \tau \rangle_A{}^\mu$, $\langle e \rangle_{A'}{}^\mu$, $\langle \phi \rangle$, $\langle b \rangle^{\mu\nu}$, $\langle \lambda_\pm \rangle$ and $\langle \psi_\pm \rangle^\mu$ are not all independent, as there exist various algebraic relations between them. The latter correspond to Noether identities for those local symmetries of the NR action \eqref{eq:NRaction}, under which none of the fundamental fields $\tau_\mu{}^A$, $e_\mu{}^{A'}$, $\phi$, $b_{\mu\nu}$, $\lambda_\pm$ and $\psi_{\mu \pm}$ transform as a gauge field. For longitudinal Lorentz transformations, transversal rotations and Galilean boosts, these Noether identities are given by
\begin{subequations}   \label{eq:Noetherids1}
\begin{align}
& \epsilon^{AB} \langle \tau \rangle_{AB} - 2 \bar{\lambda}_+ \langle \lambda_- \rangle + 2 \bar{\lambda}_- \langle \lambda_+ \rangle - 2 \bar{\psi}_{\mu +} \langle \psi_- \rangle^\mu + 2 \bar{\psi}_{\mu -} \langle \psi_+ \rangle^\mu= 0 \,, \label{eq:NoetherSO11}  \\
& \langle e \rangle_{[A'B']} - \bar{\lambda}_+ \Gamma_{A'B'} \langle \lambda_- \rangle - \bar{\lambda}_- \Gamma_{A'B'} \langle \lambda_+ \rangle - \bar{\psi}_{\mu +} \Gamma_{A'B'} \langle \psi_- \rangle^\mu - \bar{\psi}_{\mu -} \Gamma_{A'B'} \langle \psi_+ \rangle^\mu = 0 \,, \\
& \langle e \rangle_{A'A} + \epsilon_{A}{}^B \langle b \rangle_{BA'} + 2 \bar{\lambda}_+ \Gamma_{AA'} \langle \lambda_+ \rangle + 2 \bar{\psi}_{\mu +} \Gamma_{A A'} \langle \psi_+ \rangle^\mu = 0 \,.
\end{align}
\end{subequations}
Note that these 45 Noether identities imply that the 45 components $\langle \tau \rangle_{[AB]}$, $\langle e \rangle_{[A'B']}$ and $\langle e \rangle_{AA'}$ of the 100 Euler-Lagrange derivatives $\langle \tau \rangle_A{}^\mu$ and $\langle e \rangle_{A'}{}^\mu$ can be written in terms of other Euler-Lagrange derivatives. We are thus left with 55 algebraically independent components in $\langle \tau \rangle_A{}^\mu$ and $\langle e \rangle_{A'}{}^\mu$, the same number of components that is contained in the relativistic Einstein equations.

Naively, one would then say that the Euler-Lagrange derivatives $\langle \tau \rangle_A{}^\mu$, $\langle e \rangle_{A'}{}^\mu$, $\langle \phi \rangle$, $\langle b \rangle^{\mu\nu}$, $\langle \lambda_\pm \rangle$ and $\langle \psi_\pm \rangle^\mu$ have as many algebraically independent components as their relativistic counterparts $[G]^{\mu\nu}$, $[B]^{\mu\nu}$, $[\Phi]$, $[\uplambda]$ and $[\Psi]^\mu$ and that the NR action \eqref{eq:NRaction} thus leads to as many equations of motion as there are relativistic ones. This counting is however not correct as it does not yet take into account extra algebraic Noether identities that are associated to the emergent dilatation and $S$- and $T$-symmetries. In particular, the $S$- and $T$-symmetries \eqref{eq:STsymm} lead to the following relations
\begin{align}
\label{eq:NoetherST}
&  \langle \lambda_+ \rangle - \frac12 \Gamma_+ \langle \psi_- \rangle^\mu \tau_\mu{}^+ = 0 \,, \qquad \qquad \qquad \tau_\mu{}^+ \langle \psi_+ \rangle^\mu = 0 \,,
\end{align}
while the dilatation symmetry \eqref{eq:dilsymm} implies that
\begin{align}
\label{eq:Noetherdil1}
& \langle \tau \rangle_A{}^\mu \tau_\mu{}^A - 8 \langle \phi \rangle + 2 \bar{\lambda}_+ \langle \lambda_- \rangle - 2 \bar{\lambda}_- \langle \lambda_+ \rangle + 2 \bar{\psi}_{\mu +} \langle \psi_- \rangle^\mu - 2 \bar{\psi}_{\mu -} \langle \psi_+ \rangle^\mu = 0 \,.
\end{align}
In what follows, it will be useful to simplify this identity, by using the Noether identity \eqref{eq:NoetherSO11} for SO$(1,1)$ longitudinal Lorentz transformations to eliminate the last four terms. This gives
\begin{align} \label{eq:Noetherdil}
\langle \tau \rangle_A{}^\mu \tau_\mu{}^A + \epsilon^{AB} \langle \tau \rangle_{AB} - 8 \langle \phi \rangle = 0 \qquad \qquad \Leftrightarrow \qquad \qquad \tau_\mu{}^- \langle \tau \rangle_-{}^\mu = 4 \langle \phi \rangle \,.
\end{align}
From \eqref{eq:NoetherST} and \eqref{eq:Noetherdil}, we see that $\langle \psi_+ \rangle^+$ is identically zero and that e.g. $\langle \tau \rangle_-{}^{-}$ and $\langle \lambda_+ \rangle$ are not independent.

In \eqref{eq:nrEOM}, we saw that the equations of motion, derived from the action \eqref{eq:NRaction}, arise from a NR limit that consists of retaining only the leading order terms in the $c^{-2}$--expansion of the relativistic equations of motion, obtained by setting $[E]_A{}^\mu$, $[E]_{A'}{}^\mu$, $[B]^{\mu\nu}$, $[\Phi]$, $[\Psi]^\mu + \Gamma^\mu [\uplambda]$ and $[\uplambda]$ equal to zero. The Noether identities \eqref{eq:NoetherST} and \eqref{eq:Noetherdil} then tell us that taking the limit in this way leads to some of the resulting equations being the same or identically zero, so that one is left with less independent NR equations than relativistic ones. It is however also possible to take the NR limit directly at the level of the relativistic equations of motion, in such a way that it preserves the total number of algebraically independent equations. To see how this works, we note that the Noether identity for dilatations says that not all the leading order components in the $c^{-2}$--expansions of the relativistic $[E]_A{}^\mu$ and $[\Phi]$ are linearly independent. Indeed, since $\tau_\mu{}^- \langle \tau \rangle_-{}^\mu = \big(E_\mu{}^- [E]_-{}^\mu\big)^{(0)}$, the identity \eqref{eq:Noetherdil} tells us that the leading ($c^0$--)order contributions in the expansions of $E_\mu{}^- [E]_-{}^\mu$ and $[\Phi]$ are proportional to each other:
\begin{align} \label{eq:rewriteNoetherdil}
& \tau_\mu{}^- \langle \tau \rangle_-{}^\mu = 4 \langle \phi \rangle \qquad \Leftrightarrow  \qquad \big(E_\mu{}^- [E]_-{}^\mu\big)^{(0)}= 4 \big([\Phi]\big)^{(0)} \notag \\
& \qquad \Leftrightarrow \qquad E_\mu{}^- [E]_-{}^\mu = 4 \langle \phi \rangle + \mathcal{O}(c^{-2}) \qquad \text{and} \qquad [\Phi] = \langle \phi \rangle + \mathcal{O}(c^{-2}) \,.
\end{align}
Similarly, the Noether identities \eqref{eq:NoetherST} for the $S$- and $T$-symmetries are equivalent to saying that the contribution to the $c^{-2}$--expansion of certain components of $[\Psi]^\mu$ vanishes identically at the order indicated in \eqref{eq:nrEOM}:
\begin{align} \label{eq:rewriteNoetherST}
\big(\Pi_-\,E_-{}^\mu [\Psi]_\mu\big)^{(1/2)} = 0\,,\qquad\mathrm{and}\qquad \big(\Pi_+\,E_-{}^\mu [\Psi]_\mu\big)^{(3/2)} &= 0\,.
\end{align}
This then indicates how one can take the NR limit of the equations of motion, such that one ends up with as many NR equations of motion as relativistic ones. The limit of most of the relativistic equations of motion is taken as in \eqref{eq:nrEOM}. As regards the equations $\Pi_\pm\,E_-{}^\mu [\Psi]_\mu = 0$ however, one has to take into account that they vanish at the order given in \eqref{eq:nrEOM} and that one should instead retain the terms at one order lower in the $c^{-2}$--expansion. Furthermore, instead of applying the NR limit to the equations $\{E_\mu{}^- [E]_-{}^\mu = 0, [\Phi] = 0\}$, one should apply it to $\{E_\mu{}^- [E]_-{}^\mu = 0, E_\mu{}^- [E]_-{}^\mu - 4 [\Phi]=0\}$, so that one ends up with two linearly independent equations. Taking the NR limit of the equations of motion in this way, the set of NR equations of motion, obtained by setting \eqref{eq:nrEOM} to zero, is then supplemented with the following extra equations:
\begin{align}\label{eq:missingeoms}
\langle \psi^{(S)}_-\rangle &\equiv \big(\Pi_-\,E_-{}^\mu [\Psi]_\mu\big)^{(-3/2)} = 0\,, \qquad  \qquad \qquad \langle \psi^{(T)}_+\rangle \equiv \big(\Pi_+\,E_-{}^\mu [\Psi]_\mu\big)^{(-1/2)} = 0 \,, \notag \\
\langle P \rangle &\equiv \big(E_\mu{}^- [E]_-{}^\mu - 4 [\Phi] \big)^{(-2)} = 0 \,.
\end{align}
We will refer to these as `the missing equations of motion'. Although they are not derived from the NR action \eqref{eq:NRaction}, they are valid NR equations of motion, in the sense that they correspond to the leading order in the $c^{-2}$--expansion of particular components/combinations of components of the relativistic equations of motion.

Explicitly, the fermionic missing equations of motion are given by:
\begin{subequations} \label{eq:fermmissingexpl}
\begin{align}
\langle \psi^{(S)}_-\rangle  \equiv & -D_-\lambda_- +2\,\tau_-{}^\mu e_{A'}{}^\nu\,\Gamma^{A'}D_{[\mu}\psi_{\nu]-} + 2\,\tau_-{}^\mu\tau_+{}^\nu\,\Gamma^+D_{[\mu}\psi_{\nu]+} \notag\\
&+ \big(\Gamma^{A'}D_{A'}\phi + \frac{1}{24}\,\Gamma^{A'B'C'}h_{A'B'C'}\big)\tau_-{}^\mu\psi_{\mu-}-2\,\tau_{A'}{}^{++}\Gamma^{A'}\tau_+{}^\mu\psi_{\mu-}  \notag \\ & \  \text{(+ terms of higher order in the fermions)} = 0 \,,\\
\langle \psi^{(T)}_+\rangle  \equiv  & -D_-\lambda_+ +2\,\tau_-{}^\mu e_{A'}{}^\nu\,\Gamma^{A'}D_{[\mu}\psi_{\nu]+} + \big(\Gamma^{A'}D_{A'}\phi + \frac{1}{24}\,\Gamma^{A'B'C'}h_{A'B'C'}\big)\tau_-{}^\mu\psi_{\mu+}\notag\\
&+\tau_{A'}{}^{++}\big(\Gamma^{A'-}\,\lambda_- - \Gamma^{B'}\Gamma^{A'-}\psi_{B'-}\big) + \frac12\,\tau_{A'B'}{}^+\Gamma^{A'B'-}\,\tau_-{}^\mu\psi_{\mu -} \notag \\ & \ \text{(+ terms of higher order in the fermions)} = 0\,.
\end{align}
\end{subequations}
One can then explicitly check that $\langle \psi^{(S)}_-\rangle$ and $\langle \psi^{(T)}_+\rangle$ are invariant under $S$- and $T$- transformations, while they transform covariantly under dilatations, with weights $-3/2$ and $-1/2$ respectively. The explicit expression for the bosonic equation of motion reads:
\begin{align} \label{eq:Poisson}
\langle P\rangle &
\equiv \tau_A{}^\mu e_{A^\prime}{}^\nu \rmR_{\mu\nu}(G)^{AA'} + \epsilon^{AB}\tau_A{}^\mu \tau_B{}^\nu \rmR_{\mu\nu}(M) \ \, \text{(+ fermionic contributions)} = 0\,,
\end{align}
where $\rmR_{\mu\nu}(G)^{AA'}$ and $\rmR_{\mu\nu}(M)$ are defined in \eqref{eq:curv2forms}. The linearization of the bosonic part of this equation contains a term $\partial_{A'} \partial^{A'} b_{01}$. Since in \cite{Bergshoeff:2021bmc}, it was argued that $b_{01}$ can be identified as the Newton potential, we see that the missing bosonic equation can be identified with a supersymmetric generalization of the Poisson equation and we will refer to it as `the Poisson equation' in what follows. Under dilatations, $\langle P \rangle$ scales covariantly with weight $-2$, while the following transformation rules under the $S$- and $T$-symmetries \eqref{eq:STsymm} are found
\begin{align} \label{eq:STPoisson}
\delta_T \langle P \rangle &= 4\, \bar{\rho}_- \langle \psi^{(T)}_+ \rangle \,, \qquad \qquad \delta_S \langle P \rangle = 2\, \bar{\eta}_- \Gamma^- \langle \psi_-^{(S)} \rangle \,.
\end{align}
The Poisson equation thus transforms to the two fermionic missing equations of motion under the $S$- and $T$-symmetries. Since the NR action is invariant under dilatations, $S$- and $T$-symmetries, we thus find that the full set of NR field equations, including the missing ones, is invariant under these symmetries.

The above discussion can be made more transparent, by performing a field redefinition in the relativistic action \eqref{eq:Nisoneaction}. This field redefinition is such that the full set of NR field equations, including the missing ones, is obtained by retaining the leading orders of the $c^{-2}$--expansions of the equations of motion of the redefined fields (instead of of non-trivial combinations of equations of motion of different fields). We will denote the redefined fields with a tilde, as some of them correspond to rescalings of the NR fields with a tilde, that were defined in \eqref{eq:NRfermionstilde}, with a power of $c$. This field redefinition is explicitly given by:
\begin{align}\label{eq:Einsteinframe}
&\tilde E_\mu{}^- \equiv \rme^{-2\Phi}E_\mu{}^-\,, && \tilde E_\mu{}^+ \equiv E_\mu{}^+\,,&&\tilde E_\mu{}^{A'} \equiv E_\mu{}^{A'}\,,\notag\\
&\tilde \Phi \equiv \Phi\,,&&\tilde B_{\mu\nu} \equiv b_{\mu\nu} \equiv B_{\mu\nu} +\epsilon_{AB}\,E_\mu{}^A E_\nu{}^B\,, \notag\\
&\tilde\uplambda_\pm \equiv \uplambda_\pm \,, && \tilde\Psi_{\mu +} \equiv \Pi_+\Psi_\mu - \frac12\,E_\mu{}^+\Gamma_+\Pi_-\uplambda\,,\notag\\
&\tilde\Psi_{-} \equiv E_+{}^\mu\Pi_-\Psi_\mu\,, && \tilde\Psi_{\mu-} \equiv \Pi_-\Psi_\mu - E_\mu{}^+E_+{}^\nu\Pi_-\Psi_\nu\,.
\end{align}
Note that this field redefinition is invertible. We can then use \eqref{eq:rescaleinv} to write the fields with a tilde as powers of $c$ multiplied with NR fields with a tilde, that can be expressed in terms of the NR fields without a tilde:
\begin{align}\label{eq:Einsteinframec}
&\tilde E_\mu{}^- \equiv c^{-1}\, \tilde{\tau}_\mu{}^- \equiv c^{-1}\, \rme^{-2\phi} \tau_\mu{}^-\,, && \tilde E_\mu{}^+ \equiv c \, \tilde{\tau}_\mu{}^+ \equiv c\, \tau_\mu{}^+ \,,&&\tilde E_\mu{}^{A'} \equiv \tilde{e}_\mu{}^{A'} \equiv e_\mu{}^{A'}\,,\notag\\
&\tilde \Phi \equiv \tilde{\phi} + \log c = \phi + \log c \,,&&\tilde B_{\mu\nu} \equiv \tilde{b}_{\mu\nu} \equiv b_{\mu\nu}\,, \notag\\
&\tilde\uplambda_\pm \equiv c^{\pm 1/2} \tilde\lambda_\pm = c^{\pm 1/2} \lambda_\pm \,, && \tilde\Psi_{\mu +} \equiv c^{1/2} \tilde{\psi}_{\mu +} \,,\notag\\
&\tilde\Psi_{-} \equiv c^{-3/2} \tilde{\psi}_- \,, && \tilde\Psi_{\mu-} \equiv c^{-1/2} \tilde{\psi}_{\mu -}\,.
\end{align}
The expressions for $\tilde{\psi}_{\mu \pm}$ and $\tilde{\psi}_-$ in terms of NR fields without a tilde are given in \eqref{eq:NRfermionstilde}. As in \eqref{eq:rescaleinv}, the dilatation weights of the NR fields with a tilde in the above formulas coincide with the exponents of the powers of $c$ that multiply these fields. Note that \eqref{eq:Einsteinframec} contains two types of redefinition that will be used in the following. On the one hand, it expresses how the relativistic fields with a tilde are given in terms of NR fields with a tilde, multiplied with a power of $c$. On the other hand, it also indicates how the NR fields with a tilde are related to those without a tilde.

By applying \eqref{eq:Einsteinframe}, we can express the relativistic action \eqref{eq:Nisoneaction} in terms of the fields with a tilde and define functional derivatives of the resulting action $S\big[\tilde E_\mu{}^\pm,\tilde E_\mu{}^{A'},\tilde \Phi,\tilde B_{\mu\nu},\tilde \Psi_{\mu\pm},\tilde\Psi_{-},\tilde\uplambda_\pm]$ via the following variation
\begin{align} \label{eq:releom3}
\delta S = \frac{1}{2\kappa^2}\int \rmd^{10}x\,\tilde E\Big\{ & \widetilde{[E]}_-{}^\mu\delta\tilde E_\mu{}^- + \widetilde{[E]}_+{}^\mu\delta\tilde E_\mu{}^+ + \widetilde{[E]}_{A'}{}^\mu\delta\tilde E_\mu{}^{A'} + \frac12 \widetilde{[B]}^{\mu\nu} \delta \tilde{B}_{\mu\nu} -8\,\widetilde{[\Phi]}\delta\tilde\Phi \notag\\
&+ 4\,\delta\bar{\tilde \uplambda}_+\widetilde{[\uplambda_-]}  + 4\,\delta\bar{\tilde\Psi}_{\mu+}\widetilde{[\Psi_-]}^\mu + 4\,\delta\bar{\tilde\Psi}_{\mu-}\widetilde{[\Psi_+]}^\mu\notag\\
&+ 4\,\delta\bar{\tilde \uplambda}_-\widetilde{[\uplambda_+]} + 4\,\delta\bar{\tilde \Psi}_{-}\widetilde{[\Psi_{+}]}\Big\}\,,
\end{align}
where $\tilde{E} = \mathrm{det}(\tilde{E}_\mu{}^{\hat{A}})$. Using the rules \eqref{eq:Einsteinframe}, one finds that the following non-trivial relations hold between the functional derivatives with respect to the fields with tildes and those with respect to the original fields:
\begin{align}
&\widetilde{[E]}_-{}^\mu = \rme^{2\Phi}[E]_-{}^\mu\,, && \widetilde{[E]}_+{}^\mu = [E]_+{}^\mu + 4\,\bar{\tilde\Psi}_{-}\widetilde{[\Psi_+]}^\mu - 2\,\bar{\tilde\uplambda}_-\Gamma_+\widetilde{[\Psi]}^\mu\,,\notag\\
&\widetilde{[\Phi]} = [\Phi] - \frac14\,E_\mu{}^-[E]_-{}^\mu\,,&& \widetilde{[\Psi_\pm]}^\mu = \Pi_\pm\big([\Psi]^\mu + \Gamma^\mu[\uplambda]\big)\,,\\
&\widetilde{[\Psi_{+}]} = E_\mu{}^+\Pi_+[\Psi]^\mu\,, && \widetilde{[\uplambda_+]} = -\frac12\,E_\mu{}^+\Pi_+\Gamma_+[\Psi]^\mu\,.\notag
\end{align}
For all other functional derivatives, the relation is trivial, e.g., $\widetilde{[E]}_{A'}{}^\mu = [E]_{A'}{}^\mu$. From this, we see that the field equations of $\tilde{\Phi}$, $\tilde{\uplambda}_-$ and $\tilde{\Psi}_{-}$ are given by
\begin{align}
E_\mu{}^- [E]_-{}^\mu - 4 [\Phi] = 0 \,, \qquad \Pi_-\,E_-{}^\mu [\Psi]_\mu = 0\,, \qquad\Pi_+\,E_-{}^\mu [\Psi]_\mu = 0\,,
\end{align}
and, according to \eqref{eq:missingeoms} and the discussion preceding it, thus indeed reproduce the missing equations of motion in their $c^{-2}$--expansion, as was the goal of the field redefinition \eqref{eq:Einsteinframe}.

We can then revisit the NR limit of the equations of motion. First, using \eqref{eq:Einsteinframec} in \eqref{eq:releom3}, we have
\begin{align} \label{eq:releom4}
\delta S = \frac{1}{2\kappa^2}\int \rmd^{10}x\,\tilde E\Big\{ & c^{-1} \, \widetilde{[E]}_-{}^\mu\delta\tilde\tau_\mu{}^- + c\, \widetilde{[E]}_+{}^\mu\delta\tilde\tau_\mu{}^+ + \widetilde{[E]}_{A'}{}^\mu\delta\tilde e_\mu{}^{A'} + \frac12 \widetilde{[B]}^{\mu\nu} \delta \tilde{b}_{\mu\nu} -8\,\widetilde{[\Phi]}\delta\tilde\phi \notag\\
&+ 4\, c^{1/2}\, \delta\bar{\tilde \lambda}_+\widetilde{[\uplambda_-]}  + 4\, c^{1/2}\, \delta\bar{\tilde\psi}_{\mu+}\widetilde{[\Psi_-]}^\mu + 4\,c^{-1/2} \, \delta\bar{\tilde\psi}_{\mu-}\widetilde{[\Psi_+]}^\mu\notag\\
&+ 4\, c^{-1/2} \, \delta\bar{\tilde \lambda}_-\widetilde{[\uplambda_+]} + 4\, c^{-3/2} \, \delta\bar{\tilde \psi}_{-}\widetilde{[\Psi_{+}]}\Big\}\,.
\end{align}
As in \eqref{eq:releom2}, the quantities that multiply $\delta \tilde\tau_\mu{}^\pm$, $\delta \tilde e_\mu{}^{A'}$, $\delta \tilde b_{\mu\nu}$, $\delta \tilde \phi$, $\delta \bar{\tilde\lambda}_\pm$, $\delta \bar{\tilde\psi}_{\mu \pm}$ and $\delta \bar{\tilde{\psi}}_-$ have not yet been expanded in powers of $c^{-2}$. Requiring compatibility with $\delta S = \delta S_{NR} + c^{-2} \delta S^{(-2)}$ (where $S_{NR}$ and $S^{(-2)}$ are expressed in terms of the NR fields with a tilde) shows that the following expansions hold:
\begin{align}\label{eq:bulkeom}
&\widetilde{[E]}_-{}^\mu = c\,\widetilde{\langle \tau\rangle}_-{}^\mu + \mathcal O(c^{-1})\,, &&\widetilde{[E]}_+{}^\mu = c^{-1}\,\widetilde{\langle \tau\rangle}_+{}^\mu + \mathcal O (c^{-3})\,,\notag\\
&\widetilde{[E]}_{A'}{}^\mu = \widetilde{\langle e\rangle}_{A'}{}^\mu + \mathcal O(c^{-2}) && \widetilde{[B]}^{\mu\nu} = \widetilde{\langle b\rangle}^{\mu\nu} + \mathcal O(c^{-2})\,,\notag\\
&\widetilde{[\Psi_\pm]}^\mu = c^{\pm 1/2}\widetilde{\langle\psi_\pm\rangle}^\mu + \mathcal O(c^{\pm 1/2-2})\,, &&\widetilde{[\uplambda_-]} = c^{-1/2}\widetilde{\langle\lambda_-\rangle} + \mathcal{O}(c^{-5/2})\,.
\end{align}
The quantities $\widetilde{\langle \tau\rangle}_\pm{}^\mu$, $\widetilde{\langle e\rangle}_{A'}{}^\mu$, $\widetilde{\langle b\rangle}^{\mu\nu}$, $\widetilde{\langle\psi_\pm\rangle}^\mu$ and $\widetilde{\langle\lambda_-\rangle}$ then correspond to the functional derivatives of the NR action $S_{NR}$ with respect to $\tilde{\tau}_\mu{}^\pm$, $\tilde{e}_\mu{}^{A'}$, $\tilde{b}_{\mu\nu}$, $\tilde{\psi}_{\mu \mp}$ and $\tilde{\lambda}_+$ (after expressing $S_{NR}$ in terms of these tilded NR fields).

The same reasoning would lead one to think that the expansions of $\widetilde{[\Phi]}$, $\widetilde{[\uplambda_+]}$ and $\widetilde{[\Psi_+]}$ start at orders $c^0$, $c^{1/2}$ and $c^{3/2}$ respectively. This is however not correct. Indeed, if this were true, \eqref{eq:releom4} would imply that $S_{NR}$ depends on $\tilde{\phi}$, $\tilde{\lambda}_-$ and $\tilde{\psi}_-$. This can however not be the case, since these fields shift as St\"uckelberg fields under dilatations and $S$- and $T$-symmetries. Any dependence of $S_{NR}$ on $\tilde{\phi}$, $\tilde{\lambda}_-$ and $\tilde{\psi}_-$ would then imply that $S_{NR}$ is not invariant under these symmetries, contradicting what was found in section \ref{sec:limitaction}. We thus conclude that the expansions of $\widetilde{[\Phi]}$, $\widetilde{[\uplambda_+]}$ and $\widetilde{[\Psi_+]}$ have to start at one $c^{-2}$--order higher. This is indeed found explicitly:
\begin{align}\label{eq:misseom}
&\widetilde{[\Phi]} = -\frac{1}{4} c^{-2} \, \langle P\rangle + \mathcal O(c^{-4})\,, && \widetilde{[\uplambda_+]} = \frac{c^{-3/2}}{2}\,\Gamma_+\langle\psi_-^{(S)}\rangle + \mathcal O (c^{-7/2})  \,, \notag\\
&\widetilde{[\Psi_{+}]} = c^{-1/2}\langle \psi_+^{(T)}\rangle + \mathcal O(c^{-5/2}) \,.
\end{align}
From this, we then also see that $\langle P \rangle$, $\langle\psi_-^{(S)}\rangle$ and $\langle \psi_+^{(T)}\rangle$ can be interpreted as functional derivatives of $S^{(-2)}$ with respect to $\tilde{\phi}$, $\tilde{\lambda}_-$ and $\tilde{\psi}_-$. This is similar to what happens when considering the NR expansion of General Relativity, where the Poisson equation of NR gravity is seen to arise from subleading orders in the expansion of the Einstein-Hilbert action \cite{Hansen:2019pkl,Hansen:2020pqs}. Note also that there is a relation between the dilatation weights of the NR fields with a tilde and the exponent of the power of $c$ in front of their corresponding functional derivatives (either of $S_{NR}$ or $S^{(-2)}$) in \eqref{eq:bulkeom} and \eqref{eq:misseom}. The exponent of the power of $c$ in front of $\widetilde{\langle \tau\rangle}_-{}^\mu$, $\widetilde{\langle \tau\rangle}_+{}^\mu$, $\widetilde{\langle e\rangle}_{A'}{}^\mu$, $\widetilde{\langle b\rangle}^{\mu\nu}$, $\widetilde{\langle\psi_\pm\rangle}^\mu$ and $\widetilde{\langle\lambda_-\rangle}$ in \eqref{eq:bulkeom} is given by the negative of the dilatation weight of $\tilde{\tau}_\mu{}^-$, $\tilde{\tau}_\mu{}^+$, $\tilde{e}_\mu{}^{A'}$, $\tilde{b}_{\mu\nu}$, $\tilde{\psi}_{\mu \mp}$ and $\tilde{\lambda}_+$ respectively. This rule does not hold for $\langle P \rangle$, $\langle\psi_-^{(S)}\rangle$ and $\langle \psi_+^{(T)}\rangle$: the exponent of the power of $c$ in front of these quantities in \eqref{eq:misseom} is obtained by subtracting two from the negative of the dilatation weight of $\tilde{\phi}$, $\tilde{\lambda}_-$ and $\tilde{\psi}_-$ respectively.

Summarizing: after performing the field redefinition \eqref{eq:Einsteinframe}, the NR limit of the relativistic equations of motion can more easily be taken in such a way that it preserves the number of algebraically independent equations. The resulting NR field equations are given by
\begin{alignat}{3} \label{eq:allNReoms}
\widetilde{\langle \tau \rangle}_-{}^\mu &= 0 \,, \qquad \qquad & \widetilde{\langle \tau \rangle}_+{}^\mu &= 0 \,, \qquad \qquad & \widetilde{\langle e \rangle}_{A'}{}^\mu &= 0 \,, \nonumber \\
\widetilde{\langle b \rangle}^{\mu\nu} &= 0 \,, \qquad \qquad & \widetilde{\langle \psi_\pm \rangle}^\mu &= 0 \,, \qquad \qquad & \widetilde{\langle \lambda_- \rangle} &= 0 \,, \nonumber \\
\langle P \rangle &= 0 \,, \qquad \qquad & \langle\psi_-^{(S)}\rangle &= 0 \,, \qquad \qquad & \langle \psi_+^{(T)}\rangle &= 0 \,,
\end{alignat}
where each equation corresponds to the leading order terms in the $c^{-2}$--expansion of a relativistic equation of motion for a redefined field with a tilde. The first two lines correspond to equations that can be derived from the NR action \eqref{eq:NRaction}. The equations in the last line are the missing equations of motion, that do not follow from the NR action. The full set of NR field equations \eqref{eq:allNReoms} is invariant under the emergent dilatation and $S$- and $T$-symmetries. In the next subsection, we will address the question whether these equations also transform into each other under NR supersymmetry and Galilean boosts.

\subsection{Consistency of all NR Equations of Motion Under Supersymmetry and Galilean Boosts}\label{sec:consistencyofEOM}

\noindent Here, we will give a generic argument that the set of NR field equations \eqref{eq:allNReoms} is invariant under NR supersymmetry and Galilean boosts. As we will see, invariance under NR supersymmetry is not automatically guaranteed: it only holds when the (supersymmetric) self-dual DSNC constraint \eqref{eq:DSNC-} is imposed by hand.

We will argue exclusively in terms of the fields with tilde \eqref{eq:Einsteinframe} and their NR counterparts, defined in \eqref{eq:Einsteinframec}. It will then be useful to split the tilded fields \eqref{eq:Einsteinframe} in two sets, based on whether the limit of their equations of motion can be derived from the NR action \eqref{eq:NRaction} or not. The first set of fields is given by $\{\tilde{E}_\mu{}^{+}, \tilde{E}_\mu{}^{-}, \tilde{E}_\mu{}^{A'}, \tilde{B}_{\mu\nu}, \tilde{\Psi}_{\mu \pm}, \tilde{\uplambda}_+\}$. We will refer to the fields in this set as the (relativistic) `bulk fields' and we will collectively denote them as $B_i$, with the index $i$ enumerating the different bulk fields. The second set of fields is given by $\{\tilde{\Phi}, \tilde{\Psi}_-, \tilde{\uplambda}_-\}$ and their members will be referred to as the (relativistic) `missing fields'. We will collectively denote them as $M_\alpha$, where the index $\alpha$ is used to distinguish the different missing fields. The index $I$ and notation $X_I$ will be used to denote the members of the collection of bulk and missing fields: $\{X_I\} = \{B_i, M_\alpha\}$. The functional derivative of the relativistic action with respect to a field $B_i$, $M_\alpha$ or $X_I$ will be denoted by $[B]^i$, $[M]^\alpha$, $[X]^I$ respectively. We will split the NR fields with a tilde in a set of NR bulk fields and one of NR missing fields in an analogous manner. The NR bulk fields are given by $\{\tilde{\tau}_\mu{}^+, \tilde{\tau}_\mu{}^-, \tilde{e}_\mu{}^{A'}, \tilde{b}_{\mu\nu}, \tilde{\psi}_{\mu\pm}, \tilde{\lambda}_+\}$ and will be collectively denoted by $b_i$, whereas the NR missing fields $\{\tilde{\phi}, \tilde{\psi}_-, \tilde{\lambda}_-\}$ will be collectively denoted by $m_\alpha$. Equations \eqref{eq:Einsteinframec} can then be summarized as
\begin{align} \label{eq:expBM}
B_i = c^{b(i)} b_i \,, \qquad \qquad M_\alpha = c^{m(\alpha)} m_\alpha \,,
\end{align}
where $b(i)$ and $m(\alpha)$ are the dilatation weigths of the corresponding NR fields $b_i$ and $m_\alpha$. According to the remark made below \eqref{eq:misseom}, the expansions of the relativistic functional derivatives $[B]^i$ and $[M]^\alpha$ then take the form:
\begin{align} \label{eq:expBMeom}
[B]^i &= c^{-b(i)} \langle b \rangle^i + c^{-b(i) - 2} [B]_{SL}^i + \mathcal{O}(c^{-b(i) - 4}) \,, \nonumber \\ [M]^\alpha &= c^{-m(\alpha) - 2} \langle m \rangle^\alpha + c^{-m(\alpha) - 4} [M]_{SL}^\alpha + \mathcal{O}(c^{-m(\alpha) - 6})  \,.
\end{align}
Here, the collection of $\langle b \rangle^i$ corresponds to $\{\widetilde{\langle \tau\rangle}_-{}^\mu, \widetilde{\langle \tau\rangle}_+{}^\mu, \widetilde{\langle e\rangle}_{A'}{}^\mu, \widetilde{\langle b\rangle}^{\mu\nu}, \widetilde{\langle\psi_\pm\rangle}^\mu, \widetilde{\langle\lambda_-\rangle}\}$, the collection of $\langle m \rangle^\alpha$ to $\{\langle P \rangle, \langle\psi_-^{(S)}\rangle, \langle \psi_+^{(T)}\rangle\}$ and we have denoted the first subleading terms in the expansions of $[B]^i$ and $[M]^\alpha$ by $[B]^i_{SL}$ and $[M]^\alpha_{SL}$.

We then wish to show that the $\langle b \rangle^i$ and $\langle m \rangle^\alpha$ transform into each other under NR supersymmetry and Galilean boosts. To do this, we will rely on a formula, derived in \cite{Vanhecke:2017chr}, that shows how Euler-Lagrange derivatives, derived from an action, transform into each other under a symmetry of that action. Applied to the $[X]^I$, defined via \eqref{eq:releom3}, this formula reads:
\begin{align}\label{eq:simsekvanproeyen}
\delta [X]^I = \big(\tilde{E}\,\delta \tilde{E}^{-1}\big)\,[X]^I - \frac{\delta\,(\delta X_J)}{\delta X_I} [X]^J\,.
\end{align}
The last term of \eqref{eq:simsekvanproeyen} is written in the DeWitt notation \cite{DeWitt:1967ub}, i.e., the sum over $J$ also entails an integral that is not written out explicitly. Furthermore, $\delta X_I$ refers to an infinitesimal symmetry transformation of the fields $X_I$ that leaves the relativistic action \eqref{eq:Nisoneaction} (expressed in terms of the fields with tilde \eqref{eq:Einsteinframe}) invariant. The above formula then specifies that the way in which the $[X]^I$ transform into each other under the symmetry variation $\delta$, is determined by the functional derivatives $\delta (\delta X_J)/\delta X_I$ of $\delta X_J$ with respect to $X_I$.\footnote{In case $\delta X_J$ involves derivatives of the fields $X_I$, as is the case for supersymmetry, one can see that the second term of \eqref{eq:simsekvanproeyen} contributes terms that involve derivatives of the symmetry parameters, i.e., non-covariant terms. These non-covariant terms are, however, still zero on-shell. We refer to \cite{Vanhecke:2017chr} for more details.} In the following, we will start from the formula \eqref{eq:simsekvanproeyen}, applied to relativistic supersymmetry and boosts, and expand it in powers of $c^{-2}$. This will allow us to infer how $\langle b \rangle^i$ and $\langle m \rangle^\alpha$ transform into each other under NR supersymmetry and Galilean boosts.

Before showing that all NR field equations \eqref{eq:allNReoms} transform into each other under NR supersymmetry, we need to investigate the structure of the NR supersymmetry transformation rules in more detail. First, we note that by writing the $c^{-2}$--expansions \eqref{eq:bossusy}, \eqref{eq:susyexpansion} of the relativistic supersymmetry transformation rules (denoted here by $\delta_Q$) in terms of the NR fields with a tilde, the following $c^{-2}$--expansions are seen to hold:
\begin{align}
\label{eq:cexpdeltaQrel}
\delta_Q b_i &= \delta^{(0)}_Q b_i + c^{-2} \delta^{(-2)}_Q b_i \,, \nonumber \\
\delta_Q m_\alpha &= c^2 \delta_Q^{(2)} m_\alpha + \delta^{(0)}_Q m_\alpha + c^{-2} \delta^{(-2)}_Q m_\alpha \,.
\end{align}
The only non-zero $\delta_Q^{(2)} m_\alpha$ take the form of specific $S$- and $T$-transformations and are determined by \eqref{eq:susydivergencered}. Note that $\delta_Q^{(2)} m_\alpha$ vanishes when the self-dual DSNC constraint \eqref{eq:DSNC-} is imposed. The terms $\delta_Q^{(0)} b_i$ and $\delta_Q^{(0)} m_\alpha$ at order $c^0$ in \eqref{eq:cexpdeltaQrel} constitute the NR supersymmetry transformation rules. We will for simplicity ignore the $S$- and $T$-transformations \eqref{eq:STrest} that are in principle present in $\delta_Q^{(0)} m_\alpha$. Doing this will not affect our arguments significantly. Using \eqref{eq:Einsteinframec}, we can then express these NR supersymmetry rules in terms of the fields with a tilde. In what follows, it will turn out to be important that $\delta_Q^{(0)} b_i$ takes the following form
\begin{align}\label{eq:splitoffm}
\delta_Q^{(0)}b_i = \tilde\delta b_i + \delta_{\mathrm{SO}(1,1)} b_i+ \delta_{\mathrm{boost}} b_i\,,
\end{align}
where $\tilde\delta b_i$ is independent of the missing fields $m_\alpha$ (so $\delta\big(\tilde\delta b_i\big)/\delta m_\alpha = 0$) and $\delta_{\mathrm{SO}(1,1)} b_i$ and $\delta_{\mathrm{boost}} b_i$ correspond to a local longitudinal SO$(1,1)$ transformation and Galilean boost, whose parameters depend on the missing fields $m_\alpha$. Explicitly, the parameters of $\delta_{\mathrm{SO}(1,1)}$ and $\delta_{\mathrm{boost}}$ are given (up to bilinear fermion terms in the fermionic transformation rules) by\footnote{Note that none of the $\delta^{(0)}_Q b_i$ then depends on the dilaton $\tilde\phi$. A priori it is not clear why this is the case, as the dilaton appears explicitly in \eqref{eq:NRsupersymmetry}. Moreover, since the spin connections and $b_\mu$ in \eqref{eq:NRsupersymmetry} depend on $\tau_\mu{}^-=\rme^{2\tilde\phi}\tilde\tau_\mu{}^-$, an extra dependence on $\tilde{\phi}$ can be introduced, when writing these connections in terms of the NR fields with a tilde. Explicitly, one however finds that $\omega_\mu = \tilde\omega_\mu + \partial_\mu\tilde\phi$, $b_\mu = \tilde b_\mu + \partial_\mu\tilde\phi$, $\omega_\mu{}^{AA'}=\tilde\omega_\mu{}^{AA'}$ and $\omega_\mu{}^{A'B'} = \tilde\omega_\mu{}^{A'B'}$. The $\tilde\phi$-dependence in $\delta^{(0)} b_i$ due to the connections then drops out, because $\omega_\mu$ and $b_\mu$ always appear in the combination $b_\mu - \omega_\mu$ in \eqref{eq:NRsupersymmetry}. The explicit dilaton dependence in \eqref{eq:NRsupersymmetry} also drops out, when going to the NR fields with a tilde, as this dependence appears in the form $\partial_\mu \phi - b_\mu  = \partial_\mu\tilde\phi - (\tilde b_\mu + \partial_\mu\tilde\phi) = -\tilde b_\mu$. Alternatively, one can also obtain these results by noting that the redefinition from $\tau_\mu{}^\pm$ to $\tilde\tau_\mu{}^\pm$ can be viewed as a symmetry operation, namely the diagonal of an $\mathrm{SO}(1,1)$ transformation with a dilatation with parameters $\lambda_M=\lambda_D=-\phi$. The above explicit expressions for the connections then follow from this observation.}
\begin{align}\label{eq:compensating11boosts}
&\lambda_M = -\bar\epsilon_+\tilde\lambda_-\,, && \lambda^{+A'}=0\,,&& \lambda^{-A'} = \bar\epsilon_+\Gamma^{A'}\tilde\psi_{-} + \frac12\bar\epsilon_-\Gamma^{A'}\Gamma_+\tilde\lambda_-\,,
\end{align}
and the $\tilde{\delta} b_i$ are explicitly found as
\begin{align}
&\tilde\delta\,\tilde\tau_\mu{}^+ = \bar\epsilon_+\Gamma^+\tilde\psi_{\mu+}\,, && \tilde\delta\,\tilde\tau_\mu{}^- = -\bar\epsilon_-\tilde\lambda_+\,\tilde\tau_\mu{}^-\,,\notag\\
&\tilde\delta\,\tilde e_\mu{}^{A'} = \bar\epsilon_\pm\Gamma^{A'}\tilde\psi_{\mu\mp}\,,&& \tilde\delta\,\tilde b_{\mu\nu} = 4\,\tilde\tau_{[\mu}{}^+\bar\epsilon_-\Gamma_+\tilde\psi_{\nu]-} +2\,\tilde e_{[\mu}{}^{A'}\tilde\delta\,\tilde e_{\nu]}{}^{A'}\,,\notag\\
&\tilde\delta\,\tilde\psi_{\mu+} = \tilde\nabla^{(+)}_\mu\epsilon_+ + \tilde T_\mu\epsilon_- - \frac12\,\tilde\tau_\mu{}^+\left(\tilde{\slashed b} + \frac{1}{12}\,\tilde{\slashed{h}}\right)\Gamma_+\epsilon_-\,, && \tilde\delta\,\tilde\psi_{\mu-} = \left(\delta_\mu{}^\nu - \tilde\tau_\mu{}^+\tilde\tau_+{}^\nu\right)\tilde\nabla^{(+)}_\nu\epsilon_-\,,\notag\\
&\tilde\delta\,\tilde\lambda_+ = -\left(\tilde{\slashed b}  + \frac{1}{12}\,\tilde{\slashed h}\right)\epsilon_+ + \frac12\,\tilde\tau^{A'B'+}\Gamma_{A'B'+}\epsilon_-\,.
\end{align}
Here and in the following, we use the notation that field dependent quantities, such as $\tilde{\tau}_{\mu\nu}{}^A$, $\tilde{h}_{\mu\nu\rho}$, $\tilde{\omega}_\mu{}^{A'B'}$, $\cdots$ are obtained from their counterparts without a tilde, by replacing all NR fields without a tilde by ones with a tilde. Above, we have then denoted $\tilde T_\mu\epsilon_- \equiv (\tilde e_{\mu B'}\tilde\tau^{B'A'+}+\tilde\tau_\mu{}^-\tilde\tau^{A'++})\Gamma_{A'+}\epsilon_-$ and the superscript $(+)$ on the covariant derivatives means that these are defined with respect to a modified $\mathrm{SO}(8)$ spin connection $\tilde\omega_\mu{}^{(+)A'B} \equiv \tilde\omega_\mu{}^{A'B'} + 1/2\,\tilde e_{\mu C'}\tilde h^{C'A'B'}$. We have also used the Feynman-slash notation to denote complete contractions with transversal gamma matrices, i.e., $\slashed c = c_{A'B'\cdots C'}\Gamma^{A'B'\cdots C'}$.

We can now argue that $\langle b \rangle^i$ and $\langle m \rangle^\alpha$ transform into each other under NR supersymmetry $\delta_Q^{(0)}$, when the self-dual DSNC constraint \eqref{eq:DSNC-}\footnote{Note that this constraint assumes the same form in terms of the NR fields with a tilde, i.e., it is given by $\tilde{\tau}_{A'B'}{}^- = 0$ and $\tilde{\tau}_{A'+}{}^- = 0$.} is imposed. To do this, we specify \eqref{eq:simsekvanproeyen} to relativistic supersymmetry transformations $\delta_Q$ and split the $[X]^I$ into $[B]^i$ and $[M]^\alpha$:
\begin{align} \label{eq:simsekvanproeyen2}
\delta_Q [B]^i &= \big(\tilde{E}\,\delta_Q \tilde{E}^{-1}\big)\,[B]^i - \frac{\delta\,(\delta_Q B_j)}{\delta B_i} [B]^j - \frac{\delta\,(\delta_Q M_\alpha)}{\delta B_i} [M]^\alpha \,, \nonumber \\
\delta_Q [M]^\alpha &= \big(\tilde{E}\,\delta_Q \tilde{E}^{-1}\big)\,[M]^\alpha - \frac{\delta\,(\delta_Q B_i)}{\delta M_\alpha} [B]^i - \frac{\delta\,(\delta_Q M_\beta)}{\delta M_\alpha} [M]^\beta \,.
\end{align}
We then expand these equations in powers of $c^{-2}$, using \eqref{eq:expBM}, \eqref{eq:expBMeom}, \eqref{eq:cexpdeltaQrel}, as well as that generically $\delta_Q [X]^I = c^2 \delta_Q^{(2)} [X]^I + \delta_Q^{(0)} [X]^I + c^{-2} \delta_Q^{(-2)} [X]^I + \mathcal{O}(c^{-4})$. The logic behind this is very similar to how we analyzed the invariance of the non-relativistic action in section \ref{sec:limitaction}. The results on invariance of the NR field equations under supersymmetry will follow from the subleading order of these expansions.

Before discussing this subleading order however, let us first check how the leading order is satisfied and what can be learnt from it. This leading order amounts to the following equations:
\begin{subequations}
\begin{align}
\delta_Q^{(2)}\langle b\rangle^i &= 0\,,\label{eq:delta2B}\\
\delta_Q^{(2)}\langle m\rangle^\alpha &= -\frac{\delta\big(\delta_Q^{(0)}b_i\big)}{\delta m_\alpha}\langle b \rangle^i - \frac{\delta\big(\delta_Q^{(2)}m_\beta\big)}{\delta m_\alpha}\langle m\rangle^\beta\,. \label{eq:delta2M}
\end{align}
\end{subequations}
The second of these equations can be simplified, by making use of the Noether identity for relativistic Lorentz transformations, which states that for a Lorentz transformation $\delta_L$ with arbitrary parameters
\begin{align} \label{eq:NoetheridLor}
\delta_L X_I [X]^I = 0 \,.
\end{align}
This Noether identity can be expanded in powers of $c^{-2}$, by using \eqref{eq:expBM}, \eqref{eq:expBMeom} and the fact that a generic Lorentz transformation $\delta_L$ takes the form $\delta_L = \delta_L^{(0)} + c^{-2} \delta_L^{(-2)}$, with $\delta_L^{(0)}$ corresponding to NR longitudinal SO$(1,1)$ Lorentz transformations, Galilean boosts and transversal rotations. The leading order term in the expansion of \eqref{eq:NoetheridLor} then says that
\begin{align} \label{eq:NoetherExp0}
\delta^{(0)}_L b_i\,\langle b\rangle^i &= 0\,,
\end{align}
with $\delta_L^{(0)} b_i$ an arbitrary longitudinal SO$(1,1)$ transformation, Galilean boost or transversal rotation of $b_i$. According to \eqref{eq:splitoffm}, the only dependence of $\delta_Q^{(0)} b_i$ on the fields $m_\alpha$ occurs in field-dependent SO$(1,1)$ and Galilean boost transformations that act on $b_i$. As a consequence, $\delta\big(\delta_Q^{(0)} b_i\big)/\delta m_\alpha$ also corresponds to a longitudinal SO$(1,1)$ or Galilean boost transformation (with $m_\alpha$ stripped from the symmetry parameter), acting on $b_i$. The first term on the right-hand-side of \eqref{eq:delta2M} is then zero as a consequence of \eqref{eq:NoetherExp0}, so that we are left with
\begin{align} \label{eq:delta2BMsimpl}
\delta_Q^{(2)}\langle b\rangle^i &= 0\,, \qquad \qquad
\delta_Q^{(2)}\langle m\rangle^\alpha = - \frac{\delta\big(\delta_Q^{(2)}m_\beta\big)}{\delta m_\alpha}\langle m\rangle^\beta\,.
\end{align}
Note that the only contributions to $\delta_Q^{(2)} \langle b\rangle^i$ or $\delta_Q^{(2)} \langle m\rangle^\alpha$ can come from their dependence on the $m_\alpha$, since only these fields transform under $\delta_Q^{(2)}$. The first of \eqref{eq:delta2BMsimpl} is then explained by the fact that $\langle b \rangle^i$ do not depend on the fields $m_\alpha$, since the NR action \eqref{eq:NRaction} from which they are derived is independent of the missing fields $m_\alpha$. Furthermore, since $\delta_Q^{(2)} m_\alpha$ is given by field-dependent $S$- and $T$-transformations, the second of \eqref{eq:delta2BMsimpl} is consistent with the fact that the missing NR equations of motion transform among themselves under $S$- and $T$-symmetries (see, e.g., \eqref{eq:STPoisson}).

The subleading term in the $c^{-2}$--expansion of \eqref{eq:simsekvanproeyen2} gives the following equations:
\begin{subequations} \label{eq:delta0BM}
\begin{align}
\delta_Q^{(0)}\langle b\rangle^i &= \big(\tilde e\,\delta_Q^{(0)}\tilde e^{-1}\big)\,\langle b\rangle^i - \frac{\delta \big(\delta_Q^{(0)}b_j\big)}{\delta b_i}\langle b\rangle^j - \frac{\delta \big(\delta_Q^{(2)}m_\alpha\big)}{\delta b_i}\langle m \rangle^\alpha - \delta_Q^{(2)}[B]_{SL}^i\,,\label{eq:delta0B}\\
\delta_Q^{(0)}\langle m\rangle^\alpha &= \big(\tilde e\,\delta_Q^{(0)}\tilde e^{-1}\big)\,\langle m\rangle^\alpha - \frac{\delta\big(\delta_Q^{(0)}b_j\big)}{\delta m_\alpha}[B]_{SL}^j - \frac{\delta\big(\delta_Q^{(-2)}b_i\big)}{\delta m_\alpha}\langle b \rangle^i \notag\\
&\qquad - \frac{\delta\big(\delta_Q^{(0)}m_\beta\big)}{\delta m_\alpha}\langle m\rangle^\beta - \frac{\delta\big(\delta_Q^{(2)}m_\beta\big)}{\delta m_\alpha}[M]_{SL}^\beta -\delta_Q^{(2)}[M]_{SL}^\alpha\,.\label{eq:delta0M}
\end{align}
These equations tell us how the $\langle b \rangle^i$ and $\langle m \rangle^\alpha$ transform under NR supersymmetry $\delta_Q^{(0)}$.\footnote{Note again that we assume that the $S$- and $T$-transformations \eqref{eq:STrest} are not included in $\delta_Q^{(0)} m_\alpha$. The effect of including these transformations is that $\delta_Q^{(0)} \langle m \rangle^\alpha$ and $(\delta\big(\delta_Q^{(0)}m_\beta\big)/\delta m_\alpha) \, \langle m \rangle^\beta$ in \eqref{eq:delta0M} can receive extra contributions. These are however always proportional to $\langle m \rangle^\alpha$ (for $\delta_Q^{(0)} \langle m \rangle^\alpha$ this is because the $\langle m \rangle^\alpha$ transform into each other under $S$- and $T$-transformations), so that the overall conclusion that all NR field equations transform into each other under supersymmetry is not affected.} The appearance of the subleading parts $[B]_{SL}^i$ and $[M]_{SL}^\alpha$ of the expansions of $[B]^i$ and $[M]^\alpha$ is worrisome, as it implies that the $\langle b \rangle^i$ and $\langle m \rangle^\alpha$ do not form a closed set under NR supersymmetry. Note however that the second term on the right-hand-side of \eqref{eq:delta0M} is harmless, due to the Noether identity \eqref{eq:NoetheridLor}. Indeed, the subleading order of the $c^{-2}$--expansion of \eqref{eq:NoetheridLor} implies that
\end{subequations}
\begin{align} \label{eq:NoetherExp-2}
\delta^{(0)}_L b_i\,[B]_{SL}^i &= -\delta^{(0)}_L m_\alpha\,\langle m\rangle^\alpha - \delta^{(-2)}_L b_i\,\langle b\rangle^i\,,
\end{align}
so that $\delta^{(0)}_L b_i\,[B]_{SL}^i$, with $\delta^{(0)}_L b_i$ an arbitrary longitudinal SO$(1,1)$ transformation, Galilean boost or transversal rotation of $b_i$, can be written as a combination of $\langle b \rangle^i$ and $\langle m \rangle^\alpha$. As mentioned below \eqref{eq:NoetherExp0}, $\delta\big(\delta_Q^{(0)}b_j\big)/\delta m_\alpha$ takes the form of a longitudinal SO$(1,1)$ transformation or Galilean boost, acting on $b_i$, and consequently, the $[B]^i_{SL}$ in the second term on the right-hand-side of \eqref{eq:delta0M} can be traded for $\langle b \rangle^i$ and $\langle m \rangle^\alpha$. The remaining terms that involve $[B]_{SL}^i$ and $[M]_{SL}^\alpha$ in \eqref{eq:delta0BM} can not necessarily be expressed in terms of $\langle b \rangle^i$ and $\langle m \rangle^\alpha$. They however vanish, when the self-dual DSNC constraint \eqref{eq:DSNC-} is imposed. For $\delta_Q^{(2)}[B]_{SL}^i$ and $\delta_Q^{(2)}[M]_{SL}^\alpha$, this is because each contribution to these terms is proportional to $\delta_Q^{(2)} m_\alpha$ and thus vanishes when the self-dual DSNC constraint is imposed. From \eqref{eq:susydivergencered}, one can see that the functional derivatives $\delta\big(\delta_Q^{(2)}m_\beta\big)/\delta m_\alpha$ are zero when the self-dual DSNC constraint holds. The fifth term on the right-hand-side of \eqref{eq:delta0M} then also vanishes, upon imposition of this constraint.

We can thus conclude that the full set of NR field equations \eqref{eq:allNReoms} is invariant under NR supersymmetry when the self-dual DSNC constraint \eqref{eq:DSNC-} is imposed by hand. Let us stress again that this constraint is itself invariant under NR supersymmetry so that it can be imposed consistently in a supersymmetric fashion, without the need for extra constraints. It is furthermore also interesting to note that from the form of \eqref{eq:delta0B}, one can see that the supersymmetry variation of the NR field equations $\langle b \rangle^i = 0$, that follow from the NR action \eqref{eq:NRaction}, in general, gives rise to the missing equations of motion $\langle m \rangle^\alpha = 0$. One can check explicitly that this indeed happens. This phenomenon occurs because the NR action \eqref{eq:NRaction} is only invariant under NR supersymmetry up to the self-dual DSNC constraint \eqref{eq:DSNC-}. So, even though the missing equations of motion do not follow directly from the NR action \eqref{eq:NRaction}, they can be obtained indirectly from it by varying its equations of motion under NR supersymmetry.

Finally, let us finish this section, by noting that the above arguments can be adapted to show that the $\langle b \rangle^i$ and $\langle m \rangle^\alpha$ also transform into each other under Galilean boosts. First, we note that, by writing the $c^{-2}$--expansions \eqref{eq:LorsymmBOS}, \eqref{eq:LorsymmFER} of the relativistic boost transformation rules (denoted here by $\delta_B$) in terms of the NR fields with a tilde, the following $c^{-2}$--expansions are seen to hold:
\begin{align}
\label{eq:cexpdeltaBrel}
\delta_B b_i &= \delta^{(0)}_B b_i + c^{-2} \delta^{(-2)}_B b_i \,, \qquad \qquad
\delta_B m_\alpha = \delta^{(0)}_B m_\alpha + c^{-2} \delta^{(-2)}_B m_\alpha \,,
\end{align}
where $\delta^{(0)}_B$ corresponds to a Galilean boost transformation. We can then again start from \eqref{eq:simsekvanproeyen}, applied to relativistic boosts
\begin{align} \label{eq:simsekvanproeyen3}
\delta_B [B]^i &= - \frac{\delta\,(\delta_B B_j)}{\delta B_i} [B]^j - \frac{\delta\,(\delta_B M_\alpha)}{\delta B_i} [M]^\alpha \,, \quad
\delta_B [M]^\alpha =  - \frac{\delta\,(\delta_B B_i)}{\delta M_\alpha} [B]^i - \frac{\delta\,(\delta_B M_\beta)}{\delta M_\alpha} [M]^\beta \,,
\end{align}
and expand these equations in powers of $c^{-2}$, using \eqref{eq:expBM}, \eqref{eq:expBMeom}, \eqref{eq:cexpdeltaBrel}, as well as $\delta_B [X]^I = \delta_B^{(0)} [X]^I + c^{-2} \delta_B^{(-2)} [X]^I + \mathcal{O}(c^{-4})$. The leading order in the $c^{-2}$--expansion of the first of \eqref{eq:simsekvanproeyen3} and the leading and subleading order of the second of \eqref{eq:simsekvanproeyen3} then lead to the following equations:
\begin{subequations} \label{eq:deltaBbm}
\begin{align}
& \delta_B^{(0)} \langle b \rangle^i = - \frac{\delta \big(\delta_B^{(0)} b_j\big)}{\delta b_i} \langle b \rangle^j \,, \label{eq:deltaBb} \\
& \frac{\delta \big(\delta_B^{(0)} b_i\big)}{\delta m_\alpha} \langle b \rangle^i = 0 \,, \label{eq:deltaBm1} \\
& \delta_B^{(0)} \langle m \rangle^\alpha = - \frac{\delta \big( \delta_B^{(0)} b_i\big)}{\delta m_\alpha} [B]_{SL}^i - \frac{\delta \big( \delta_B^{(-2)} b_i\big)}{\delta m_\alpha} \langle b \rangle^i - \frac{\delta \big( \delta_B^{(0)} m_\beta\big)}{\delta m_\alpha} \langle m \rangle^\beta \,. \label{eq:deltaBm2}
\end{align}
\end{subequations}
The second of these equations is identically satisfied, as a consequence of \eqref{eq:NoetherExp0} (applied to a Galilean boost) and the fact that $\delta \big(\delta_B^{(0)} b_i\big)/\delta m_\alpha$ corresponds to a Galilean boost transformation (with the field $m_\alpha$ stripped off its parameter), acting on $b_i$. Similarly, by using \eqref{eq:NoetherExp-2} (applied to a Galilean boost), we see that the first term on the right-hand-side of \eqref{eq:deltaBm2} can be rewritten as a combination of $\langle b \rangle^i$ and $\langle m \rangle^\alpha$. We then see that Galilean boost transformations transform the $\langle b \rangle^i$ and $\langle m \rangle^\alpha$ among themselves. Note that \eqref{eq:deltaBbm} shows that the Galilean boost transformation of the missing equations of motion $\langle m \rangle^\alpha = 0$ includes the equations $\langle b \rangle^i = 0$, while the latter only transform among themselves since they are derived from the Galilean boost invariant action \eqref{eq:NRaction}. As a representation of the non-semisimple group that consists of longitudinal SO$(1,1)$ transformations, Galilean boosts, and transversal rotations, the NR field equations \eqref{eq:allNReoms} are then seen to form a so-called reducible indecomposable representation. Note that the appearance of reducible indecomposable representations is quite common when discussing finite-dimensional representations of non-semisimple groups \cite{George-Levy-Nahas}.

\section{Conclusions} \label{sec:conclusions}

\noindent In this paper we extended our previous work on taking the NR limit of ten-dimensional NS-NS gravity to the supersymmetric case. This leads to a NR minimal supergravity theory that is common to all NR superstring theories.  In doing so we encountered two complications that were absent in the bosonic case. First of all, the relation between the string sigma model and the target space effective action is less direct than in the bosonic case. This had the effect that we could not independently check the two fermionic St\"uckelberg symmetries that we found in the target space effective action at the level of the sigma model description of the superstring. The second complication is that we found, upon taking the NR limit, divergent terms in the supersymmetry rules of the fermionic fields  that were all proportional to the torsion components that define a self-dual DSNC geometry, see \eqref{selfdualintro}. These divergent terms could be controlled by (i) using the fact that there are two emergent fermionic St\"uckelberg symmetries and (ii) the constraints defining a self-dual DSNC geometry are invariant under all the symmetries of the NR theory and therefore can be imposed by hand without the need to impose more constraints. An important simplifying feature that we used is that,  after making a particular field redefinition, see \eqref{eq:Einsteinframe},  all NR equations of motion occurred as the leading term in the expansion of a corresponding relativistic equation of motion, without the need to take special combinations of equations of motion like we did for the bosonic case. Similar to the bosonic case we found that  all divergent terms in the action vanished due to cancellations of different contributions and that the NR action  did not give rise to all equations of motion. In particular, the Poisson equation for the Newton potential and two fermionic  equations were missing. A  difference with the bosonic case is that the so-called missing equations of motion, that do not follow from a variation of the NR action, can be obtained by a supersymmetry variation of the equations of motion that do follow from a variation of the NR action.

%
%
%
%

In hindsight, it is a good thing that we found divergent terms in the supersymmetry rules. Would such terms have been absent we would have found a NR action that is exactly supersymmetric without the need to impose any constraint. In that case the results of \cite{Vanhecke:2017chr} would apply with the consequence that the equations of motion corresponding to the NR action would transform to each other forming a supermultiplet but that
none of these equations of motion would transform under supersymmetry to the missing Poisson equation and hence the Poisson equation would not be part of this supermultiplet. Such a situation would be hard to reconcile with the closure of the supersymmetry algebra.

Given the fact that we found an emerging dilatation symmetry and two emerging superconformal symmetries one could wonder in which sense the NR minimal supergravity multiplet we found defines a conformal supergravity multiplet. Apart from a few similarities there are  important differences. First of all conformal supergravity is usually presented as an off-shell multiplet whereas the NR minimal supergravity multiplet is on-shell. Secondly, the NR minimal supergravity multiplet lacks the special conformal symmetries. This is a consequence of the fact that, unlike in conformal supergravity, all  components of the dilatation gauge field are dependent, see the first equation in \eqref{eq:galspinconn}.

In appendix \ref{sec:NRSYM}, we give some initial results on the Yang-Mills sector needed to discuss the case of the heterotic superstring. Assuming that in the flat spacetime limit we can define an independent NR Yang-Mills supermultiplet, we showed that, starting from  a particular field redefinition, there is a unique NR string limit of the relativistic ten-dimensional super-Yang-Mills theory that avoids the occurrence of infinities in the supersymmetry rules. We give the expressions for the action and supersymmetry rules in appendix \ref{sec:NRSYM}. It is interesting to compare our results with those of \cite{Lambert:2019nti} (see also \cite{Lambert:2019fne}) where a similar analysis has been made of Yang-Mills systems in flat spacetime involving the scaling of fields with a parameter and taking the limit that this parameter goes to infinity.
One difference is that  we take the different scalings such that the limit does not lead to divergences in the supersymmetry rules and in the action. This is related to the fact that we scale {\it two} of the flat coordinates different from the rest. We did this because we had a NR string limit in mind whereas the discussion of \cite{Lambert:2019nti} is more general. It would be interesting to apply the general analysis of \cite{Lambert:2019nti} to a matter coupled to gravity system and see under which conditions the global scale symmetry, observed in \cite{Lambert:2019nti}, extends to a local scale symmetry like in this work.

After a dimensional  reduction of the spatial worldsheet direction the bosonic part of the NR ten-dimensional super-Yang-Mills theory seems to coincide with the one given in \cite{Santos:2004pq,Bergshoeff:2015sic} and, more recently, in \cite{Gomis:2020fui} where it was identified with the low energy dynamics of open strings ending on $N$ coincident D-branes in flat spacetime. The same lower-dimensional bosonic Yang-Mills theory also follows from a null reduction of a relativistic Yang-Mills theory in one dimension higher \cite{Festuccia:2016caf}. These results seem to be consistent with the T-duality of the bosonic NR open string theory as discussed in \cite{Gomis:2020izd}. Here instead, we are interested in the occurrence of Yang-Mills within the heterotic superstring theory. To construct the Yang-Mills coupled to supergravity system, i.e.~the NR heterotic supergravity theory, one may proceed in two ways. Either one couples, via a Noether procedure like in the relativistic case \cite{Bergshoeff:1981um}, the NR Yang-Mills theory to the minimal supergravity theory that we already constructed in this work or one takes the NR limit of the relativistic Yang-Mills coupled to supergravity theory. Approaching the problem from a sigma model point of view, a new complication, not encountered in the pure supergravity or flat spacetime Yang-Mills theory, occurs, namely  the occurrence of a  worldsheet anomaly in the  sigma model giving rise to a NR Chern-Simons term. We hope to show in a follow-up work how these two  different approaches lead to the same NR heterotic supergravity theory with its characteristic NR Yang-Mills Chern-Simons term.

It would be interesting to approach the construction of a NR heterotic supergravity theory from a Double Field Theory point of view where in the bosonic case a construction of the NS-NS gravity theory using Double Field Theory with  a degenerate geometry has been given \cite{Gallegos:2020egk}. An intriguing issue arises in the supergravity case. Although we took a NR string limit of the ${\mathcal {N}} =1$  supergravity theory in this paper, it is not clear what the dual null reduction  would correspond to. The reason for this is that the null Killing condition corresponds to a constraint on the relativistic supergravity theory that is not invariant under supersymmetry. In fact, we are not aware of any relativistic ten-dimensional supergravity theory exhibiting a null isometry direction without breaking supersymmetry.\,\footnote{We stress that this is different from looking to {\sl solutions} of the 10D supergravity theory with a null isometry direction. This leads to supersymmetric Killing spinor conditions that do exist and in general break part of the supersymmetry. }
We hope to come back to this issue in a forthcoming publication.

It is natural to generalize the results of this paper to IIA and IIB supergravity and eleven-dimensional supergravity corresponding to the IIA and IIB  superstring theories and M-theory.\,\footnote{For the bosonic part of M-theory, see \cite{Blair:2021ycc}.} In particular, it would be interesting to see what happens with the potential divergent terms in the supersymmetry rules and the occurrence of emergent fermionic St\"uckelberg symmetries. Once the finite supersymmetry rules have been constructed one could study  compactifications and look for interesting NR supersymmetric solutions by analyzing the Killing spinor equations. For the case of minimal supergravity dealt with in this paper we have collected the bosonic equations of motion (with the fermions set equal to zero) and the Killing spinor equations in a separate appendix. This appendix is a convenient starting point for discussing NR supersymmetric NS-NS solutions. In particular, it would be interesting to see whether one can find NR supersymmetric solutions (probably with a horizon) that can take over the role that black holes play in the AdS/CFT correspondence.  Such solutions could play an important role in setting up a NR holographic principle independent of the relativistic one. We hope to return to these issues in the nearby future.

\section*{Acknowledgements}

\noindent We would like to thank  Jaume Gomis, Joaquim Gomis, Neil Lambert, Ziqi Yan, and Utku Zorba for useful discussions.  The work of C\c S is part of the research programme of the Foundation for Fundamental Research on Matter (FOM), which is financially supported by the Netherlands Organisation for Science Research (NWO). The work of LR is supported by the FOM/NWO free program {\sl Scanning New Horizons}.

\appendix

\section{Notation and Conventions}\label{sec:conventions}

\subsection{Bosonic Conventions} \label{ssec:bosconventions}

\noindent Ten-dimensional flat Lorentz indices are denoted by $\hat A$ and split into $(A,A')$, where $A=0,1$ and $A'=2,\cdots,9$. We refer to this as a splitting into longitudinal, resp. transversal directions. We use the `mostly plus' form of the Minkowski metric, i.e., $(\eta_{\hat A\hat B}) = (-++\cdots +)$. The ten-dimensional Levi-Civita epsilon symbol is normalized by $\epsilon_{01\cdots 9}=+1$ ($\epsilon^{01\cdots 9} = -1$). We also use a two-dimensional longitudinal epsilon symbol $\epsilon_{AB}$ that is normalized as $\epsilon_{01}=+1$ ($\epsilon^{01}=-1$). 
The following identities then hold
\begin{equation}
\label{eq:epsids}
\epsilon_{AC} \epsilon_{BD} = -\eta_{AB} \eta_{CD} + \eta_{AD} \eta_{BC} \,, \qquad \qquad \epsilon_A{}^C \epsilon_{CB} = \eta_{AB} \,.
\end{equation}
Instead of writing longitudinal components of tensors with respect to coordinates $x^A$ ($A=0,1$), we will also often write them with respect to light-cone coordinates $x^\pm$ that are defined as follows:
\begin{align}
x^\pm = \frac{1}{\sqrt2}\lr x^0 \pm x^1\rr\,.
\end{align}
The longitudinal $\eta_{AB}$-part of the Minkowski metric then has $\eta_{+-}= \eta_{-+} = -1$ as its non-zero components and one similarly has that $\epsilon_{+-}= - \epsilon_{-+} = -1$. It is straightforward to check that
\begin{align}
X^A\eta_{AB}Y^B =  - X^- Y^+-X^+Y^-\,&,& &X^A\epsilon_{AB}Y^B = X^-Y^+ - X^+Y^-\,,\\
\frac12\,X^A\lr\eta_{AB}+\epsilon_{AB}\rr Y^B = X^+ Y_+\,&, & &\frac12\,X^A\lr\eta_{AB}-\epsilon_{AB}\rr Y^B = X^- Y_-\,.\label{eq:+-proj}
\end{align}
A curved, lower $\mu$ index is turned into flat $A$ or $A'$ indices, using the projective inverse Vielbeine $\tau_A{}^\mu$, $e_{A'}{}^\mu$ via the rule
\begin{align} \label{eq:curvedflatrule}
X_A = \tau_A{}^\mu X_\mu \,, \qquad X_{A'} = e_{A'}{}^\mu X_\mu \qquad \Leftrightarrow \qquad X_\mu = \tau_\mu{}^A X_A + e_\mu{}^{A'} X_{A'} \,.
\end{align}

\subsection{Spinor and Clifford Algebra Conventions} \label{ssec:fermconventions}

\noindent The defining relation for the Clifford algebra that is generated by the ten-dimensional gamma matrices $\Gamma_{\hat{A}}$ is taken to have the following normalization
\begin{align} \label{eq:Cliffalgdef}
\{\Gamma_{\hat{A}}, \Gamma_{\hat{B}}\} = 2 \eta_{\hat{A}\hat{B}} \mathds{1} \,.
\end{align}
The gamma matrices satisfy the following hermiticity property:
\begin{align}
\label{eq:hermgammas}
\Gamma_{\hat{A}}^\dag = \Gamma_0 \Gamma_{\hat{A}} \Gamma_0 \,.
\end{align}
The charge conjugation matrix $C$ is unitary and satisfies
\begin{align}\label{eq:ChargeConj}
C^T = -C\,,\qquad \qquad \Gamma_{\hat{A}}^T = - C\Gamma_{\hat A} C^{-1} \,.
\end{align}
All spinors are assumed to satisfy a Majorana-Weyl condition, so that the conjugate $\bar{\psi}$ of a spinor $\psi$ can be interpreted as $\bar{\psi} = \psi^T C$. Using \eqref{eq:ChargeConj}, one can show that spinor bilinears obey the following symmetry property
\begin{align}\label{eq:flippingrulez}
\bar\chi_1 \Gamma_{\hat A_1\cdots \hat A_n}\chi_2 =
\begin{cases}
\quad+\bar\chi_2 \Gamma_{\hat A_1\cdots \hat A_n} \chi_1 \hskip 1.5truecm \mathrm{for}\quad n=0,3\,\mod 4 \,, \\
\quad-\bar\chi_2 \Gamma_{\hat A_1\cdots \hat A_n} \chi_1 \hskip 1.5truecm \mathrm{for}\quad n=1,2\,\mod 4\,.
\end{cases}
\end{align}
Left-, resp. right-handed spinors are obtained by projecting with the Weyl projectors $P_L$, resp. $P_R$, where
\begin{align}
\label{eq:Weylproj}
P_L = \frac12 \left(\mathds{1} + \Gamma_* \right) \,, \qquad  P_R = \frac12 \left(\mathds{1} - \Gamma_* \right) \,, \qquad \qquad \text{with }\ \Gamma_* = - \Gamma_0 \Gamma_1 \cdots \Gamma_9 \,.
\end{align}

In this paper, we frequently work with spinors that are projected, using the following `worldsheet chirality' orthogonal projection operators $\Pi_\pm$
\begin{align}
\label{eq:Piproj}
\Pi_\pm = \frac12 \left(\mathds{1} \pm \Gamma_{01}\right) \qquad \qquad (\text{obeying }\, \Pi_\pm^2 = \Pi_\pm \ \ \text{and} \ \ \Pi_\pm \Pi_\mp = 0) \,.
\end{align}
The projection of a spinor $\chi$ with $\Pi_+$, resp. $\Pi_-$ will be denoted as $\chi_+ = \Pi_+ \chi$, resp. $\chi_- = \Pi_- \chi$. One thus has
\begin{align}
\chi_\pm = \Pi_\pm\chi \,,\qquad \qquad \Pi_\pm \chi_\mp = 0\,, \qquad \qquad \Gamma_{01}\chi_\pm = \pm\chi_\pm\,.
\end{align}
Since
\begin{align}
\Gamma_*\Gamma_{01} = \Gamma_{01}\Gamma_*\,,\qquad \Gamma_A\Gamma_{01} = -\Gamma_{01}\Gamma_A\,,\qquad \Gamma_{A'}\Gamma_{01} = \Gamma_{01}\Gamma_{A'}\,,\qquad \Gamma_{01}^T C = -C\Gamma_{01} \,,
\end{align}
worldsheet chirality projection is compatible with the Majorana-Weyl condition. The conjugate of a worldsheet chirality projected spinor is defined as $\bar\chi_\pm \equiv \chi_\pm^TC$. Note that one then has that
\begin{align}
\bar\chi_\pm \Gamma_{01} = \mp \bar\chi_\pm\,.
\end{align}
As a consequence, bilinears of the form
\begin{align}
\label{eq:projbil}
\bar{\chi}_\pm\Gamma_{A_1^\prime \cdots A_m^\prime}\psi_\pm \,, \qquad \quad \bar{\chi}_\pm\Gamma_{A B A_1^\prime \cdots A_m^\prime}\psi_\pm \,, \qquad \quad \bar{\chi}_\pm\Gamma_{A A_1^\prime \cdots A_m^\prime}\psi_\mp \,,
\end{align}
are identically zero.

When working with purely longitudinal gamma matrices $\Gamma_A$, the following duality relations are often handy:
\begin{align}
\Gamma_{AB}=\epsilon_{AB}\,\Gamma_{01}\,,\hskip 1.5truecm \Gamma_A = -\epsilon_{AB}\Gamma^B\Gamma_{01}\,.
\end{align}
These can for instance be used to show that
\begin{alignat}{2}\label{eq:etaepsilongammagamma}
\lr\eta_{AB} + \epsilon_{AB}\rr\chi_+ &= \Gamma_A\Gamma_B\chi_+\,,\hskip 1.5truecm & \lr\eta_{AB} - \epsilon_{AB}\rr\chi_+ &= \Gamma_B\Gamma_A\chi_+\,, \nonumber \\
\lr\eta_{AB} - \epsilon_{AB}\rr\chi_- &= \Gamma_A\Gamma_B\chi_-\,,\hskip 1.5truecm & \lr\eta_{AB} + \epsilon_{AB}\rr\chi_- &= \Gamma_B\Gamma_A\chi_-\,.
\end{alignat}
Instead of using indices $A,B = 0,1$ for longitudinal gamma matrices, it is sometimes useful to work with longitudinal gamma matrices with light-cone indices
\begin{align}
\Gamma_\pm = \frac{1}{\sqrt{2}}\left(\Gamma_0\pm\Gamma_1\right) \,.
\end{align}
These satisfy $\Gamma_\pm\Gamma_{01} = \mp\Gamma_\pm$ and one thus has
\begin{align}\label{eq:selfduality}
\Gamma_+\chi_+=0\,,\qquad\mathrm{and}\qquad \Gamma_-\chi_-=0\,,
\end{align}
implying that e.g. $Y^A\Gamma_A\psi_+ = Y^-\Gamma_-\psi_+$.\\

\section{Torsional String Newton-Cartan Geometry} \label{sec:TSNC}

\noindent In this section, we give some details on the non-Lorentzian geometric structures appearing in this paper. We will use the name torsional string Newton-Cartan (TSNC) for generic geometric structures without any further geometric constraints on the torsion. The self-dual DSNC geometry that is relevant in this paper is a special case where the torsion tensor satisfies $T_{\mu\nu}^\rho\,\tau_\rho{}^-=0$, or equivalently \eqref{selfdualintro}. We refer the reader to appendices B and C of \cite{Bergshoeff:2021bmc} for more details.

The main novel feature of these structures is the occurrence of a $2-$form $b_{\mu\nu}$ and a scalar $\phi$ as part of the geometric structure, next to the longitudinal $\tau_\mu{}^A$ $(A=0,1)$, and transversal Vielbeinen $e_\mu{}^{A'}$ $(A'=2,\cdots,9)$. As explained in \cite{Bergshoeff:2021bmc}, one can introduce spin connections for local $\mathrm{SO}(1,1)\times\mathrm{SO}(8)-$rotations $(\omega_\mu,\omega_\mu{}^{A'B'})$, Galilean boosts $\omega_\mu{}^{AA'}$, dilatations $b_\mu$, together with an affine connection $\Gamma_{\mu\nu}^\rho$, by imposing \footnote{Note that it is straightforward to derive supercovariant versions of the above expressions by adding the appropriate gravitino bilinears to the anholonomy coefficients $\tau_{\mu\nu}{}^A/e_{\mu\nu}{}^{A'}/h_{\mu\nu\rho}/\partial_\mu\phi$, for example $\hat\tau_{\mu\nu}{}^A = \tau_{\mu\nu}{}^A - 1/2\,\bar\psi_{\mu+}\Gamma^A\psi_{\nu+}$. In this work we choose to write out the fermion bilinears explicitly when necessary, and not use supercovariant expressions.}
\begin{subequations}\label{eq:metcomp}
\begin{align}
\nabla_\mu\phi &\equiv \partial_\mu\phi - b_\mu = e_\mu{}^{A'}\nabla_{A'}\phi\,,\\
H_{\mu\nu\rho} &\equiv h_{\mu\nu\rho} + 6\,\epsilon_{AB}\,\omega_{[\mu}{}^{AB'}\tau_{\nu}{}^B\,e_{\rho]B'} = e_\mu{}^{A'}e_\nu{}^{B'}e_\rho{}^{C'}h_{A'B'C'}\,,\\
\nabla_\mu\tau_\nu{}^A &\equiv \partial_\mu\tau_\nu{}^A - \omega_\mu\,\epsilon^{AB}\tau_{\nu B} - b_\mu\,\tau_\nu{}^A - \Gamma_{\mu\nu}^\rho\tau_\rho{}^A = 0\,,\label{eq:taucomp}\\
\nabla_\mu e_\nu{}^{A'} &\equiv \partial_\mu e_\nu{}^{A'} - \omega_\mu{}^{A'B'} e_{\nu B'} + \omega_\mu{}^{AA'}\tau_{\nu A} - \Gamma_{\mu\nu}^\rho e_\rho{}^{A'}=0\,,\label{eq:ecomp}
\end{align}
\end{subequations}
where $h_{\mu\nu\rho}=3\,\partial_{[\mu}b_{\nu\rho]}$. When considering the antisymmetric part of \eqref{eq:taucomp} and \eqref{eq:ecomp} we observe that not all components depend on the spin connections and hence cannot be used as conventional constraints. This is different from usual (semi-)Riemannian geometry where the affine connection can be chosen to be symmetric without loss of generality. Here, however, this implies that generic TSNC geometries have intrinsic torsion \cite{Figueroa-OFarrill:2020gpr}
\begin{align}
T_{\mu\nu}^\rho &= 2\,\Gamma_{[\mu\nu]}^\rho = \rmR_{\mu\nu}(H^A)\tau_A{}^\rho\,,\notag\\
\mathrm{with}\qquad \rmR_{\mu\nu}(H^A) &  = e_\mu{}^{A'}e_\nu{}^{B'}\,\tau_{A'B'}{}^A + 2\,e_{[\mu}{}^{A'}\tau_{\nu]B}\,\tau_{A'}{}^{\{BA\}}\,,\label{eq:torsion2Form}
\end{align}
where $\rmR_{\mu\nu}(H^A)\equiv 2\,\partial_{[\mu}\tau_{\nu]}{}^A - 2\,\big(\epsilon^A{}_B\,\omega_{[\mu} +\delta^A{}_{B}\,b_{[\mu}\big)\tau_{\nu]}{}^B$ is the covariant version of $\tau_{\mu\nu}{}^A=\partial_{[\mu}\tau_{\nu]}{}^A$. Note that the independent components in the torsion tensor are equivalent to what we refer to as the DSNC-torsion components in the main text. Phrased differently---setting $T_{\mu\nu}^\rho=0$ is equivalent to imposing DSNC geometry \eqref{stringconstraint2}. Similarly, the self-dual DSNC geometry \eqref{selfdualintro} that is relevant to this paper is equivalent to imposing $T_{\mu\nu}^\rho\tau_\rho{}^-=0$ (and, analogously, anti self-dual DSNC $\Leftrightarrow~T_{\mu\nu}^\rho\tau_\rho{}^+=0$). By using the above and solving the remaining conventional contraints in \eqref{eq:metcomp}, one finds the following explicit expressions for the spin connections
\begin{subequations}\label{eq:galspinconn}
\begin{align}
b_\mu &= e_\mu{}^{A'}\,\tau_{A'A}{}^A +\tau_\mu{}^A\partial_A\phi \,,\\
\omega_\mu &= \big(\,\tau_\mu{}^{AB}-\frac12\,\tau_\mu{}^C\tau^{AB}{}_C \big)\epsilon_{AB} - \tau_\mu{}^A\,\epsilon_{AB}\partial^B\phi\,,\\
\omega_\mu{}^{AA'} &= -e_\mu{}^{AA'}+e_{\mu B'}e^{AA'B'} + \frac12\,\epsilon^A{}_B\,h_\mu{}^{BA'}  + \tau_{\mu B} W^{BAA'} \,,\\
\omega_\mu{}^{A'B'} &= -2\, e_{\mu}{}^{[A'B']}+e_{\mu C'}e^{A'B'C'} - \frac12\,\tau_\mu{}^A\,\epsilon_{AB}\,h^{BA'B'} \,,
\end{align}
\end{subequations}
where $e_{\mu\nu}{}^{A'} = \partial_{[\mu} e_{\nu]}{}^{A'}$. However, not all components can be solved for, which is reflected by the undetermined $W^{ABA'}$ which is traceless symmetric in the $(A,B)$ indices, but otherwise arbitrary. Since all the relevant expressions---such as action, equations of motion, and symmetry transformation rules---follow from a limit it is clear that nothing depends on $W$, see also \cite{Bergshoeff:2021bmc}. The constraints \eqref{eq:metcomp} can furthermore be used to give the explicit expression for the ``affine'' connection\footnote{Note that this expression is not invariant under Galilean boosts. Hence it is a slight abuse of language to call it an \emph{affine} connection. It would be interesting to find a boost invariant connection for torsional string Newton-Cartan geometries following the procedures outlined in \cite{Geracie:2015dea}, \cite{Hartong:2015zia} or \cite{Festuccia:2016awg}.}
\begin{align}
\Gamma_{\mu\nu}^\rho = \tau_A{}^\rho \big( \partial_\mu\tau_\nu{}^A - \omega_\mu\,\epsilon^{AB}\tau_{\nu B} - b_\mu\,\tau_\nu{}^A\big) + e_{A'}{}^\rho \big(\partial_\mu e_\nu{}^{A'} - \omega_\mu{}^{A'B'} e_{\nu B'} + \omega_\mu{}^{AA'}\tau_{\nu A}\big)\,.
\end{align}
Apart from the ``metric-compatible'' covariant derivative $\nabla_\mu$ as defined in \eqref{eq:taucomp} and \eqref{eq:ecomp} we use two more covariant derviatives for the local Galilean and dilatation symmetries. It is useful to distinguish between a derivative $\mathcal D_\mu = \mathcal D_\mu(\omega, \omega^{A'B'},b)$ that is covariant with respect to $\mathrm{SO}(1,1)\times \mathrm{SO}(8)$ and dilatations, and one $D_\mu=D_\mu(\omega, \omega^{A'B'}, \omega^{AA'},b)$ that is covariant with respect to all local Galilean symmetries and dilatations. In other words, the two derivatives differ by a covariantization with respect to Galilean boosts, schematically
\begin{align}
D_\mu = \mathcal D_\mu - \delta_G\big(\omega_\mu{}^{AA'}\big)\,.
\end{align}
To clarify the difference, let us write \eqref{eq:ecomp} in the following equivalent ways: $\nabla_\mu e_\nu{}^{A'} = D_\mu e_\nu{}^{A'} - \Gamma_{\mu\nu}^\rho e_\rho{}^{A'} = \mathcal D_\mu e_\nu{}^{A'} + \omega_\mu{}^{AA'}\tau_{\nu A}- \Gamma_{\mu\nu}^\rho e_\rho{}^{A'}$.

Finally, we give the full covariant curvatures corresponding to the gauge field of Dilatations $(D)$, $\mathrm{SO}(1,1)-$rotations $(M)$, Galilean boosts $(G)$, and $\mathrm{SO}(8)-$rotations $(J)$. These are the expressions that appear in the bosonic equations of motion:
\begin{subequations}\label{eq:curv2forms}
\begin{align}
\rmR_{\mu\nu}(D)         &=
2\,\partial_{[\mu}b_{\nu]} {+ 2\,e_{[\mu}{}^{A'}\omega_{\nu]}{}^{BB'}}\,\tau_{A'B'B} + 2\,\tau_{[\mu}{}^A\,\omega_{\nu]A}{}^{A'}\,\nabla _{A'}\phi\,,\\
\rmR_{\mu\nu}(M)         &=
2\,\partial_{[\mu}\omega_{\nu]}+ 2\,\epsilon_{AB}\,e_{[\mu}{}^{A'}\omega_{\nu]}{}^{AB'}\,\tau_{A'B'}{}^{B}\notag\\
&\quad - 4\,\tau_{[\mu}{}^A\omega_{\nu]}{}^{BB'}\,\epsilon_B{}^C\tau_{B'\{AC\}} + 2\,\epsilon_{AB}\,\tau_{[\mu}{}^A\omega_{\nu]}{}^{BB'}\,\nabla _{B'}\phi\,,\\
\rmR_{\mu\nu}(G)^{AA'}   &=
2\,\partial_{[\mu}\omega_{\nu]}{}^{AA'}-2\,\epsilon^A{}_B\,\omega_{[\mu}{}\omega_{\nu]}{}^{BA'} - 2\,\omega_{[\mu}{}^{A'B'}\omega_{\nu]}{}^A{}_{B'} + 2\,b_{[\mu}{} \omega_{\nu]}{}^{AA'}\notag\\
&\quad - 4\,e_{[\mu}{}^{B'}\big(\omega_{\nu]B}{}^{[A'}\tau^{B']\{AB\}} -\frac14\,\epsilon^{AB}\,\omega_{\nu]BC'}\,h^{A'B'C'}\big) \,,\\
\rmR_{\mu\nu}(J)^{A'B'}  &=
2\,\partial_{[\mu}\omega_{\nu]}{}^{A'B'} + 2\,\omega_{[\mu}{}^{A'C'}\omega_{\nu]}{}^{B'}{}_{C'}\notag\\
&\quad +2\,e_{[\mu}{}^{C'}\big( 2\,\omega_{\nu]}{}^{C[A'}\tau^{B']}{}_{C'C}{-\omega_{\nu]CC'}\tau^{A'B'C}}\big) \notag\\
&\quad+8\,\tau_{[\mu}{}^A\big(\omega_{\nu]}{}^{B[A'}\tau^{B']}{}_{\{AB\}}-\frac18\,\epsilon_{A}{}^B\,\omega_{\nu]BC'}h^{A'B'C'}\big)\,.
\end{align}
\end{subequations}
These curvatures satisfy a number of non-trivial Bianchi identities, some of which can be found in \cite{Bergshoeff:2021bmc}. There are several $\mathrm{SO}(1,1)\times \mathrm{SO}(8)-$singlets that one can define from the above, for example
\begin{align}\label{eq:RJ}
\rmR(J) \equiv -e_{A'}{}^\mu e_{B'}{}^\nu\,\rmR_{\mu\nu}(J)^{A'B'}\,,
\end{align}
which appears explicitly in the pseudo-action \eqref{eq:BosAction}.

\section{Bosonic Equations of Motion and Killing Spinor Equations} \label{sec:KSeqs}

\noindent Here, we provide the bosonic truncation of both the NR bosonic field equations and the transformation rules of the fermionic fields. The truncated bosonic field equations comprise the field equations that follow from the variation of the NR action in \eqref{eq:actionexpansion},
\begin{subequations}\label{eq:nrBEOM}
\begin{align}
\langle \phi\rangle   &
=\nabla^{A'}\nabla_{A'}\phi - \big(\nabla_{A'}\phi \big)^2 + \frac14\,\rmR(J) - \frac{1}{48}\,h_{A'B'C'}h^{A'B'C'} - \tau_{A'\{AB\}}\tau^{A'\{AB\}}\,,\label{eq:phi}\\
\langle \tau\rangle _{\{AB\}} &
= 4\,\big(\nabla_{B'} - 2\,(\nabla_{B'}\phi)\big)\tau^{B'}{}_{\{AB\}}\,,\label{eq:STL}\\
\langle \tau \rangle _{AA'} &
=2 \, \rmR_{AC'}(J)_{A'}{}^{C'} -4 \,\nabla_A\nabla_{A'}\phi -4\,\nabla^B\tau_{A'\{AB\}}\,,\label{eq:V-}\\
\langle e\rangle _{A'B'} &
= -2\, \rmR_{C'(A'}(J)_{B')}{}^{C'} -4\,\nabla_{(A'}\nabla_{B')}\phi + \frac12\,h_{A'C'D'}h_{B'}{}^{C'D'}\nonumber \\&\quad\, + 8\,\tau_{A'\{AB\}}\tau_{B'}{}^{\{AB\}}+4\,\delta_{A'B'}\langle\phi \rangle \,,\label{eq:GA'B'}\\
\langle e\rangle _{A'A} &
= 2\,\big(\nabla_{B'} -2\,( \nabla_{B'}\phi)\big)\,\tau^{B'}{}_{A'A} + 4\,\tau^{B'}{}_{\{AB\}}\tau_{B'A'}{}^B + \epsilon_{AB}\,h_{A'B'C'}\tau^{B'C'B}\,,\label{eq:V+}\\
\langle b\rangle_{AB}  &
= 2\,\epsilon_{AB}\,\tau_{A'B'C}\tau^{A'B'C}\,,\label{eq:S-}\\
\langle b\rangle _{A'B'} &
= \big(\nabla_{C'} - 2\,(\nabla_{C'}\phi)\big) h^{C'}{}_{A'B'} + 2\,\epsilon^{AB}\,\nabla_A\tau_{A'B'B}\,,\label{eq:BA'B'}
\end{align}   \end{subequations}
and the Poisson equation that follows from the supersymmetry transformation of the missing fermionic field equations, as explained in section \ref{sec:consistencyofEOM},
\begin{align} 
\langle P\rangle &
\equiv \rmR_{AA'}(G)^{AA'} + \epsilon^{AB} \rmR_{AB}(M) \ \, = 0\,,
\end{align}
where the metric-compatible covariant derivative $\nabla_\mu$ and the covariant curvatures with respect to the NR bosonic symmetries are defined in appendix \ref{sec:TSNC}.

The Killing spinor equations of the NR minimal supergravity are given by \begin{subequations}
\begin{align}
\delta \lambda_{-}&=\nabla_{A'}\phi\Gamma^{A'}\epsilon_{-}-\frac{1}{12}\,h_{A'B'C'}\Gamma^{A'B'C'}\epsilon_{-}+\eta_-\,=0\,,\\
\delta\lambda_{+}&=\nabla_{A'}\phi\Gamma^{A'}\epsilon_{+}-\frac{1}{12}\,h_{A'B'C'}\Gamma^{A'B'C'}\epsilon_{+}+\frac{1}{2}\,\tau^{A'B'+}\Gamma_{A'B'+}\epsilon_{-}\,=0\,,\\
\delta \psi_{\mu -}&=\mathcal D_{\mu}\epsilon_{-}-\frac12\,\omega_{\mu}{}^{-A'}\Gamma_{-A'}\epsilon_+-\frac{1}{8}\,e_{\mu}{}^{C'}h_{A'B'C'}\Gamma^{A'B'}\epsilon_{-}+\tau_{\mu}{}^+ \rho_-\,=0\,,\\
\delta \psi_{\mu +}&=\mathcal D_{\mu}\epsilon_{+}-\frac{1}{8}\,e_{\mu}{}^{C'}h_{A'B'C'}\Gamma^{A'B'}\epsilon_{+}+(e_{\mu}{}^{B'}\tau{}_{B'}{}^{A'+}+\tau_{\mu}{}^-\tau^{A'++})\Gamma_{A'+}\epsilon_{-}+\frac12\,\tau_{\mu}{}^+\Gamma_{+}\eta_-\,=0\,.
\end{align}
\end{subequations}
The fermionic completion of the bosonic field equations in \eqref{eq:nrBEOM} can be easily obtained from the NR action and does not play a role in defining background solutions. However, the complete Poisson equation can only be derived from the NR action indirectly. Thus, we give terms quadratic in fermions in the Poisson equation for completeness:
\begin{subequations}
\begin{align}
\langle P\rangle \bigg\vert_{\lambda\lambda}&=-2\,\bar{\lambda}_{-}\Gamma^{A}D_{A}\lambda_{-}\,,\\
\nonumber\\
\langle P\rangle\bigg\vert_{\lambda\psi}&=4\,\bar{\lambda}_{-}\Gamma^{AB'}\tau{}_{A}{}^{\mu}e_{B'}{}^{\nu}D_{[\mu}\psi_{\nu ]-}
+3\,\bar{\lambda}_{-}\Gamma^{AB}\tau{}_{A}{}^{\mu}\tau{}_{B}{}^{\nu}D_{[\mu}\psi_{\nu] +}+\bar{\lambda}_{+}\Gamma^{AB}\tau{}_{A}{}^{\mu}\tau{}_{B}{}^{\nu}D_{[\mu}\psi_{\nu] -}\nonumber\\
&\quad\,+2\,\bar{\psi}_{B'-}\Gamma^{AB'}D_{A}\lambda_{-}-2\,\bar{\psi}_{A-}\Gamma^{AB'}D_{B'}\lambda_{-}+3\,\bar{\psi}_{B-}\Gamma^{AB}D_{A}\lambda_{+}
+\bar{\psi}_{B+}\Gamma^{AB}D_{A}\lambda_{-}\nonumber\\
&\quad\,+D^{A}(\bar{\lambda}_{\pm}\psi_{A\mp})+\frac{1}{6}\,h_{A'B'C'}(\bar{\psi}_{D-}\Gamma^{DA'B'C'}\lambda_{-})\,,\\
\nonumber\\
\langle P\rangle\bigg\vert_{\psi\psi}&= D^{A}(\bar{\psi}_{B'\pm}\Gamma^{B'}\psi_{A\mp}+2\bar{\psi}_{B+}\Gamma^{B}\psi_{A+})+D^{C'}(2\,\bar{\psi}_{A-}\Gamma^{A}\psi_{C'-}+\bar{\psi}_{A+}\Gamma^{C'AB}\psi_{B-})\nonumber\\
&\quad\,+2\,\bar{\psi}_{A-}\Gamma^{AB'C'}e{}_{B'}{}^{\mu}e{}_{C'}{}^{\nu}D_{[\mu}\psi_{\nu]-}+4\,\bar{\psi}_{C'-}\Gamma^{AB'C'}\tau{}_{A}{}^{\mu}e{}_{B'}{}^{\nu}D_{[\mu}\psi_{\nu]-}\nonumber\\
&\quad\,+\bar{\psi}_{C'+}\Gamma^{ABC'}\tau{}_{A}{}^{\mu}\tau{}_{B}{}^{\nu}D_{[\mu}\psi_{\nu]-}+2\,\bar{\psi}_{B+}\Gamma^{ABC'}e{}_{C'}{}^{\mu}\tau{}_{A}^{\nu}D_{[\mu}\psi_{\nu]-}\nonumber\\
&\quad\,+3\,\bar{\psi}_{C'-}\Gamma^{ABC'}\tau{}_{A}{}^{\mu}\tau{}_{B}{}^{\nu}D_{[\mu}\psi_{\nu]+}+6\,\bar{\psi}_{A-}\Gamma^{ABC'}\tau{}_{B}{}^{\mu}e_{C'}{}^\nu D_{[\mu}\psi_{\nu]+}\nonumber\\
&\quad\,+4\,\epsilon^{AB} D_{C'}\phi(\bar{\psi}_{A+}\Gamma^{C'}\psi_{B-})-4\,\tau_{C'}{}^{\{AB\}}(\bar{\psi}_{C'-}\Gamma_{A}\psi_{B-}) \nonumber\\
&\quad\,+\frac{1}{4}\,h_{A'B'C'}(\bar{\psi}_{D'-}\Gamma^{A'B'C'D'E}\psi_{E-})+\frac{1}{6}\,h_{A'B'C'}(\bar{\psi}_{D+}\Gamma^{A'B'C'DE}\psi_{E-})\,.
\end{align}
\end{subequations}

\section{Closure of the Non-Relativistic Super-Algebra}\label{sec:algebra}
\noindent This section gives some details on how the algebra is realized on fields by describing some commutation relations involving the zeroth order of relativistic supersymmetry rules (denoted with $\delta_{Q}^{(0)}$) and some commutators involving $S$ and $T$-symmetries. We recall that $\delta_{Q}^{(0)}$ coincides with non-relativistic supersymmetry for the bosonic fields while it contains also extra $S$- and $T$-symmetry transformations for fermionic fields, see \eqref{eq:deltasusy0}. We will not list the fully field-dependent parameters. It is to be understood that the realizations below hold provided the self-dual DSNC constraint is imposed.

\paragraph[Boost-$\delta^{(0)}$]{Boost-\boldmath $\delta^{(0)}$}
The commutator between boosts and  $\delta_{Q}^{(0)}$ closes on symmetries as follow
\begin{subequations}
\begin{align}
[\delta_{G}(\lambda^{AB'}),\delta^{(0)}_{Q}(\epsilon_{-})]&=\delta_{T}(\rho'_{-})\ , \\
[\delta_{G}(\lambda^{AB'}),\delta^{(0)}_{Q}(\epsilon_{+})]&=\delta^{(0)}_{Q}(\epsilon'_{-})+\delta_{S}(\eta'_{-})+\delta_{T}(\rho''_{-})\ ,
\end{align}\label{eq:boostsusy0}
\end{subequations}
with parameters
\begin{subequations}
\begin{align}
\epsilon'_{-}&=\frac{1}{2}\lambda^{A'-}\Gamma_{-A'}\epsilon_{+}\ , \\
\rho'_{-}&=\frac{1}{2}\Big(\partial_{A'}\lambda_{B'}{}^{-}+\lambda_{C'}{}^{-}e_{A'B'}{}^{C'}+2\lambda^{-}{}_{A'}\tau_{-B'}{}^{-}-\lambda^{-C'}\overline{\psi}_{A' -}\Gamma_{C'}\psi_{B' +}\Big)\Gamma^{A'B'}\epsilon_{-}\ , \\
\rho''_{-}&=\frac{1}{2}\Big(\partial^{-}\lambda^{A'-}+2\lambda^{B'-}e_{+}{}^{A'B'}+2\lambda^{A'-}\tau_{+-}{}^{-}+2\lambda^{-B'}\tau^{\nu}{}_{+}e^{\rho A'}\overline{\psi}_{[\nu -}\Gamma_{B'}\psi_{\rho] +}\Big)\Gamma_{-A'}\epsilon_{+}\ , \\
\eta'_{-}&=\frac{1}{2}\Big(\partial_{A'}\lambda_{B'}{}^{-}  +\lambda_{C'}{}^{-}e_{A'B'}{}^{C'}+2\lambda^{-}{}_{A'}\tau_{-B'}{}^{-}-\lambda^{-C'}\overline{\psi}_{A'-}\Gamma_{C'}\psi_{B'+}\Big)\Gamma^{+A'B'}\epsilon_{+}\ .
\end{align}
\end{subequations}
Only one component of the boost parameter, $\lambda^{-A'}$ appears in the RHS of \eqref{eq:boostsusy0}. If $\lambda^{-A'}=0$ the commutators above reduce to an Abelian algebra.

\paragraph[Dilatation-$\delta^{(0)}$]{Dilatation-\boldmath $\delta^{(0)}$}
The commutator between the emerging dilatation symmetry and  $\delta_{Q}^{(0)}$ closes as follows
\begin{subequations}
\begin{align}
[\delta_{D}(\lambda_{D}),\delta^{(0)}_{Q}(\epsilon_{+})]&=\delta^{(0)}_{Q}(\epsilon'_{+})+\delta_{S}(\eta'_{-})\ , \\
[\delta_{D}(\lambda_{D}),\delta^{(0)}_{Q}(\epsilon_{-})]&=\delta^{(0)}_{Q}(\epsilon'_{-})+\delta_{S}(\eta''_{+})+\delta_{T}(\rho_{-})\ ,
\end{align}
\end{subequations}
where
\begin{subequations}
\begin{align}
\epsilon'_{\pm}&=\mp\frac{1}{2}\lambda_{D} \epsilon_{\pm}\ , & \rho_{-}&=-\partial_{+}\lambda_{D}\epsilon_{-}\ , \\
\eta'_{-}&=\Gamma^{A}\tau^{\mu}{}_{A}\partial_{\mu}\lambda_{D} \epsilon_{+}\ , & \eta''_{-}&=2\Gamma^{A'}\epsilon_{-}\partial_{A'}\lambda_{D}\ .
\end{align}
\end{subequations}


\paragraph[S-symmetry-$\delta^{(0)}$]{S-symmetry-\boldmath $\delta^{(0)}$}
The commutator between S-symmetry and  $\delta_{Q}^{(0)}$ gives
\begin{subequations}
\begin{align}
[\delta_{S}(\eta_{-}),\delta_{Q}^{(0)}(\epsilon_{-})]&=\delta_{G}(\lambda_{(S-)}^{AB'})+\delta_{S}(\eta_{(S-)})+\delta_{T}(\rho_{(S-)})\ , \\
[\delta_{S}(\eta_{-}),\delta_{Q}^{(0)}(\epsilon_{+})]&=\delta_{L}(\lambda_{(S+)}^{AB})+\delta_{D}(\lambda_{D(S+)})+\delta_{S}(\eta_{(S+)})+\delta_{T}(\rho_{(S+)})\ ,\label{eq:Ssusy}
\end{align}
\end{subequations}
where
\begin{subequations}
\begin{align}
\lambda_{(S-)}^{A'+}&=0\ ,& \lambda_{(S-)}^{A'-}&=\frac{1}{2}\overline{\epsilon}_{-}\Gamma^{A'-}\eta_{-}\ ,\\
\lambda_{(S+)}^{AB}&=-\frac{1}{2}\epsilon^{AB}(\overline{\epsilon}_{+}\eta_{-})\ ,& \lambda_{D(S+)}&=\frac{1}{2}(\overline{\epsilon}_{+}\eta_{-})
\end{align}
\end{subequations}
We note that dilatation appears only in the right hand side of \eqref{eq:Ssusy}.


\paragraph[T-symmetry-$\delta^{(0)}$]{T-symmetry-\boldmath$\delta^{(0)}$}
Commutation relations between T-symmetry and  $\delta_{Q}^{(0)}$ close as follow
\begin{subequations}
\begin{align}
[\delta_{T}(\rho_{-}),\delta^{(0)}_{Q}(\epsilon_{+})]&=\delta_{G}(\lambda_{(T+)}^{AB'})+\delta_{S}(\eta_{(T+)-})+\delta_{T}(\rho_{(T+)-})\ , \\
[\delta_{T}(\rho_{-}),\delta^{(0)}_{Q}(\epsilon_{-})]&=\delta_{T}(\rho_{(T-)-})\ ,
\end{align}
\end{subequations}
with
\begin{align}
\lambda_{(T+)}^{A'+}=0\ ,\qquad \lambda_{(T+)}^{A'-}=-\overline{\epsilon}_{+}\Gamma^{A'}\rho_{-}\ .
\end{align}


\paragraph[$\delta^{(0)}$-$\delta^{(0)}$]{\boldmath$\delta^{(0)}$-$\delta^{(0)}$}
The commutators between two  $\delta_{Q}^{(0)}$ transformations close on the algebra as
\begin{subequations}
\begin{align}
[\delta_{Q}^{(0)}(\eta_{+}),\delta_{Q}^{(0)}(\epsilon_{+})]&=\mathcal{L}_{\xi_{(++)}}+\delta_Q^{(0)}(\epsilon_{(++)+})+\delta_Q^{(0)}(\epsilon_{(++)-})+\delta_{L}(\lambda_{(++)})+\nonumber\\
&+\delta_{\theta}(\theta_{(++)})+\langle EOM\rangle\ , \label{eq:commutatorQ+Q+2}\\
[\delta_{Q}^{(0)}(\eta_{+}),\delta_{Q}^{(0)}(\epsilon_{-})]&=\mathcal{L}_{\xi_{(+-)}}+\delta_Q^{(0)}(\epsilon_{(+-)+})+\delta_Q^{(0)}(\epsilon_{(+-)-})+\delta_{L}(\lambda_{(+-)})+\nonumber\\
&+\delta_{\theta}(\theta_{(+-)})+\delta_{T}(\rho_{(+-)-})+\delta_{S}(\eta_{(+-)-})+\langle EOM\rangle\ , \label{eq:commutatorQ+Q-2}\\
[\delta_{Q}^{(0)}(\eta_{-}),\delta_{Q}^{(0)}(\epsilon_{-})]&=\delta_Q^{(0)}(\epsilon_{(--)-})+\delta_{L}(\lambda_{(--)})+\delta_{\theta}(\theta_{(--)})+\delta_{T}(\rho_{(--)-})+\langle EOM\rangle\ , \label{eq:commutatorQ-Q-2}
\end{align}
\end{subequations}
where $\delta_\theta$ denotes the $1-$form gauge symmetry $\delta_\theta b_{\mu\nu}=2\,\partial_{[\mu}\theta_{\nu]}$. The symbol $\delta_{L}$ collectively denotes longitudinal $\mathrm{SO}(1,1)$ Lorentz, transverse $\mathrm{SO}(8)$ rotations, and Galilean boost transformations,
\begin{subequations}
\begin{align}
\xi^{\mu}_{(++)}&=\xi^{A}_{(++)}\tau_A{}^{\mu}\ , &\xi^{\mu}_{(+-)}&=\xi^{A'}_{(+-)}e_{A'}{}^{\mu}\ , &\lambda^{AB}_{(--)}&=0\ , \\
\theta_{(++)\mu}&=b_{\mu\nu}\xi^\nu_{(++)}\ , &
\theta_{(--)\mu}&=-2\,\tau_{\mu A}\xi^{A}_{(--)}\ , &
\theta_{(+-)\mu}&=b_{\mu\nu}\xi^\nu_{(+-)}-e_{\mu A'}\xi^{A'}_{(+-)}\ , \
\end{align}
\end{subequations}

\noindent and we have used $\mathcal{L}_{X}$ to denote Lie derivative, $\langle EOM\rangle$ to define terms proportional to the equations of motion, and the notation
\begin{align}
&\xi^A_{(++)} = \bar\epsilon_+\Gamma^A\eta_+\,,&&\xi^A_{(--)}=\bar\epsilon_-\Gamma^A\eta_-\,,&&\xi^{A'}_{(+-)} = \bar\epsilon_-\Gamma^{A'}\eta_+\,.
\end{align}
\paragraph{Dilatation-S/T Symmetry}
The commutation rules between dilatation and S- and T-Symmetries are
\begin{align}
[\delta_{D}(\lambda_{D}),\delta_{T}(\rho_{-})]&=\delta_{T}\bigg(\frac{3}{2}\lambda_{D}\rho_{-}\bigg)\ , &
[\delta_{D}(\lambda_{D}),\delta_{S}(\eta_{-})]&=\delta_{S}\bigg(\frac{1}{2}\lambda_{D}\eta_{-}\bigg)\ .
\end{align}
\paragraph{Boost-S/T Symmetry}
The commutation rules between boost and S and T-symmetries are
\begin{align}
[\delta_{G}(\lambda_{AB'}),\delta_{T}(\rho_{-})]&=0\ , &
[\delta_{G}(\lambda_{AB'}),\delta_{S}(\eta_{-})]&=\delta_{T}\bigg(-\frac{1}{4}\lambda_{AB'}\Gamma^{AB'}\Gamma_{+}\eta_{-}\bigg)\ .
\end{align}
\paragraph{S/T Symmetry-S/T Symmetry}\label{sec:algstst}
S and T-symmetries define an Abelian algebra:
\begin{align}
[\delta_{T}(\rho_{-}),\delta_{T}(\rho'_{-})]&=0\ , &
[\delta_{S}(\eta_{-}),\delta_{S}(\eta'_{-})]&=0\ , &
[\delta_{T}(\rho_{-}),\delta_{S}(\eta_{-})]&=0\ .
\end{align}

\section{\boldmath Non-Relativistic $D=10$ Super-Yang-Mills}\label{sec:NRSYM}

\noindent In this section, we present a consistent non-relativistic string limit of ten-dimensional super Yang-Mills (SYM) theory \cite{Brink:1976bc} in flat space. Moreover, we will show that the multiplet structure is compatible with the soft algebra derived in section \ref{sec:algebra}. Thus it is suggestive that one can, in principle, couple this theory via a Noether procedure to $\mathcal N=1$ supergravity along the lines of \cite{Bergshoeff:1981um}. Essentially, this section is independent of the rest of the paper. We do, however, use the same notation and conventions as explained in \ref{sec:conventions}.\\
We will show that there is one unique re-scaling Ansatz giving rise to a regular limit $c\to\infty$. Remarkably, this requires an an-isotropic re-scaling of the gauge field. Typically, such an Ansatz would break the spacetime symmetries of the theory. However, since we are working with a 2-foliation structure, we can introduce an-isotropic re-scalings in the longitudinal lightcone directions, which is consistent with the diagonal $\mathrm{SO}(1,1)\times\mathrm{SO}(8)$ part of Lorentzian symmetries. We show that this leads to a theory with $16$ supercharges, defined on flat space with a two-dimensional foliation. We give the re-scaling Ansatz \eqref{eq:genericAnsatz}/\eqref{eq:parchoice}, the non-relativistic multiplet structure \eqref{eq:nrymsusy}, and the action for NR SYM \eqref{eq:NRSYM}. \\
Let us start by reviewing the relativistic theory \cite{Brink:1976bc}. The ten-dimensional SYM on-shell multiplet is described by a gauge field $A_\mu{}^I$ and a gaugino $\mathfrak X^I$, both in the adjoint representation of some gauge group $\mathcal G$. The gaugino is a right-handed Majorana-Weyl fermion. The action
\begin{align}\label{eq:SYMaction}
S = \frac{1}{g^2}\int\,d^{10}X\,\Big\{-\frac14\,F_{\mu\nu}{}^I F^{\mu\nu I} - \bar{\mathfrak X}^I\slashed D\mathfrak X^I\Big\}
\end{align}
is invariant under $16$ supercharges with parameter $\varepsilon$
\begin{align}
\delta A_\mu{}^I = \bar\varepsilon\,\Gamma_\mu\,\mathfrak X^I\,,\hskip 2.5 truecm \delta \mathfrak X^I = -\frac14\,\Gamma^{\mu\nu}\varepsilon\,F_{\mu\nu}{}^I\,,
\end{align}
where $D_\mu\mathfrak X^I = \partial_\mu\mathfrak X^I + f_{JK}^I\,A_\mu{}^J\mathfrak X^K$, and $F_{\mu\nu}{}^I = 2\,\partial_{[\mu}A_{\nu]}{}^I + f^I_{JK}A_\mu{}^J A_\nu{}^K$. \\
Let us now define a non-relativistic limit by introducing a (dimensionless) contraction parameter $c$. We choose this Ansatz such that the limit $c\to\infty$ is well-defined both in the symmetry transformation rules and the action. In order to get Galilean boost symmetries with parameter $\lambda^{AA'}=c^{-1}\Lambda^{AA'}$ we rescale the flat coordinates $X^\mu = (X^0,X^1,X^{A'=2,\cdots,9})$ as $X^\pm = (X^0\pm X^1)/\sqrt 2 = c\,x^\pm$ and $X^{A'}=x^{A'}$. (This can be seen as the flat space limit of \eqref{eq:rescale}.) It is not a priori clear how to choose a consistent Ansatz for the fields $(A_\mu{}^I,\mathfrak X^I)$---hence we parametrize different choices with four to-be determined parameters $\alpha/\beta/\gamma/\delta$
\begin{align}\label{eq:genericAnsatz}
A_\mu{}^I\,\rmd X^\mu = c^\alpha\,a^I\,\rmd x^+ + c^\beta\,b{}^I\,\rmd x^- + c^\gamma\,c_{A'}{}^I\rmd x^{A'}\,,\qquad \mathfrak X^I = c^{\delta + 1/2}\,\chi_+^I + c^{\delta-1/2}\,\chi_-^I\,,
\end{align}
which is equivalent to expressing $A_\mu{}^I = (A_+{}^I,A_-{}^I,A_{A'}{}^I)$ as $a^I = c^{1-\alpha}A_+{}^I$, $b=c^{1-\beta}A_-{}^I$, $c_{A'}{}^I = c^{-\gamma}\,A_{A'}{}^I$, and $\chi_\pm^I = c^{-\delta\mp 1/2}\Pi_{\pm}\mathfrak X^I$. Note that the re-scaling of $\chi_+^I$ and $\chi_-^I$ differ by a relative power of $c^1$. This has been chosen so that the fermions transform appropriately under Galilean boosts, see \eqref{eq:ymboost}.\\
Let us now constrain the parameters $\alpha/\beta/\gamma/\delta$ by requiring that the limit $c\to\infty$ is well-defined. In other words, we choose the parameters such that there are no positive powers of $c$ in the symmetry rules. This gives a number of constraints---such as $\beta\geq 2$ from $\delta b^I = c^{2-\beta}\,\bar\epsilon_+\Gamma_-\chi_+^I$---that are enough to uniquely fix
\begin{align}\label{eq:parchoice}
&\alpha =0\,, &&\beta = 2\,, && \gamma=0\,, &&\delta =0\,,
\end{align}
It is remarkable that we are led to an an-isotropic limit, where the gauge field in the $X^-/x^-$ direction $b^I = c^{-1} A_-^I$ is re-scaled differently from the rest---see \eqref{eq:genericAnsatz}. Note, that this is only compatible with the bosonic symmetries of the theory due to the lightlike nature of this direction since under longitudinal $\mathrm{SO}(1,1)$ rotations $\delta A_\pm{}^I = \pm\,\Lambda_M\,A_\pm{}^I$. It would be interesting to see whether similar limits can be taken for other foliation structures, too.\\
Let us now consider the non-relativistic symmetries that follow after taking the limit $c\to\infty$. First of all, the an-isotropic rescaling implies that the $b^I = c^{-1}\,A_-^I$ field becomes a matter field under gauge transformations
\begin{align}\label{eq:nrymtrafo}
&\delta_\theta a^I = \partial_+\theta^I - f_{JK}^I \,\theta^J a^K\,, && \delta_\theta c_{A'} = \partial_{A'}\theta^I - f_{JK}^I\theta^J c_{A'}{}^K\,,\notag\\
&\delta_\theta b^I =  -f_{JK}^I \,\theta^J a^K\,, && \delta_\theta \chi_\pm^I = -f^I_{JK}\theta^J\chi_\pm^K\,,
\end{align}
which motivates the introduction of the following covariant expressions
\begin{subequations}
\begin{align}
f_{A'B'}{}^I &\equiv 2\,\partial_{[A'}c_{B']}{}^I + f_{JK}^I\,c_{A'}{}^J c_{B'}{}^K\\
f_{A'}{}^I &\equiv \partial_+ c_{A'}{}^I - \partial_{A'} a^I + f_{JK}^Ia^Jc_{A'}{}^K \,,\\
D_{A'} b^I &\equiv \partial_{A'} b^I +f_{JK}^Ic_{A'}{}^J b^K\,,\\
D_+ b^I &\equiv \partial_+ b^I + f_{JK}^I a^J b^K\,,
\end{align}
\end{subequations}
and similarly for the fermions. Here and in the following we take $\partial_{A'}=\partial/\partial x^{A'}$ and $\partial_\pm = \partial/\partial x^\pm$. Similarly, it is not hard to see that the limit of the action \eqref{eq:SYMaction} is well defined with $S_{NR}=\lim_{c\to\infty} c^{-2}\,S$,\footnote{Alternatively one could define an effective non-relativistic coupling $g_{YM}=c^{-1}g$.} and explicitly given by
\begin{align}\label{eq:NRSYM}
S_{NR} = \frac{1}{g^2}\int\,\rmd^{10} x\,\Big\{&-\frac14\,f_{A'B'}{}^If^{A'B'I} -f_{A'}{}^I D_{A'}b^I + D_+b^I\,D_+b^I\notag\\
&-2\,\bar\chi_-^I\gamma^{A'}D_{A'}\chi_+^I - \bar\chi_+^I\Gamma^+D_+\chi_+^I + f_{IJK}b^J\,\bar\chi_-^I\Gamma^-\chi_-^K\Big\}\,.
\end{align}
Remarkably this action (and, relatedly, none of the equations of motion) contains a derivative in the $x^-$ direction. This observation makes it tempting to perform a dimensional reduction along the $\partial/\partial x^1$ direction leading to a theory in nine dimensions with a one-dimensional foliation structure. Some aspects of such Galilean gauge theories have been studied in \cite{Bergshoeff:2015sic, Festuccia:2016caf,Gomis:2020fui,Chapman:2020vtn,Lambert:2019jwi}. \\
Let us now study the symmetries of the non-relativistic SYM action \eqref{eq:NRSYM}. The theory is manifestly invariant under Yang-Mills tranformations \eqref{eq:nrymtrafo} and Galilean symmetries. The action of the diagonal part $\mathrm{SO}(1,1)\times\mathrm{SO}(8)$ is as expected from the index structure. The Galilean boosts, however, act non-trivially
\begin{align}\label{eq:ymboost}
&\delta_G a^I = -\lambda^{-A'}c_{A'}{}^I\,,&& \delta_G c_{A'}{}^I = -\lambda^{-A'}b^I\,,&& \delta_G b^I =0\,, \notag\\
&\delta_G \chi_-^I = \frac12\,\lambda^{-}{}_{A'}\Gamma_{-A'}\chi_+^I\,,&& \delta_G \chi_+^I=0\,,
\end{align}
showing that the multiplet forms a reducible, yet in-decomposable representation of the Galilei algebra. Furthermore, just as the relativistic parent theory, the non-relativistic action \eqref{eq:NRSYM} is invariant under $16$ supercharges with parameters $(\epsilon_+,\epsilon_-)$, and transformation rules
\begin{subequations}
\label{eq:nrymsusy}
\begin{align}
\delta a^I &= \bar{\epsilon}_-\Gamma_+\chi_-^I\,,\\
\delta b^I &= \bar\epsilon_+\Gamma_-\chi_+^I\,,\\
\delta c_{A'}^I &= \bar\epsilon_+\Gamma_{A'}\chi_-^I + \bar\epsilon_-\Gamma_{A'}\chi_+^I\,,\\
\delta \chi_+^I &= -\frac{1}{4}\,\Gamma^{A'B'}\epsilon_+ f_{A'B'}{}^I + \frac{1}{2}\Gamma^{-A'}\epsilon_-\,D_{A'}b^I - \frac{1}{2}\,\epsilon_+\,D_+b^I \,,\\
\delta\chi_-^I &= -\frac{1}{4}\,\Gamma^{A'B'}\epsilon_- f_{A'B'}{}^I - \frac{1}{2}\Gamma^{+A'}\epsilon_+\,f_{A'}{}^I +\frac12\,\epsilon_-\,D_+b^I\,.
\end{align}
\end{subequations}
It is not hard to see that these transformations close on the symmetries of the theory. Boost commute with supersymmetry as follows
\begin{align}
\big[\delta_G(\lambda^{AA'}),\delta(\epsilon_+)\big] = \delta\big(\epsilon_-'=\frac12\,\lambda^{-A'}\Gamma_{-A'}\epsilon_+\big)\,,\qquad \big[\delta_G(\lambda^{AA'}),\delta(\epsilon_-)\big] = 0\,,
\end{align}
which shows that one can in principle truncate to a theory with $8$ supercharges by setting $\epsilon_+=0$. However, this contraction does not allow for an interesting superalgebra closing on spacetime translations, as can be seen from
\begin{subequations}
\begin{align}
\big[\delta(\eta_+),\delta(\epsilon_+)\big] &= \xi_{(++)}\partial_+ + \delta_\theta(-\xi_{(++)}a^I)\,,\\
\big[\delta(\eta_-),\delta(\epsilon_-)\big] &=\delta_\theta(-\xi_{(--)}b^I)\,,\\
\big[\delta(\eta_+),\delta(\epsilon_-)\big] &= \xi_{(+-)}^{A'}\partial_{A'} + \delta_\theta(-\xi_{(+-)}^{A'}c_{A'}{}^I)\,,
\end{align}
\end{subequations}
where $\xi_{(++)}=\bar\epsilon_+\Gamma^+\eta_+$, $\xi_{(--)}=\bar\epsilon_-\Gamma^-\eta_-$, and $\xi^{A'}_{(+-)}=\bar\epsilon_-\Gamma^{A'}\eta_+$. Just as in the relativistic theory these commutators close off-shell on the bosonic fields and on-shell on the fermionic ones. Remarkably this superalgebra is compatible with the $\mathcal N=1$ superalgebra derived in section \ref{sec:algebra}. As mentioned above, this is an encouraging observation suggesting that the non-relativistic SYM multiplet can be coupled to the minimal supergravity theory studied in this paper along the lines of \cite{Bergshoeff:1981um}. We hope to return to these questions soon.
%

\providecommand{\href}[2]{#2}\begingroup\raggedright\endgroup

\end{document}